\documentclass[a4paper]{article}

\usepackage{booktabs}
\usepackage[T1]{fontenc}
\usepackage[latin1]{inputenc}
\usepackage{graphicx,fancybox,wrapfig,ifthen,fancyhdr,longtable,color}
\usepackage{ulem}
\usepackage{lscape}
\usepackage{rotating}
\usepackage{cite}
\usepackage{graphicx}
\usepackage{epsfig}
\usepackage{amssymb}
\usepackage{amsmath}
\usepackage{caption}
\usepackage{axodraw}
\usepackage{feynarts}
\usepackage{pstricks}
\usepackage{pst-node}
\usepackage{pst-blur}
\definecolor{Pink}{rgb}{1.,0.75,0.8}
\makeatletter   
\newsavebox{\tempbox}
\renewcommand{\@makecaption}[2]{                         %
  \vspace{10pt}\sbox{\tempbox}{\footnotesize\textbf{#1:}\ #2}         %
  \ifthenelse{\lengthtest{\wd\tempbox > \linewidth}}     %
    {\footnotesize\textbf{#1:} #2\par}                                
    {\begin{center}\footnotesize\textbf{#1:} #2\end{center}}          %
  }

\newcommand{\be}{\begin{equation}}
\newcommand{\ee}{\end{equation}}
\newcommand{\nn}{\nonumber}
\newcommand{\bea}{\begin{eqnarray}}
\newcommand{\eea}{\end{eqnarray}}
\newcommand{\bfig}{\begin{figure}}
\newcommand{\efig}{\end{figure}}
\newcommand{\bc}{\begin{center}}
\newcommand{\ec}{\end{center}}
\newcommand{\bd}{\begin{displaymath}}
\newcommand{\ed}{\end{displaymath}}

\newcommand{\dsl}[1]{\not \hspace{-0.7mm}#1}
\makeatletter
\def\dsl{\mathpalette\make@slash}
\def\make@slash#1#2{\setbox\z@\hbox{$#1#2$}%
  \hbox to 0pt{\hss$#1/$\hss\kern-\wd0}\box0}
\makeatother

\newcommand{\ri}{{\mathrm{i}}}
\newcommand{\rd}{{\mathrm{d}}}

\def\mysp[#1,#2,#3,#4]{\left({#1}_{#2}{#3}_{#4}\right)}

\newcommand{\indexsep}{,}
\def\SP[#1,#2]{\langle{#1}{#2}\rangle}
\def\SPS[#1,#2]{\langle{#1}{#2}\rangle^*}
\allowdisplaybreaks[1]
\newcommand{\Pe}{\mathrm{e}}
\newcommand{\Pg}{\mathrm{g}}
\newcommand{\PZ}{\mathrm{Z}}
\newcommand{\PW}{\mathrm{W}}

\newcommand{\alphas}{\alpha_{\mathrm{s}}}

\newcommand{\fc}{F_{\mathrm{C}}}
\newcommand{\citere}[1]{Ref.~\cite{#1}}
\newcommand{\citeres}[1]{Refs.~\cite{#1}}
\newcommand{\refeq}[1]{(\ref{#1})}
\newcommand{\refeqs}[1]{(\ref{#1})}
\newcommand{\reffig}[1]{Fig.~\ref{#1}}
\newcommand{\reffigs}[1]{Figs.~\ref{#1}}
\newcommand{\refsec}[1]{Section~\ref{#1}}
\newcommand{\refsecs}[1]{Sections~\ref{#1}}
\newcommand{\reftab}[1]{Table~\ref{#1}}

\newcommand{\lrb}{\left(}
\newcommand{\rrb}{\right)}
\newcommand{\lcb}{\left\{}
\newcommand{\rcb}{\right\}}
\newcommand{\lsb}{\left[}
\newcommand{\rsb}{\right]}
\newcommand{\TeV}{\unskip\,\mathrm{TeV}}
\newcommand{\GeV}{\unskip\,\mathrm{GeV}}
\newcommand{\MeV}{\unskip\,\mathrm{MeV}}

\newcommand{\nb}{\unskip\,\mathrm{nb}}

\def\mathswitch#1{\relax\ifmmode#1\else$#1$\fi}
\def\mathswitchr#1{\relax\ifmmode{\mathrm{#1}}\else$\mathrm{#1}$\fi}

\newcommand{\MW}{\mathswitch {M_\PW}}
\newcommand{\PH}{\mathswitchr H}

\newcommand{\MZ}{\mathswitch {M_\PZ}}
\newcommand{\MH}{\mathswitch {M_\PH}}
\newcommand{\Me}{\mathswitch {m_\Pe}}

\newcommand{\Mt}{\mathswitch {m_\Pt}}
\newcommand{\GW}{\Gamma_{\PW}}

\newcommand{\GZ}{\Gamma_{\PZ}}

\newcommand{\LEP}{{\mathrm{LEP}}}

\newcommand{\Pt}{\mathswitchr t}

\def\Ga{\Gamma}
\def\d{\hbox{d}}

\def\e{\epsilon}

\def\Re{\mathop{\mathrm{Re}}\nolimits}
\def\Li{\mathop{\mathrm{Li}_2}\nolimits}

\newcommand{\muF}{\mu_{\mathrm{F}}}

\def\draftdate{\relax}
\def\mda{\relax}
\def\mua{\relax}
\def\mla{\relax}
\def\draft{
\def\thtystars{******************************}
\def\sixtystars{\thtystars\thtystars}
\typeout{}
\typeout{\sixtystars**}
\typeout{* Draft mode!
         For final version remove \protect\draft\space in source file *}
\typeout{\sixtystars**}
\typeout{}
\def\draftdate{\today}
\def\mua{\marginpar[\boldmath\hfil$\uparrow$]%
                   {\boldmath$\uparrow$\hfil}%
                    \typeout{marginpar: $\uparrow$}\ignorespaces}
\def\mda{\marginpar[\boldmath\hfil$\downarrow$]%
                   {\boldmath$\downarrow$\hfil}%
                    \typeout{marginpar: $\downarrow$}\ignorespaces}
\def\mla{\marginpar[\boldmath\hfil$\rightarrow$]%
                   {\boldmath$\leftarrow $\hfil}%
                    \typeout{marginpar: $\leftrightarrow$}\ignorespaces}
\def\Mua{\marginpar[\boldmath\hfil$\Uparrow$]%
                   {\boldmath$\Uparrow$\hfil}%
                    \typeout{marginpar: $\Uparrow$}\ignorespaces}
\def\Mda{\marginpar[\boldmath\hfil$\Downarrow$]%
                   {\boldmath$\Downarrow$\hfil}%
                    \typeout{marginpar: $\Downarrow$}\ignorespaces}
\def\Mla{\marginpar[\boldmath\hfil$\Rightarrow$]%
                   {\boldmath$\Leftarrow $\hfil}%
                    \typeout{marginpar: $\Leftrightarrow$}\ignorespaces}
\overfullrule 5pt
\oddsidemargin -15mm
\marginparwidth 29mm
}

\begin{document}
\begin{titlepage}
\vspace*{-1cm}
\begin{flushright}

FR-PHENO-2010-15 \\
PSI-PR-10-06 \\
ZU-TH 04/10
\end{flushright}                                
\vskip 3.5cm

\begin{center}
\boldmath
{\Large\bf Electroweak corrections to hadronic event shapes\\[2mm] and jet 
production in $\Pe^+\Pe^-$ annihilation}
\unboldmath
\vskip 2.cm
{\large Ansgar Denner$^a$, Stefan Dittmaier$^{b}$, Thomas Gehrmann$^c$,
Christian Kurz$^{a,c}$}
\vskip .7cm
{\it $^a$ Paul Scherrer Institut, CH-5232 Villigen PSI,
Switzerland} 
\vskip .4cm
{\it $^b$ Physikalisches Institut, Albert-Ludwigs-Universit\"at
Freiburg, D-79104 Freiburg, Germany}
\vskip .4cm
{\it $^c$ Institut f\"ur Theoretische Physik,
Universit\"at Z\"urich, CH-8057 Z\"urich, Switzerland}
\end{center}
\vskip 2.0cm

\begin{abstract}
  We present a complete calculation of the electroweak ${\cal
    O}(\alpha^3\alpha_{\mathrm{s}})$ corrections to three-jet
  production and related event-shape observables at electron--positron
  colliders. The Z-boson resonance is described within the
  complex-mass scheme, rendering the calculation valid both in the
  resonance and off-shell regions.  Higher-order initial-state
  radiation is included in the leading-logarithmic approximation. We
  properly account for the corrections to the total hadronic cross
  section and for the experimental photon isolation criteria.  To this
  end we implement contributions of the quark-to-photon fragmentation
  function both in the slicing and subtraction formalism. The effects
  of the electroweak corrections on various event-shape distributions
  and on the three-jet rate are studied.  They are typically at the
  few-per-cent level, and remnants of the radiative return are found
  even after inclusion of appropriate cuts.
\end{abstract}
\vfill
March 2010

\end{titlepage}                                                                
\newpage

\renewcommand{\theequation}{\mbox{\arabic{section}.\arabic{equation}}}

\section{Introduction}
\setcounter{equation}{0}

Jet production in $\Pe^+\Pe^-$ annihilation provides an ideal
environment for studies of the dynamics of the strong interaction,
described by the theory of quantum chromodynamics (QCD)~\cite{qcd}.
The kinematical distribution of jets closely reflects the parton-level
kinematics of the event. Consequently, the first observation of
three-jet final states at DESY PETRA~\cite{tasso}, produced through
quark--antiquark--gluon final states~\cite{ellis}, provided conclusive
evidence for the existence of the gluon. Jets are defined through a
jet algorithm, which is a procedure to recombine individual hadrons
into jets using a distance measure, resolution criterion and
recombination prescription. The theoretical description applies the
same jet algorithm to partons in the final state.  Closely related to
jet cross sections are event-shape distributions. Event-shape
variables measure certain geometrical properties of hadronic final
states, and can equally be calculated in perturbative QCD from
partonic final states.

Jet cross sections and event-shape distributions were studied very
extensively at $\Pe^+\Pe^-$ colliders~\cite{reviews}, and
high-precision data are available from the LEP experiments
ALEPH~\cite{alephqcd}, OPAL~\cite{opal}, L3~\cite{l3},
DELPHI~\cite{delphi}, from SLD~\cite{sld} at SLAC and from
JADE~\cite{jade} at DESY PETRA.  The theoretical description of these
data in perturbative QCD contains only a single parameter: the strong
coupling constant $\alphas$. By comparing experimental results with
the theoretical description, one can thus perform a measurement of
$\alphas$ from jet cross sections and event shapes. For a long
period, the perturbative description of these observables was based on
next-to-leading order (NLO)~\cite{ERT,kunszt,event} in perturbative
QCD, improved by the resummation of next-to-leading-logarithmic
corrections~\cite{nlla} to all orders. The uncertainty on these
theoretical predictions from missing higher-order terms results in a
theory error on the extraction of $\alphas$, which was quantified to
be around five per cent, and thus larger than any source of
experimental uncertainty.

Owing to recent calculational progress, the QCD predictions for event
shapes~\cite{ourevent,weinzierlevent} and three-jet
production~\cite{our3j,weinzierl3j} are now accurate to
next-to-next-to-leading order (NNLO, $\alpha^2\alpha_\mathrm{s}^3$) in
QCD perturbation theory.  Inclusion of these corrections results in an
estimated residual uncertainty of the QCD prediction from missing
higher orders at the level of well below five per cent for the
event-shape distributions, and around one per cent for the three-jet
cross section.  Using these results (combined~\cite{gionata} with the
previously available resummed expressions), new determinations of
$\alphas$ from event-shape and jet production data were performed,
resulting in a considerable improvement of the theory uncertainty to
three per cent from event shapes~\cite{asevent} and below two per cent
from jet rates~\cite{asjets}. A further improvement can be anticipated
for the event shapes from the resummation of subleading logarithmic
corrections~\cite{becherschwartz}.

At this level of theoretical precision, higher-order electroweak
effects could be of comparable magnitude. Until recently, only partial
calculations of electroweak corrections to three-jet production and
event shapes have been available~\cite{CarloniCalame:2008qn}, which
can not be compared with experimental data directly. In a previous
work~\cite{Denner:2009gx}, we briefly reported our results on the
first calculation of the NLO electroweak ($\alpha^3\alphas$)
corrections to three-jet observables in $\Pe^+\Pe^-$ collisions
including the quark--antiquark--photon ($q\bar{q}\gamma$) final
states.
Here, we describe this calculation in detail and perform extensive
numerical studies on the impact of the electroweak corrections to
three-jet-like observables at different $\Pe^+\Pe^-$ collider
energies.

The full NLO electroweak ($\alpha^3\alphas$) corrections to jet
observables contain four types of contributions: genuine weak
corrections from virtual exchanges, photonic corrections
to quark--antiquark--gluon ($q\bar qg$) final states, gluonic
corrections to $q\bar{q}\gamma$ final states, and QCD/electroweak
interference effects in $q\bar q q\bar q$ final states of identical
quark flavour.  The latter were not included in our previous work, but
turn out to be numerically negligible as anticipated.

Any jet-like observable at NLO in the electroweak theory receives
virtual one-loop corrections and contributions from real photon
radiation. Experimental cuts on isolated hard photons in the final
state allow to suppress these real photon contributions. However,
photons radiated inside hadronic jets can often not be distinguished
from hadrons (like neutral pions), and are thus not removed by
experimental cuts. The real photon contribution at NLO thus results
from a complicated interplay of jet reconstruction and photon
isolation cuts.
Through the isolated-photon veto, these observables are sensitive to
final-state particle identification, and thus to fragmentation
processes.  In our case, we must include a contribution from
quark-to-photon fragmentation~\cite{Koller:1978kq} to obtain a
well-defined and infrared-safe observable. Since our calculation is
among the very first to perform electroweak corrections to jet
observables with realistic photon isolation cuts, we describe the
relevant calculational aspects in detail below.

To define the observables considered here, we describe in
Section~\ref{jetrate} the jet-clustering algorithms used in
$\Pe^+\Pe^-$ annihilation and the standard set of event-shape
variables. In this section, we also review the current description of
these observables in perturbative QCD. The calculation of NLO
electroweak corrections is outlined in Section~\ref{sec:struc}, where
we describe the calculation of the virtual and real corrections in
detail. The real corrections contain infrared
divergences from unresolved photon and 
gluon radiation. These infrared divergences cancel against similar
divergences in the virtual corrections. To accomplish this
cancellation, it is, however, necessary to extract them analytically
from the real corrections, which is done by a slicing or subtraction
procedure (described in Section~\ref{sec:realcorr}). Our results are
implemented into a parton-level event generator (described in
Section~\ref{sec:num}), which allows the simultaneous evaluation of
all event-shape variables and jet cross sections. Numerical results
for jet production and event-shape distributions for $\Pe^+\Pe^-$
collision energies at LEP1, LEP2 and a future linear collider are
presented in Section~\ref{sec:results}. At energies above the
$\PZ$ peak, we observe non-trivial kinematical structures in the
distributions. It is shown that these structures are a remnant of the
radiative-return phenomenon, resulting from a complicated interplay of
event-selection and photon isolation cuts applied in the experimental
definition of the observables. Finally, we conclude with an outlook on
the impact of electroweak effects on future precision QCD studies at
$\Pe^+\Pe^-$ colliders in Section~\ref{sec:conc}.

\section{Jet observables}
\setcounter{equation}{0}
\label{jetrate}
A commonly used method for reconstructing jets was originally
introduced by the JADE group \cite{Bethke:1988zc}. The algorithm is
based on successive combinations. In a first step, each observed
particle is listed as a jet. In the next step, a resolution parameter
$y_{ij}$ is calculated for each particle pair, and the particle pair
leading to the smallest value of $y_{ij}$ is combined into a single
pseudo-particle. This yields a new list of jets, and the algorithm
proceeds with step two. The procedure is repeated until no pair of
particles is left with $y_{ij}<y_{\mathrm{cut}}$, where
$y_{\mathrm{cut}}$ is a preset cut-off.

Different proposals exist in the literature in how to define $y_{ij}$
(see e.g.\ \citere{Dissertori:2003pj}).
The original JADE definition reads
\be
y_{ij,\mathrm{J}}=\frac{2E_iE_j\lrb1-\cos\theta_{ij}\rrb}{E_{\mathrm{vis}}^2},
\label{yijJ}
\ee
where $E_i$ is the energy of the $i$-th particle, $\cos\theta_{ij}$
the angle between the particles, and $E_{\mathrm{vis}}$ the total
visible hadronic energy in the event. Improving upon this definition,
different jet resolution parameters have been proposed. Most widely
used at LEP was the $k_\mathrm{T}$ or Durham algorithm
\cite{Brown:1990nm,Catani:1991hj}, which defines
\be
y_{ij,\mathrm{D}}=\frac{2\min\lrb E_i^2,E_j^2\rrb\lrb1-\cos\theta_{ij}\rrb}{E_{\mathrm{vis}}^2}\,.
\label{yijD}
\ee
In addition to the choice of jet resolution parameter, there also
exist different ways of combining the four-momenta of the two
particles with the lowest $y_{ij}$ to one four-momentum $p_{ij}$.  In
the so-called $E$-scheme one simply adds the two four-momenta, leading
to $p_{ij}=p_i+p_j$. In the $P$-scheme the invariant mass of the
pseudo-particle is set to zero by rescaling the energy
\be
{\vec{p}}_{ij}={\vec{p}}_{i}+{\vec{p}}_{j},\qquad
E_{ij}=\vert{\vec{p}}_{i}+{\vec{p}}_{j}\vert.
\label{Pscheme}
\ee
In the $P_0$-scheme, \refeq{Pscheme} is used to construct the
resulting four-momentum, however, after each recombination
$E_{\mathrm{vis}}$ is recalculated.  Finally, in the $E_0$-scheme the
  three-momentum rather than the energy is rescaled.

Since an event containing three jets is due to the emission of a gluon
off an \mbox{(anti-)quark} at a large angle and with significant
energy, the ratio of the number of observed three-jet to two-jet
events is, in leading order, proportional to the strong coupling
constant. In general, the $n$-jet rate $R_n(y)$, which depends on the
choice of the jet resolution parameter $y=y_{\mathrm{cut}}$, is
defined through the respective cross sections for $n\ge 2$ jets
\be
R_n(y,\sqrt{s})=\frac{\sigma_{\mbox{\scriptsize$n$-jet}}}{\sigma_{\mathrm{had}}},
\label{n-jetrate}
\ee
such that
\be
\sum_{n=1}^\infty R_n(y) = 1
\ee
and $\sqrt{s}$ is the centre-of-mass (CM) energy.
In order to characterise the topology of an event a large number of
observables have been developed. Most of them require at least three
momenta of final-state particles to be non-trivial. In the following
we introduce six variables which have been extensively used in
experimental analyses: thrust $T$ \cite{Brandt:1964sa,Farhi:1977sg},
the normalised heavy-jet mass $\rho$ \cite{Clavelli:1981yh}, the wide
and total jet broadenings $B_\mathrm{W}$ and $B_\mathrm{T}$
\cite{Rakow:1981qn,Catani:1992jc},
the $C$-parameter \cite{Parisi:1978eg,Donoghue:1979vi}, and the
transition from three-jet to two-jet final-state using
$y_{ij,\mathrm{D}}$
\cite{Catani:1991hj,Brown:1990nm,Brown:1991hx,Bethke:1991wk}.
\begin{itemize}
\item[$\bullet$]
Thrust is defined through
\be
T=\max_{\vec{n}}\frac{\sum_i\vert \vec{p}_i\cdot \vec{n}\vert}{\sum_i\vert \vec{p}_i\vert},
\label{Thrust}
\ee
where $\vec{p}_i$ is the three-momentum of the $i$-th particle, and
$\vec{n}$ is varied to maximise the momentum flow in its direction,
yielding the thrust axis.
\item[$\bullet$]
  Every event can be divided into two hemispheres $H_1$ and $H_2$ by a
  plane perpendicular to the thrust axis. In each hemisphere $H_i$ one
  can calculate the invariant mass $M_i^2$, the larger of which yields
  the heavy-jet mass
\be
M_\mathrm{had}^2=\max\lrb M_1^2,M_2^2\rrb,
\ee
and the normalised heavy-jet mass
\be
\rho=\frac{M_\mathrm{had}^2}{E_{\mathrm{vis}}^2}.
\ee
\item[$\bullet$]
Using the definition of the hemispheres from above, one can calculate the hemisphere broadenings
\be
B_i=\frac{\sum_{j\in H_i}\vert \vec{p}_j\times \vec{n}\vert}
{2\sum_j\vert \vec{p}_j\vert},
\quad i=1,2.
\ee
The wide and total jet broadenings $B_\mathrm{W}$ and $B_\mathrm{T}$ are then obtained through
\be
B_\mathrm{W}=\max\lrb B_1,B_2\rrb, \qquad
B_\mathrm{T}=B_1+B_2.
\ee
\item[$\bullet$]
Starting from the linearised momentum tensor
\be
\Theta^{\alpha\beta}=\frac{1}{\sum_i\vert \vec{p}_i\vert}\sum_j\frac{p_j^\alpha p_j^\beta}{\vert
\vec{p}_j\vert}
,\quad \alpha,\beta=1,2,3,
\ee
and its three eigenvalues $\lambda_1,\lambda_2,\lambda_3$, the $C$-parameter is defined through
\be
C=3\lrb \lambda_1\lambda_2 +\lambda_2\lambda_3 +\lambda_3\lambda_1 \rrb.
\ee
\item[$\bullet$]
The jet transition variable $Y_3$ is defined as the value of the jet resolution parameter for which
an event changes from a three-jet-like to a two-jet-like configuration.
\end{itemize}
In the following we often denote event-shape observables generically $y$.
Taking $(1-T)$ instead of $T$ the two-particle configuration is located
at $y=0$ for all the event-shape variables defined above.
\subsection{Event shapes and jet rates in perturbation theory}
\label{esinpt}
At leading order (LO), $\mathcal{O}{\lrb\alpha^2\alphas\rrb}$, the first
process that occurs at tree level in $\Pe^+\Pe^-$ annihilation is
gluon radiation off a quark or antiquark (see
\reffig{fi:borndiags_qqg}).
\begin{figure}                                                
\centerline{\footnotesize
\begin{feynartspicture}(82,82)(1,1)
\FADiagram{}
\FAProp(0.,15.)(5.5,10.)(0.,){/Straight}{1}
\FALabel(1,13)[tr]{$\Pe$}
\FAProp(0.,5.)(5.5,10.)(0.,){/Straight}{-1}
\FALabel(-.2,8)[tl]{$\Pe$}
\FAProp(20.,17.)(15.5,13.5)(0.,){/Straight}{-1}
\FALabel(17.5,16)[br]{$q$}
\FAProp(20.,10.)(15.5,13.5)(0.,){/Cycles}{0}
\FALabel(15,9)[bl]{$\Pg$}
\FAProp(20.,3.)(12.,10.)(0.,){/Straight}{1}
\FALabel(16,5.5)[tr]{$q$}
\FAProp(5.5,10.)(12.,10.)(0.,){/Sine}{0}
\FALabel(8.75,8.93)[t]{$\gamma,\PZ$}
\FAProp(15.5,13.5)(12.,10.)(0.,){/Straight}{-1}
\FALabel(13.134,12.366)[br]{$q$}
\FAVert(5.5,10.){0}
\FAVert(15.5,13.5){0}
\FAVert(12.,10.){0}
\end{feynartspicture}
\hspace{2em}
\begin{feynartspicture}(82,82)(1,1)                                  
\FADiagram{}
\FAProp(0.,15.)(5.5,10.)(0.,){/Straight}{1}
\FALabel(1,13)[tr]{$\Pe$}
\FAProp(0.,5.)(5.5,10.)(0.,){/Straight}{-1}
\FALabel(-.2,8)[tl]{$\Pe$}
\FAProp(20.,17.)(11.5,10.)(0.,){/Straight}{-1}
\FALabel(15,14)[br]{$q$}
\FAProp(20.,10.)(15.5,6.5)(0.,){/Cycles}{0}
\FALabel(15.5,8.5)[bl]{$\Pg$}
\FAProp(20.,3.)(15.5,6.5)(0.,){/Straight}{1}
\FALabel(17,4)[tr]{$q$}
\FAProp(5.5,10.)(11.5,10.)(0.,){/Sine}{0}
\FALabel(8.75,8.93)[t]{$\gamma,\PZ$}
\FAProp(11.5,10.)(15.5,6.5)(0.,){/Straight}{-1}
\FALabel(12.9593,7.56351)[tr]{$q$}
\FAVert(5.5,10.){0}
\FAVert(11.5,10.){0}
\FAVert(15.5,6.5){0}
\end{feynartspicture}
                     }
\vspace*{-2em}
\caption{Lowest-order diagrams for $\Pe^+\Pe^-\rightarrow q\bar{q}\Pg$.} 
\label{fi:borndiags_qqg} 
\end{figure}%
As mentioned above, by comparing the measured three-jet rate and
event-shape observables with theoretical predictions, one can
determine $\alphas$.

In perturbation theory up to next-to-next-to-leading order (NNLO) in
QCD, the expansion of a distribution in the generic observable $y$ at
CM energy $\sqrt{s}$ for renormalisation scale $\mu=\sqrt{s}$ and
$\alphas=\alphas(s)$, normalised to the Born cross section
$\sigma_0(s)$ of the process $\Pe^+\Pe^-\rightarrow q\bar{q}$ is given
by
\be
\frac{1}{\sigma_0}\frac{\rd\sigma}{\rd y}=\lrb\frac{\alphas}{2\pi}\rrb\frac{\rd A}{\rd y}+
\lrb\frac{\alphas}{2\pi}\rrb^2\frac{\rd B}{\rd y}+\lrb\frac{\alphas}{2\pi}\rrb^3\frac{\rd C}{\rd y}
+\mathcal{O}\lrb\alphas^4\rrb,
\ee
where $A$, $B$, and $C$ denote the QCD contributions of LO,
next-to-leading order (NLO), and NNLO.  The experimentally measured
event-shape distribution is normalised to the total hadronic cross
section $\sigma_{\mathrm{had}}$, which for massless quarks reads
\be
\sigma_{\mathrm{had}}=\sigma_0\lrb1+\lrb\frac{\alphas}{2\pi}\rrb K_1+\lrb\frac{\alphas}{2\pi}\rrb^2 K_2
+\mathcal{O}\lrb\alphas^3\rrb\rrb,
\ee
such that
\be
\frac{1}{\sigma_{\mathrm{had}}}\frac{\rd\sigma}{\rd y}=\lrb\frac{\alphas}{2\pi}\rrb\frac{\rd
\bar{A}}{\rd y}+
\lrb\frac{\alphas}{2\pi}\rrb^2\frac{\rd \bar{B}}{\rd y}+\lrb\frac{\alphas}{2\pi}\rrb^3\frac{\rd \bar{C}}{\rd y}
+\mathcal{O}\lrb\alphas^4\rrb,
\ee
where
\bea
\bar{A}=A,
\qquad
\bar{B}=B - A K_1,
\qquad
\bar{C}=C - B K_1 + A K_1^2 - A K_2.
\label{QCDycoeff}
\eea
The coefficients in \refeq{QCDycoeff} up to NLO have been calculated
in
\citeres{ERT,kunszt,event}.
Furthermore, kinemati\-cal\-ly-dominant leading and next-to-leading
logarithms have been resummed \cite{Catani:1991kz,nlla}, and
non-perturba\-ti\-ve models of power-suppressed hadronisation effects
have been included
\cite{Korchemsky:1994is,Dokshitzer:1995zt,Dokshitzer:1997ew,Dokshitzer:1998pt}
to increase the theoretical accuracy. Recently the first NNLO
calculations have been completed \cite{ourevent,weinzierlevent,our3j,weinzierl3j},
and the matching of next-to-leading logarithms and
next-to-next-to-leading logarithms to the fixed-order NNLO calculation
has been performed \cite{becherschwartz,gionata}. These results
have subsequently been used in precision
determinations~\cite{asevent,asjets,becherschwartz}
of the strong coupling constant $\alphas$.

With regard to jet rates, fixed-order calculations are known up to
next-to-next-to-next-to-leading order ($\mathrm{N^3LO}$) in QCD for
the two-jet rate
\cite{Anastasiou:2004qd,GehrmannDeRidder:2004tv,Weinzierl:2006ij,our3j},
up to NNLO for the three-jet rate
\cite{ERT,kunszt,event,our3j,weinzierl3j},
and up to NLO for the four-jet rate
\cite{Signer:1996bf,Dixon:1997th,Nagy:1997yn,Campbell:1998nn,Weinzierl:1999yf}.

NLO electroweak (EW) corrections could be of comparable magnitude as
the NNLO QCD corrections and are therefore worth further
consideration.  The factorisable EW corrections have been calculated
in \citere{Maina:2002wz} and a further step towards the full NLO EW
corrections has been made in \citere{CarloniCalame:2008qn}.  In this
work we describe the first calculation of the complete NLO EW
corrections to the normalised event-shape distributions. First results
of this calculation on the thrust distribution and the three-jet rate
at $\sqrt{s} = M_{{\rm Z}}$ have been presented in
\citere{Denner:2009gx}.

In analogy to the QCD corrections, we write the total hadronic cross
section including $\mathcal{O}{\lrb\alpha\rrb}$ corrections as
\be
\sigma_{\mathrm{had}}=\sigma_0\lrb1+\lrb\frac{\alpha}{2\pi}\rrb\delta_{\sigma,1}
+\mathcal{O}\lrb\alpha^2\rrb\rrb,
\label{sig0NLO}
\ee
and the expansion of the observable $\rd\sigma/\rd y$ as
\be
\frac{1}{\sigma_0}\frac{\rd\sigma}{\rd y}=\lrb\frac{\alphas}{2\pi}\rrb\frac{\rd A}{\rd y}+
\lrb\frac{\alpha}{2\pi}\rrb\frac{\rd \delta_\gamma}{\rd y}+
\lrb\frac{\alpha}{2\pi}\rrb\lrb\frac{\alphas}{2\pi}\rrb\frac{\rd \delta_A}{\rd y}+\mathcal{O}\lrb\alpha^2\rrb,
\label{dsdy_EW}
\ee
where the LO purely electromagnetic contribution $\delta_\gamma$
arises from tree-level quark--antiquark--photon ($q\bar{q}\gamma$)
final states without a gluon%
\footnote{Since the event-shape observables are calculated from parton
  momenta, the $q\bar{q}\gamma$ final states contribute if the photon
  is clustered with a quark into a jet and the event is no longer
  removed by the photon cuts.}  and $\delta_A$ comprises the NLO EW
corrections to the distribution $\rd\sigma/\rd y$.

Normalising \refeq{dsdy_EW} to $\sigma_{\mathrm{had}}$ yields
\be
\frac{1}{\sigma_{\mathrm{had}}}\frac{\rd\sigma}{\rd y}=
\lrb\frac{\alphas}{2\pi}\rrb
\frac{\rd A}{\rd y}+\lrb\frac{\alpha}{2\pi}\rrb
\frac{\rd \delta_\gamma}{\rd y}+ \lrb\frac{\alpha}{2\pi}\rrb
\lrb\frac{\alphas}{2\pi}\rrb\lrb\frac{\rd \delta_A}{\rd y}-
\frac{\rd A}{\rd y}\delta_{\sigma,1}\rrb+\mathcal{O}\lrb\alpha^2\rrb.
\label{dsdyhad_EW}
\ee
Hence, the full $\mathcal{O}{\lrb\alpha\rrb}$ EW corrections%
\footnote{In \citere{Denner:2009gx} the definition of $\delta_{\mathrm{EW}}$
was somewhat different and, in particular, did not explicitly contain
the effect of $\delta_\gamma$.}
are given by
\be
\frac{\rd\delta_{\mathrm{EW}}}{\rd y}=\lrb\frac{\alpha}{2\pi}\rrb\frac{\rd \delta_\gamma}{\rd y}+ \lrb\frac{\alpha}{2\pi}\rrb\lrb\frac{\alphas}{2\pi}\rrb
\lrb\frac{\rd \delta_A}{\rd y}-
\frac{\rd A}{\rd y}\delta_{\sigma,1}\rrb.
\label{deltaEW}
\ee
In order to obtain a sensible ratio, all three contributions have to
be evaluated using the same event-selection cuts.
 
The EW corrections to both $\sigma_{\mathrm{had}}$ and the
distribution in $y$ contain large corrections due to initial-state
radiation (ISR). Since these are universal, they partially cancel in
the third term in \refeq{dsdyhad_EW}, leaving only a small remainder.
If we include higher-order leading-logarithmic (LL) ISR effects in both
$\sigma_{\mathrm{had}}$ and the distribution in $y$, this leads to
\be
\sigma_{\mathrm{had}}=\sigma_0\lrb1+\lrb\frac{\alpha}{2\pi}\rrb\delta_{\sigma,1}
+\lrb\frac{\alpha}{2\pi}\rrb^2\delta_{\sigma,\ge 2,\mathrm{LL}}
+\mathcal{O}\lrb\alpha^2\rrb\rrb,
\label{sig0LL}
\ee
and
\be
\frac{1}{\sigma_0}\frac{\rd\sigma}{\rd y}=\lrb\frac{\alphas}{2\pi}\rrb
\frac{\rd A}{\rd y}+
\lrb\frac{\alpha}{2\pi}\rrb\frac{\rd \delta_\gamma}{\rd y}+
\lrb\frac{\alpha}{2\pi}\rrb\lrb\frac{\alphas}{2\pi}\rrb\frac{\rd \delta_{A}}{\rd y}+
\lrb\frac{\alpha}{2\pi}\rrb^2\lrb\frac{\alphas}{2\pi}\rrb\frac{\rd \delta_{A,\ge 2,\mathrm{LL}}}{\rd y}
+\mathcal{O}\lrb\alpha^2\rrb,
\label{dsdy_EW_log}
\ee
where $\delta_{\sigma,\ge 2,\mathrm{LL}}$ and $\delta_{A,\ge
  2,\mathrm{LL}}$ contain leading-logarithmic (LL) terms proportional to
$\alpha^n\ln^n({s}/{m_\Pe^2})$ with $n\ge 2$, as defined in
\refsec{hoisr}. 
Here $\mathcal{O}\lrb\alpha^2\rrb$ stands for
two-loop electroweak effects without the enhancement of leading ISR logarithms.
For the normalised distribution this results in
\bea
\frac{1}{\sigma_{\mathrm{had}}}\frac{\rd\sigma}{\rd y}&=&
\lrb\frac{\alphas}{2\pi}\rrb\frac{\rd A}{\rd y}+\lrb\frac{\alpha}{2\pi}\rrb \frac{\rd \delta_\gamma}{\rd y}+\lrb\frac{\alpha}{2\pi}\rrb\lrb\frac{\alphas}{2\pi}\rrb\lrb\frac{\rd \delta_A}{\rd y}-
\frac{\rd A}{\rd y}\delta_{\sigma,1}\rrb
\\
&&{}+\lrb\frac{\alpha}{2\pi}\rrb^2\lrb\frac{\alphas}{2\pi}\rrb\lsb
\lrb\frac{\rd \delta_{A,\ge 2,\mathrm{LL}}}{\rd y}-\frac{\rd A}{\rd y}\delta_{\sigma,\ge 2,\mathrm{LL}}
\rrb
-\frac{\rd \delta_{A,1,\mathrm{LL}}}{\rd y}\delta_{\sigma,1,\mathrm{LL}}+\frac{\rd A}{\rd
y}\delta_{\sigma,1,\mathrm{LL}}^2
\rsb
+\mathcal{O}\lrb\alpha^2\rrb,\nn
\label{dsdyhad_EW_log}
\eea
where $\delta_{A,1,\mathrm{LL}}$ and $\delta_{\sigma,1,\mathrm{LL}}$
denote the LL contributions contained in the NLO results. 
Their second-order effect results from taking the ratio of two NLO-corrected
quantities.
Therefore, the higher-order LL corrections read
\be
\frac{\rd\delta_{\mathrm{EW,LL}}}{\rd y}=
\lrb\frac{\alpha}{2\pi}\rrb^2\lrb\frac{\alphas}{2\pi}\rrb\lsb
\lrb\frac{\rd \delta_{A,\ge 2,\mathrm{LL}}}{\rd y}-\frac{\rd A}{\rd y}\delta_{\sigma,\ge 2,\mathrm{LL}}
\rrb
+\lrb\frac{\rd A}{\rd
y}\delta_{\sigma,1,\mathrm{LL}}^2-\frac{\rd \delta_{A,1,\mathrm{LL}}}{\rd y}\delta_{\sigma,1,\mathrm{LL}}\rrb\rsb.
\label{deltaEWLL}
\ee
Due to the universality of ISR, the terms in the first and in the
second parenthesis in \refeq{deltaEWLL} separately cancel each other
numerically to a large extent.

The same decomposition as applied here for event-shape distributions
holds also for the three-jet rate,  normalised to $\sigma_\mathrm{had}$.
\subsection{Particle identification}
\label{PI}
One of the virtues of $\Pe^+\Pe^-$ colliders is the precise knowledge
of the energy of the initial state. However, ISR of photons can lead
to difficulties in the determination of the total energy of the final
state. Therefore event-selection cuts have been devised to suppress
effects due to ISR.  In the following we describe the procedure
employed by the ALEPH collaboration at LEP \cite{Barate:1996fi}, which
we use in our numerical evaluations.

First, particles are clustered into jets according to the Durham
algorithm with $y_{\mathrm{cut,D}}=0.002$ and $E$-scheme recombination. 
Jets where the fraction of
energy carried by charged hadrons is less than 10\% are identified as
dominantly electromagnetic and are removed. In the next step the
remaining particles are clustered into two jets and the visible
invariant mass $M_{\mathrm{vis}}$ of the two-jet system is calculated.
Using total momentum conservation the reduced CM energy
$\sqrt{s'}$ is calculated.  The event is rejected if $s'/s<0.81$. This
two-step procedure is later referred to as hard-photon cut procedure
(note that it is called anti-ISR cut procedure in
\citere{Barate:1996fi}).

Removing events where the photonic energy in a jet is higher than a
certain value, as it is done in the hard-photon cuts, causes potential
problems when perturbatively calculating EW corrections.
There one relies on the cancellation of infrared (IR) singularities between
virtual and real corrections when calculating an IR-safe
observable. Removing events where a photon is close to a final-state
charged fermion leads to non-IR-safe observables and spoils this
cancellation in the collinear region. This feature is common to all
observables with identified particles in the final state, and
IR-finiteness is restored by taking into account a contribution
from fragmentation processes, in our case from the quark-to-photon
fragmentation function.

\section{Structure of the calculation}
\setcounter{equation}{0}
\label{sec:struc}

To obtain the full EW ${\cal O}(\alpha^3\alpha_{\mathrm{s}})$
corrections to normalised event-shape distributions and jet cross
sections we need to derive the NLO EW corrections to the total
hadronic cross section and to the three-jet production cross section.
The total hadronic cross section is decomposed as: 
\bea 
\sigma_{{\rm
    had}} = \int\rd\sigma^{\Pe^+\Pe^-\rightarrow q\bar{q}}_{\mathrm{Born}}+
\int\rd\sigma^{\Pe^+\Pe^-\rightarrow q\bar{q}}_{\mathrm{virtual,EW}}+
\int\rd\sigma_{\mathrm{real}}^{\Pe^+\Pe^-\rightarrow q\bar{q}\gamma},
\label{master_had}
\eea
where the first and second terms are the Born and one-loop EW 
contributions to the process $\Pe^+\Pe^-\rightarrow q\overline{q}$,
while the last term is the real radiation contribution from the process 
$\Pe^+\Pe^-\rightarrow q\overline{q}\gamma$.
Likewise, we decompose the total cross section for three-jet production 
according to
\bea
\int\rd\sigma&=&
\int\rd\sigma^{\Pe^+\Pe^-\rightarrow q\bar{q}g}_{\mathrm{Born}}+
\int\rd\sigma^{\Pe^+\Pe^-\rightarrow q\bar{q}\gamma}_{\mathrm{Born}}\nn\\
&&{}+
\int\rd\sigma^{\Pe^+\Pe^-\rightarrow q\bar{q}g}_{\mathrm{virtual,EW}}+
\int\rd\sigma^{\Pe^+\Pe^-\rightarrow q\bar{q}\gamma}_{\mathrm{virtual,QCD}}
+\int\rd\sigma_{\mathrm{real}}^{\Pe^+\Pe^-\rightarrow q\bar{q}g\gamma}
+\int\rd\sigma_{\mathrm{interference}}^{\Pe^+\Pe^-\rightarrow q\bar{q}q\bar q},
\label{master_eq}
\eea
where the first and third terms are the Born and NLO EW
contributions of the process $\Pe^+\Pe^-\rightarrow q\overline{q}g$,
the second and fourth terms are the Born and one-loop QCD
contributions of the process $\Pe^+\Pe^-\rightarrow
q\overline{q}\gamma$, the fifth term is the contribution from the real
radiation process $\Pe^+\Pe^-\rightarrow q\overline{q}g\gamma$, and
the sixth term results from the contribution of the real radiation process
$\Pe^+\Pe^-\rightarrow q\overline{q}q\bar q$ with identical quark
flavours.  In this work, we are interested in the virtual and real
radiation corrections of $\mathcal{O}\lrb\alpha^3\alphas\rrb$, which
lead to the production of three or four jets, when treating photons
and hadrons democratically. For the $q\bar q q\bar q$ final state,
this order corresponds to the interference of the EW amplitude
with the QCD amplitude. We do not include the squares of the
EW and the QCD amplitudes for this process.
The former is of $\mathcal{O}\lrb\alpha^4\rrb$ and thus beyond the 
considered accuracy, the latter is part of the NLO QCD corrections not 
considered in this work.

We have performed two independent calculations each for the virtual
and real corrections, the results of which are in mutual numerical 
agreement.

\subsection{Conventions and lowest-order cross section}
At the parton level we consider the processes
\bea
\Pe^+(k_1,\sigma_1)+\Pe^-(k_2,\sigma_2)&\rightarrow& 
q(k_3,\sigma_3)+\bar{q}(k_4,\sigma_4),\label{born_qq}\\
\Pe^+(k_1,\sigma_1)+\Pe^-(k_2,\sigma_2)&\rightarrow& 
q(k_3,\sigma_3)+\bar{q}(k_4,\sigma_4)+\Pg(k_5,\lambda),\label{born_qqg}\\
\Pe^+(k_1,\sigma_1)+\Pe^-(k_2,\sigma_2)&\rightarrow& 
q(k_3,\sigma_3)+\bar{q}(k_4,\sigma_4)+\gamma(k_5,\lambda),
\label{born_qqa}
\eea
where $q$ can be an up, down, charm, strange, or bottom quark.  The
momenta $k_i$ of the corresponding particles as well as their
helicities $\sigma_i$ and $\lambda$ are given in parentheses. The
helicities of the fermions take the values $\sigma_i=\pm1/2$, and the
helicity of the gluon or the photon assumes the values $\lambda=\pm
1$.
We neglect the masses of the external fermions wherever possible and
keep them only as regulators of the mass-singular logarithms.
Therefore all amplitudes vanish unless $\sigma_1=-\sigma_2$ and
$\sigma_3=-\sigma_4$, and we define $\sigma=\sigma_2=-\sigma_1$ and
$\sigma'=\sigma_3=-\sigma_4$.

\begin{figure}
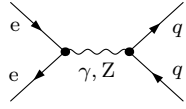
                                                
\centerline{\footnotesize
\begin{feynartspicture}(82,82)(1,1)
\FADiagram{}
\FAProp(0.,15.)(5.5,10.)(0.,){/Straight}{1}
\FALabel(1,13)[tr]{$\Pe$}
\FAProp(0.,5.)(5.5,10.)(0.,){/Straight}{-1}
\FALabel(-.2,8)[tl]{$\Pe$}
\FAProp(5.5,10.)(12.,10.)(0.,){/Sine}{0}
\FALabel(8.75,8.93)[t]{$\gamma,\PZ$}
\FAProp(17.5,15)(12,10.)(0.,){/Straight}{-1}
\FALabel(17.5,11.5)[br]{$q$}
\FAProp(17.5,5)(12,10.)(0.,){/Straight}{1}
\FALabel(17.5,8.5)[tr]{$q$}
\FAVert(5.5,10.){0}
\FAVert(12.,10.){0}
\end{feynartspicture}
                     }
\vspace*{-2.5em}
\caption{Lowest-order diagrams for $\Pe^+\Pe^-\rightarrow q\bar{q}$.} 
\label{fi:borndiags_qq} 
\end{figure}

For later use, the following set of kinematical invariants is introduced:
\be
s=\left(k_1+k_2\right)^2,\quad s_{ij}=\left(k_i+k_j\right)^2,\quad 
s_{ai}=s_{ia}=\left(k_a -k_i\right)^2,\quad a=1,2,\quad i,j=3,4,5.
\ee
For later convenience we employ the convention that indices $a,b=1,2$
refer to the initial and indices $i,j=3,4,5$ to the
final state, while the generic indices $I,J=1,\dots,5$
label all external particles.
The tree-level Feynman diagrams contributing to the process
\refeq{born_qq} are shown in \reffig{fi:borndiags_qq}, those
contributing to the process \refeq{born_qqg} in
\reffig{fi:borndiags_qqg}, and the ones contributing to the process
\refeq{born_qqa} in \reffig{fi:borndiags_qqa}.

\begin{figure}                                                
\centerline{\footnotesize
\begin{feynartspicture}(82,82)(1,1)
\FADiagram{}
\FAProp(0.,15.)(5.5,10.)(0.,){/Straight}{1}
\FALabel(1,13)[tr]{$\Pe$}
\FAProp(0.,5.)(5.5,10.)(0.,){/Straight}{-1}
\FALabel(-.2,8)[tl]{$\Pe$}
\FAProp(20.,17.)(15.5,13.5)(0.,){/Straight}{-1}
\FALabel(17.5,16)[br]{$q$}
\FAProp(20.,10.)(15.5,13.5)(0.,){/Sine}{0}
\FALabel(15,9)[bl]{$\gamma$}
\FAProp(20.,3.)(12.,10.)(0.,){/Straight}{1}
\FALabel(16,5.5)[tr]{$q$}
\FAProp(5.5,10.)(12.,10.)(0.,){/Sine}{0}
\FALabel(8.75,8.93)[t]{$\gamma,\PZ$}
\FAProp(15.5,13.5)(12.,10.)(0.,){/Straight}{-1}
\FALabel(13.134,12.366)[br]{$q$}
\FAVert(5.5,10.){0}
\FAVert(15.5,13.5){0}
\FAVert(12.,10.){0}
\end{feynartspicture}
\hspace{2em}
\begin{feynartspicture}(82,82)(1,1)                                  
\FADiagram{}
\FAProp(0.,15.)(5.5,10.)(0.,){/Straight}{1}
\FALabel(1,13)[tr]{$\Pe$}
\FAProp(0.,5.)(5.5,10.)(0.,){/Straight}{-1}
\FALabel(-.2,8)[tl]{$\Pe$}
\FAProp(20.,17.)(11.5,10.)(0.,){/Straight}{-1}
\FALabel(15,14)[br]{$q$}
\FAProp(20.,10.)(15.5,6.5)(0.,){/Sine}{0}
\FALabel(15.5,8.5)[bl]{$\gamma$}
\FAProp(20.,3.)(15.5,6.5)(0.,){/Straight}{1}
\FALabel(17,4)[tr]{$q$}
\FAProp(5.5,10.)(11.5,10.)(0.,){/Sine}{0}
\FALabel(8.75,8.93)[t]{$\gamma,\PZ$}
\FAProp(11.5,10.)(15.5,6.5)(0.,){/Straight}{-1}
\FALabel(12.9593,7.56351)[tr]{$q$}
\FAVert(5.5,10.){0}
\FAVert(11.5,10.){0}
\FAVert(15.5,6.5){0}
\end{feynartspicture}
\hspace{2em}
\begin{feynartspicture}(82,82)(1,1)                                  
\FADiagram{}
\FAProp(0.,17.)(4.25,13.5)(0.,){/Straight}{1}
\FALabel(1,15)[tr]{$\Pe$}
\FAProp(4.25,13.5)(8.5,10.)(0.,){/Straight}{1}
\FALabel(5.5,11)[tr]{$\Pe$}
\FAProp(0.,3.)(8.5,10.)(0.,){/Straight}{-1}
\FALabel(-.2,6)[tl]{$\Pe$}
\FAProp(20.,15)(14.5,10.)(0.,){/Straight}{-1}
\FALabel(20,11.5)[br]{$q$}
\FAProp(4.25,13.5)(8.75,17)(0.,){/Sine}{0}
\FALabel(7.5,13)[bl]{$\gamma$}
\FAProp(20,5)(14.5,10.)(0.,){/Straight}{1}
\FALabel(20,8.5)[tr]{$q$}
\FAProp(8.5,10.)(14.5,10.)(0.,){/Sine}{0}
\FALabel(10.75,8.93)[t]{$\gamma,\PZ$}
\FAVert(8.5,10.){0}
\FAVert(14.5,10.){0}
\FAVert(4.25,13.5){0}
\end{feynartspicture}
\hspace{2em}
\begin{feynartspicture}(82,82)(1,1)                                  
\FADiagram{}
\FAProp(0.,17.)(8.5,10)(0.,){/Straight}{1}
\FALabel(1,15)[tr]{$\Pe$}
\FAProp(0.,3.)(4.25,6.5)(0.,){/Straight}{-1}
\FALabel(-.2,6)[tl]{$\Pe$}
\FAProp(4.25,6.5)(8.5,10.)(0.,){/Straight}{-1}
\FALabel(5.5,10)[tr]{$\Pe$}
\FAProp(20.,15)(14.5,10.)(0.,){/Straight}{-1}
\FALabel(20,11.5)[br]{$q$}
\FAProp(4.25,6.5)(8.75,3)(0.,){/Sine}{0}
\FALabel(4.5,1.5)[bl]{$\gamma$}
\FAProp(20,5)(14.5,10.)(0.,){/Straight}{1}
\FALabel(20,8.5)[tr]{$q$}
\FAProp(8.5,10.)(14.5,10.)(0.,){/Sine}{0}
\FALabel(10.75,8.93)[t]{$\gamma,\PZ$}
\FAVert(8.5,10.){0}
\FAVert(14.5,10.){0}
\FAVert(4.25,6.5){0}
\end{feynartspicture}
                     }
\vspace{-2em}
\caption{Lowest-order diagrams for $\Pe^+\Pe^-\rightarrow q\bar{q}\gamma$.} 
\label{fi:borndiags_qqa} 
\end{figure}
The lowest-order partonic cross section for the processes given in
\refeqs{born_qqg} and (\ref{born_qqa}) reads
\be
\int\rd\sigma_{\mathrm{Born}} = \frac{1}{2s} 
\fc\sum_{\substack{\sigma,\sigma'=\pm\frac{1}{2}\\ \lambda=\pm 1}}
\frac{1}{4}(1-2P_1\sigma)(1+2P_2\sigma)\,
\int\mathrm{d}\Phi_3 \,
\vert\mathcal{M}_{0}^{\sigma\sigma'\lambda}\vert
^2\,\Theta_{\mathrm{cut}}(\Phi_3),
\label{eq:sigma0}
\ee
where $\fc$ is a colour factor, $P_{1,2}$ are the degrees of beam
polarisation of the incoming $\Pe^+$ and $\Pe^-$,
$\mathcal{M}_{0}^{\sigma\sigma'\lambda}$ is the colour-stripped Born
matrix element of the respective process, and the integral over the
three-particle phase space is defined by
\be
\int\mathrm{d}\Phi_3=
\left( \prod_{i=3}^5 \int\frac{\mathrm{d}^3 \vec{k}_i}{(2\pi)^3 2k_i^0} \right)\,
(2\pi)^4 \delta\Biggl(k_1+k_2-\sum_{j=3}^5 k_j\Biggr).
\label{eq:dG3}
\ee
For the process \refeq{born_qqa}, $\fc=3$, and for the process
\refeq{born_qqg} $\fc=4$. The dependence of the cross section on the
event-selection cuts is reflected by the step function
$\Theta_{\mathrm{cut}}\lrb \Phi_3
\rrb$. For the lowest-order cross sections of \refeq{born_qqg} and
\refeq{born_qqa} and the virtual corrections,
$\Theta_{\mathrm{cut}}$ depends on three-particle kinematics. It is
equal to $1$ if the event passes the cuts and equal to $0$ otherwise.

The formula corresponding to the process \refeq{born_qq} can be
obtained from \refeq{eq:sigma0} by omitting the dependence on and the
sum over the polarisation $\lambda$ of photon or gluon, using only the
two-particle phase space $\Phi_2$, and setting $\fc=3$.

\subsection{Virtual corrections}
\label{sec:virt}

\subsubsection{Survey of diagrams and setup of the loop calculation}

We calculate the one-loop EW corrections to the processes given in
\refeqs{born_qq} and (\ref{born_qqg}), and the one-loop QCD
corrections to the process given in \refeq{born_qqa}. For
\refeq{born_qqg} and \refeq{born_qqa} their contributions to the cross
section are generically given by
\be
\int\rd\sigma_{\mathrm{virtual}} = \frac{1}{2s}
\fc\sum_{\substack{\sigma,\sigma'=\pm\frac{1}{2}\\ \lambda=\pm 1}}
\frac{1}{4}(1\!-\!2P_1\sigma)(1\!+\!2P_2\sigma)
\int\mathrm{d}\Phi_3\, 
2\Re\lsb\mathcal{M}_{0}^{\sigma\sigma'\lambda}\lrb\mathcal{M}_{1}^{\sigma\sigma'\lambda}\rrb^*\rsb\!
\Theta_{\mathrm{cut}}\lrb \Phi_3 \rrb,
\label{eq:sigma_virtual}
\ee
where the notation is the same as in the previous section and
$\mathcal{M}_{1}^{\sigma\sigma'\lambda}$ denotes the contributions of
the virtual corrections to the matrix element after splitting off
the colour factor of the corresponding lowest-order amplitude.

\begin{figure}
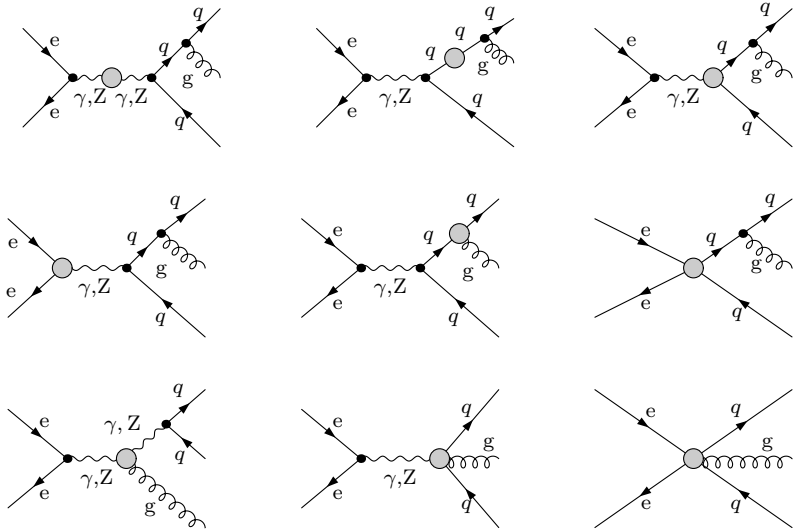
                                                 
\centerline{\footnotesize
\begin{tabular}{ccc}
\begin{feynartspicture}(82,82)(1,1)
\FADiagram{}
\FAProp(0.,15.)(5.,10.)(0.,){/Straight}{1}
\FALabel(2.69678,13.1968)[bl]{$\Pe$}
\FAProp(0.,5.)(5.,10.)(0.,){/Straight}{-1}
\FALabel(2.69678,6.80322)[tl]{$\Pe$}
\FAProp(20.,17.)(16.5,13.5)(0.,){/Straight}{-1}
\FALabel(18.3032,15.6968)[br]{$q$}
\FAProp(20.,10.)(16.5,13.5)(0.,){/Cycles}{0}
\FALabel(17.3032,10.3032)[tr]{$\Pg$}
\FAProp(20.,3.)(13.,10.)(0.,){/Straight}{1}
\FALabel(16.48,5.98)[tr]{$q$}
\FAProp(9.,10.)(5.,10.)(0.,){/Sine}{0}
\FALabel(7.,8.93)[t]{$\!\gamma,\!\PZ$}
\FAProp(9.,10.)(13.,10.)(0.,){/Sine}{0}
\FALabel(11.,8.93)[t]{$\gamma,\!\PZ$}
\FAProp(13.,10.)(16.5,13.5)(0.,){/Straight}{1}
\FALabel(14.8032,12.1968)[br]{$q$}
\FAVert(5.,10.){0}
\FAVert(16.5,13.5){0}
\FAVert(13.,10.){0}
\FAVert(9.,10.){-1}
\end{feynartspicture}
&
\hspace{1em}
\begin{feynartspicture}(82,82)(1,1)
\FADiagram{}
\FAProp(0.,15.)(5.,10.)(0.,){/Straight}{1}
\FALabel(3.11602,13.116)[bl]{$\Pe$}
\FAProp(0.,5.)(5.,10.)(0.,){/Straight}{-1}
\FALabel(3.11602,6.88398)[tl]{$\Pe$}
\FAProp(20.,16.)(17.,14.)(0.,){/Straight}{-1}
\FALabel(18.1202,15.8097)[br]{$q$}
\FAProp(20.,11.5)(17.,14)(0.,){/Cycles}{0}
\FALabel(17.5,11.6)[tr]{$\Pg$}
\FAProp(20.,3.)(11.,10.)(0.,){/Straight}{1}
\FALabel(15.8181,7.04616)[bl]{$q$}
\FAProp(14.,12.)(17.,14.)(0.,){/Straight}{0}
\FALabel(15.1202,13.8097)[br]{$q$}
\FAProp(14.,12.)(11.,10.)(0.,){/Straight}{0}
\FALabel(12.1202,11.8097)[br]{$q$}
\FAProp(5.,10.)(11.,10.)(0.,){/Sine}{0}
\FALabel(8.,8.93)[t]{$\gamma,\!\PZ$}
\FAVert(5.,10.){0}
\FAVert(17.,14.){0}
\FAVert(11.,10.){0}
\FAVert(14.,12.){-1}
\end{feynartspicture}
\hspace{1em}
&
\begin{feynartspicture}(82,82)(1,1)
\FADiagram{}
\FAProp(0.,15.)(6.,10.)(0.,){/Straight}{1}
\FALabel(3.18005,13.2121)[bl]{$\Pe$}
\FAProp(0.,5.)(6.,10.)(0.,){/Straight}{-1}
\FALabel(3.18005,6.78794)[tl]{$\Pe$}
\FAProp(20.,17.)(16.,13.5)(0.,){/Straight}{-1}
\FALabel(17.8154,16.2081)[br]{$q$}
\FAProp(20.,10.)(16.,13.5)(0.,){/Cycles}{0}
\FALabel(17,10.7919)[tr]{$\Pg$}
\FAProp(20.,3.)(12.,10.)(0.,){/Straight}{1}
\FALabel(15.98,5.48)[tr]{$q$}
\FAProp(6.,10.)(12.,10.)(0.,){/Sine}{0}
\FALabel(9.,8.93)[t]{$\gamma,\!\PZ$}
\FAProp(16.,13.5)(12.,10.)(0.,){/Straight}{-1}
\FALabel(13.4593,12.4365)[br]{$q$}
\FAVert(6.,10.){0}
\FAVert(16.,13.5){0}
\FAVert(12.,10.){-1}
\end{feynartspicture}
\\[-1.5em]
\begin{feynartspicture}(82,82)(1,1)
\FADiagram{}
\FAProp(0.,15.)(5.5,10.)(0.,){/Straight}{1}
\FALabel(1,13)[tr]{$\Pe$}
\FAProp(0.,5.)(5.5,10.)(0.,){/Straight}{-1}
\FALabel(-.2,8)[tl]{$\Pe$}
\FAProp(20.,17.)(15.5,13.5)(0.,){/Straight}{-1}
\FALabel(17.5,16)[br]{$q$}
\FAProp(20.,10.)(15.5,13.5)(0.,){/Cycles}{0}
\FALabel(15,9)[bl]{$\Pg$}
\FAProp(20.,3.)(12.,10.)(0.,){/Straight}{1}
\FALabel(16,5.5)[tr]{$q$}
\FAProp(5.5,10.)(12.,10.)(0.,){/Sine}{0}
\FALabel(8.75,8.93)[t]{$\gamma,\!\PZ$}
\FAProp(15.5,13.5)(12.,10.)(0.,){/Straight}{-1}
\FALabel(13.134,12.366)[br]{$q$}
\FAVert(5.5,10.){-1}
\FAVert(15.5,13.5){0}
\FAVert(12.,10.){0}
\end{feynartspicture}
\hspace{1em}
&
\begin{feynartspicture}(82,82)(1,1)
\FADiagram{}
\FAProp(0.,15.)(6.,10.)(0.,){/Straight}{1}
\FALabel(3.18005,13.2121)[bl]{$\Pe$}
\FAProp(0.,5.)(6.,10.)(0.,){/Straight}{-1}
\FALabel(3.18005,6.78794)[tl]{$\Pe$}
\FAProp(20.,17.)(16.,13.5)(0.,){/Straight}{-1}
\FALabel(17.8154,16.2081)[br]{$q$}
\FAProp(20.,10.)(16.,13.5)(0.,){/Cycles}{0}
\FALabel(17.5,10.)[tr]{$\Pg$}
\FAProp(20.,3.)(12.,10.)(0.,){/Straight}{1}
\FALabel(15.98,5.48)[tr]{$q$}
\FAProp(6.,10.)(12.,10.)(0.,){/Sine}{0}
\FALabel(9.,8.93)[t]{$\gamma,\!\PZ$}
\FAProp(16.,13.5)(12.,10.)(0.,){/Straight}{-1}
\FALabel(13.4593,12.4365)[br]{$q$}
\FAVert(6.,10.){0}
\FAVert(12.,10.){0}
\FAVert(16.,13.5){-1}
\end{feynartspicture}
\hspace{1em}
&
\begin{feynartspicture}(82,82)(1,1)
\FADiagram{}
\FAProp(0.,15.)(10.,10.)(0.,){/Straight}{1}
\FALabel(4.6318,13.2436)[bl]{$\Pe$}
\FAProp(0.,5.)(10.,10.)(0.,){/Straight}{-1}
\FALabel(4.6318,6.75639)[tl]{$\Pe$}
\FAProp(20.,17.)(15.,13.5)(0.,){/Straight}{-1}
\FALabel(18.3366,16.2248)[br]{$q$}
\FAProp(20.,10.)(15.,13.5)(0.,){/Cycles}{0}
\FALabel(17,10.7752)[tr]{$\Pg$}
\FAProp(20.,3.)(10.,10.)(0.,){/Straight}{1}
\FALabel(14.98,5.98)[tr]{$q$}
\FAProp(15.,13.5)(10.,10.)(0.,){/Straight}{-1}
\FALabel(12.3366,12.2248)[br]{$q$}
\FAVert(15.,13.5){0}
\FAVert(10.,10.){-1}
\end{feynartspicture}
\\[-1.5em]
\begin{feynartspicture}(82,82)(1,1)
\FADiagram{}
\FAProp(0.,15.)(6.,10.)(0.,){/Straight}{1}
\FALabel(3.18005,13.2121)[bl]{$\Pe$}
\FAProp(0.,5.)(6.,10.)(0.,){/Straight}{-1}
\FALabel(3.18005,6.78794)[tl]{$\Pe$}
\FAProp(20.,17.)(16.,13.5)(0.,){/Straight}{-1}
\FALabel(17.8154,16.2081)[br]{$q$}
\FAProp(20.,10.)(16.,13.5)(0.,){/Straight}{1}
\FALabel(18,10.7919)[tr]{$q$}
\FAProp(20.,3.)(12.,10.)(0.,){/Cycles}{0}
\FALabel(15.,5.48)[tr]{$\Pg$}
\FAProp(6.,10.)(12.,10.)(0.,){/Sine}{0}
\FALabel(9.,8.93)[t]{$\gamma,\!\PZ$}
\FAProp(16.,13.5)(12.,10.)(0.,){/Sine}{0}
\FALabel(13.4593,12.4365)[br]{$\gamma,\PZ$}
\FAVert(6.,10.){0}
\FAVert(16.,13.5){0}
\FAVert(12.,10.){-1}
\end{feynartspicture}
\hspace{1em}
&
\begin{feynartspicture}(82,82)(1,1)
\FADiagram{}
\FAProp(0.,15.)(6.,10.)(0.,){/Straight}{1}
\FALabel(3.18005,13.2121)[bl]{$\Pe$}
\FAProp(0.,5.)(6.,10.)(0.,){/Straight}{-1}
\FALabel(3.18005,6.78794)[tl]{$\Pe$}
\FAProp(20.,17.)(14.,10.)(0.,){/Straight}{-1}
\FALabel(17.2902,14.1827)[br]{$q$}
\FAProp(20.,3.)(14.,10.)(0.,){/Straight}{1}
\FALabel(17.2902,5.8173)[tr]{$q$}
\FAProp(20.,10.)(14.,10.)(0.,){/Cycles}{0}
\FALabel(19.,10.72)[b]{$\Pg$}
\FAProp(6.,10.)(14.,10.)(0.,){/Sine}{0}
\FALabel(10.,8.93)[t]{$\gamma,\!\PZ$}
\FAVert(6.,10.){0}
\FAVert(14.,10.){-1}
\end{feynartspicture}
\hspace{1em}
&
\begin{feynartspicture}(82,82)(1,1)
\FADiagram{}
\FAProp(0.,17.)(10.,10.)(0.,){/Straight}{1}
\FALabel(5.16337,14.2248)[bl]{$\Pe$}
\FAProp(0.,3.)(10.,10.)(0.,){/Straight}{-1}
\FALabel(4.66337,5.77519)[tl]{$\Pe$}
\FAProp(20.,17.)(10.,10.)(0.,){/Straight}{-1}
\FALabel(14.8366,14.2248)[br]{$q$}
\FAProp(20.,3.)(10.,10.)(0.,){/Straight}{1}
\FALabel(14.8366,5.77519)[tr]{$q$}
\FAProp(20.,10.)(10.,10.)(0.,){/Cycles}{0}
\FALabel(17.5,10.72)[b]{$\Pg$}
\FAVert(10.,10.){-1}
\end{feynartspicture}
\end{tabular}
}
\vspace{-1.5em}
\caption{Contributions of all possible vertex functions to $\Pe^+\Pe^-\rightarrow q\bar{q}\Pg$.} 
\label{fi:genericvertex} 
\end{figure}

\begin{figure}
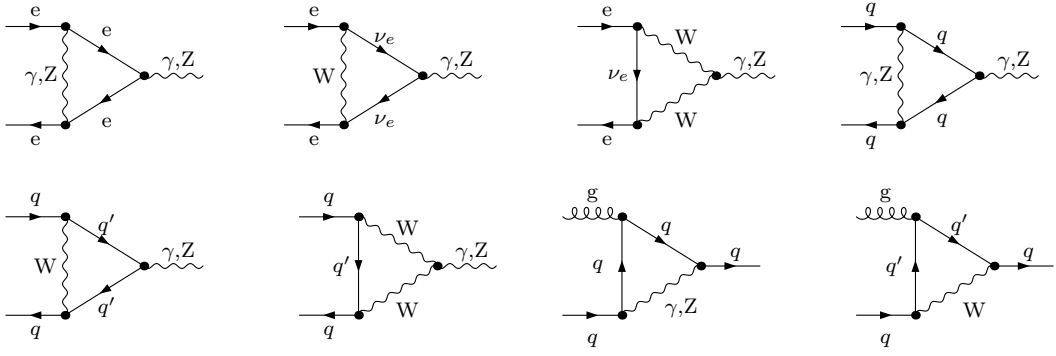
                                                  
\centerline{\footnotesize
\begin{tabular}{cccc}
\begin{feynartspicture}(82,82)(1,1)
\FADiagram{}
\FAProp(0.,15.)(6.,15.)(0.,){/Straight}{1}
\FALabel(3.,16.07)[b]{$\Pe$}
\FAProp(0.,5.)(6.,5.)(0.,){/Straight}{-1}
\FALabel(3.,3.93)[t]{$\Pe$}
\FAProp(20.,10.)(14.,10.)(0.,){/Sine}{0}
\FALabel(17.5,10.52)[b]{$\gamma,\!\PZ$}
\FAProp(6.,15.)(6.,5.)(0.,){/Sine}{0}
\FALabel(5.23,10.)[r]{$\gamma,\!\PZ$}
\FAProp(6.,15.)(14.,10.)(0.,){/Straight}{1}
\FALabel(10.2451,14)[b]{$\Pe$}
\FAProp(6.,5.)(14.,10.)(0.,){/Straight}{-1}
\FALabel(10.2451,4.81991)[b]{$\Pe$}
\FAVert(6.,15.){0}
\FAVert(6.,5.){0}
\FAVert(14.,10.){0}
\end{feynartspicture}
\hspace{1em}
&
\begin{feynartspicture}(82,82)(1,1)
\FADiagram{}
\FAProp(0.,15.)(6.,15.)(0.,){/Straight}{1}
\FALabel(3.,16.07)[b]{$\Pe$}
\FAProp(0.,5.)(6.,5.)(0.,){/Straight}{-1}
\FALabel(3.,3.93)[t]{$\Pe$}
\FAProp(20.,10.)(14.,10.)(0.,){/Sine}{0}
\FALabel(17.5,10.52)[b]{$\gamma,\!\PZ$}
\FAProp(6.,15.)(6.,5.)(0.,){/Sine}{0}
\FALabel(5.23,10.)[r]{$\PW$}
\FAProp(6.,15.)(14.,10.)(0.,){/Straight}{1}
\FALabel(10.2451,13.1801)[b]{$\nu_e$}
\FAProp(6.,5.)(14.,10.)(0.,){/Straight}{-1}
\FALabel(10.2451,4.81991)[b]{$\nu_e$}
\FAVert(6.,15.){0}
\FAVert(6.,5.){0}
\FAVert(14.,10.){0}
\end{feynartspicture}
\hspace{1em}
&
\begin{feynartspicture}(82,82)(1,1)
\FADiagram{}
\FAProp(0.,15.)(6.,15.)(0.,){/Straight}{1}
\FALabel(3.,16.07)[b]{$\Pe$}
\FAProp(0.,5.)(6.,5.)(0.,){/Straight}{-1}
\FALabel(3.,3.93)[t]{$\Pe$}
\FAProp(20.,10.)(14.,10.)(0.,){/Sine}{0}
\FALabel(17.5,10.52)[b]{$\gamma,\!\PZ$}
\FAProp(6.,15.)(6.,5.)(0.,){/Straight}{1}
\FALabel(5.23,10.)[r]{$\nu_e$}
\FAProp(6.,15.)(14.,10.)(0.,){/Sine}{0}
\FALabel(11,13.1801)[b]{$\PW$}
\FAProp(6.,5.)(14.,10.)(0.,){/Sine}{0}
\FALabel(11,4.81991)[b]{$\PW$}
\FAVert(6.,15.){0}
\FAVert(6.,5.){0}
\FAVert(14.,10.){0}
\end{feynartspicture}
&
\begin{feynartspicture}(82,82)(1,1)
\FADiagram{}
\FAProp(0.,15.)(6.,15.)(0.,){/Straight}{1}
\FALabel(3.,16.07)[b]{$q$}
\FAProp(0.,5.)(6.,5.)(0.,){/Straight}{-1}
\FALabel(3.,3.93)[t]{$q$}
\FAProp(20.,10.)(14.,10.)(0.,){/Sine}{0}
\FALabel(17.5,10.52)[b]{$\gamma,\!\PZ$}
\FAProp(6.,15.)(6.,5.)(0.,){/Sine}{0}
\FALabel(5.23,10.)[r]{$\gamma,\!\PZ$}
\FAProp(6.,15.)(14.,10.)(0.,){/Straight}{1}
\FALabel(10.2451,13.1801)[b]{$q$}
\FAProp(6.,5.)(14.,10.)(0.,){/Straight}{-1}
\FALabel(10.2451,4.81991)[b]{$q$}
\FAVert(6.,15.){0}
\FAVert(6.,5.){0}
\FAVert(14.,10.){0}
\end{feynartspicture}
\hspace{1em}
\\[-1.5em]
\begin{feynartspicture}(82,82)(1,1)
\FADiagram{}
\FAProp(0.,15.)(6.,15.)(0.,){/Straight}{1}
\FALabel(3.,16.07)[b]{$q$}
\FAProp(0.,5.)(6.,5.)(0.,){/Straight}{-1}
\FALabel(3.,3.93)[t]{$q$}
\FAProp(20.,10.)(14.,10.)(0.,){/Sine}{0}
\FALabel(17.5,10.52)[b]{$\gamma,\!\PZ$}
\FAProp(6.,15.)(6.,5.)(0.,){/Sine}{0}
\FALabel(5.23,10.)[r]{$\PW$}
\FAProp(6.,15.)(14.,10.)(0.,){/Straight}{1}
\FALabel(10.2451,13.1801)[b]{$q'$}
\FAProp(6.,5.)(14.,10.)(0.,){/Straight}{-1}
\FALabel(10.2451,4.81991)[b]{$q'$}
\FAVert(6.,15.){0}
\FAVert(6.,5.){0}
\FAVert(14.,10.){0}
\end{feynartspicture}
\hspace{1em}
&
\begin{feynartspicture}(82,82)(1,1)
\FADiagram{}
\FAProp(0.,15.)(6.,15.)(0.,){/Straight}{1}
\FALabel(3.,16.07)[b]{$q$}
\FAProp(0.,5.)(6.,5.)(0.,){/Straight}{-1}
\FALabel(3.,3.93)[t]{$q$}
\FAProp(20.,10.)(14.,10.)(0.,){/Sine}{0}
\FALabel(17.5,10.52)[b]{$\gamma,\!\PZ$}
\FAProp(6.,15.)(6.,5.)(0.,){/Straight}{1}
\FALabel(5.23,10.)[r]{$q'$}
\FAProp(6.,15.)(14.,10.)(0.,){/Sine}{0}
\FALabel(11,13.1801)[b]{$\PW$}
\FAProp(6.,5.)(14.,10.)(0.,){/Sine}{0}
\FALabel(11,4.81991)[b]{$\PW$}
\FAVert(6.,15.){0}
\FAVert(6.,5.){0}
\FAVert(14.,10.){0}
\end{feynartspicture}
&
\begin{feynartspicture}(82,82)(1,1)
\FADiagram{}
\FAProp(0.,15.)(6.,15.)(0.,){/Cycles}{0}
\FALabel(3.,16.5)[b]{$\Pg$}
\FAProp(0.,5.)(6.,5.)(0.,){/Straight}{1}
\FALabel(3.,1.8)[b]{$q$}
\FAProp(20.,10.)(14.,10.)(0.,){/Straight}{-1}
\FALabel(17.5,10.52)[b]{$q$}
\FAProp(6.,15.)(6.,5.)(0.,){/Straight}{-1}
\FALabel(3,10.)[l]{$q$}
\FAProp(6.,15.)(14.,10.)(0.,){/Straight}{1}
\FALabel(10.5,13.1801)[b]{$q$}
\FAProp(6.,5.)(14.,10.)(0.,){/Sine}{0}
\FALabel(12,4.81991)[b]{$\gamma,\!\PZ$}
\FAVert(6.,15.){0}
\FAVert(6.,5.){0}
\FAVert(14.,10.){0}
\end{feynartspicture}
\hspace{1em}
&
\begin{feynartspicture}(82,82)(1,1)
\FADiagram{}
\FAProp(0.,15.)(6.,15.)(0.,){/Cycles}{0}
\FALabel(3.,16.5)[b]{$\Pg$}
\FAProp(0.,5.)(6.,5.)(0.,){/Straight}{1}
\FALabel(3.,1.8)[b]{$q$}
\FAProp(20.,10.)(14.,10.)(0.,){/Straight}{-1}
\FALabel(17.5,10.52)[b]{$q$}
\FAProp(6.,15.)(6.,5.)(0.,){/Straight}{-1}
\FALabel(3,10.)[l]{$q'$}
\FAProp(6.,15.)(14.,10.)(0.,){/Straight}{1}
\FALabel(10.5,13.1801)[b]{$q'$}
\FAProp(6.,5.)(14.,10.)(0.,){/Sine}{0}
\FALabel(12,4.81991)[b]{$\PW$}
\FAVert(6.,15.){0}
\FAVert(6.,5.){0}
\FAVert(14.,10.){0}
\end{feynartspicture}
\end{tabular}
}
\vspace{-1.5em}
\caption{Diagrams for the $\gamma q\bar{q}$, $\PZ q\bar{q}$, $\gamma \Pe^+\Pe^-$, $\PZ \Pe^+\Pe^-$,
and $\Pg q\bar{q}$ vertex functions.} 
\label{fi:three-point} 
\end{figure}

\begin{figure}
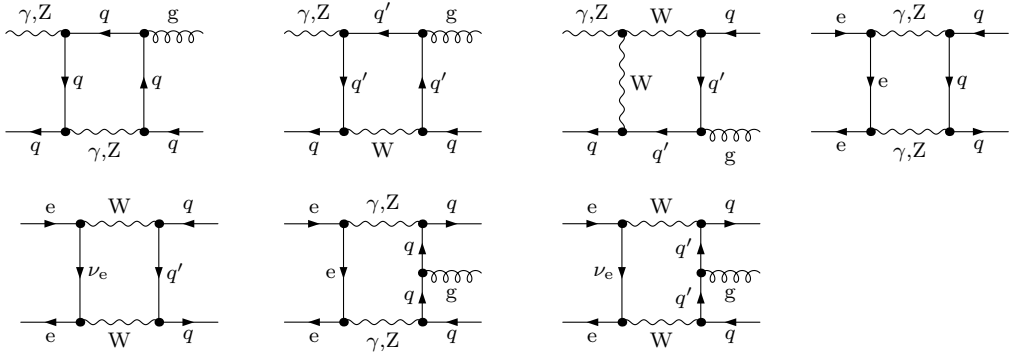
                                                   
\centerline{\footnotesize
\begin{tabular}{cccc}
\begin{feynartspicture}(82,82)(1,1)
\FADiagram{}
\FAProp(0.,15.)(6.,15.)(0.,){/Sine}{0}
\FALabel(3.,16.07)[b]{$\gamma,\!\PZ$}
\FAProp(0.,5.)(6.,5.)(0.,){/Straight}{-1}
\FALabel(3.,3.93)[t]{$q$}
\FAProp(20.,15.)(14.,15.)(0.,){/Cycles}{0}
\FALabel(17.,16.07)[b]{$\Pg$}
\FAProp(20.,5.)(14.,5.)(0.,){/Straight}{1}
\FALabel(17.,4.18)[t]{$q$}
\FAProp(6.,15.)(6.,5.)(0.,){/Straight}{1}
\FALabel(6.77,10.)[l]{$q$}
\FAProp(6.,15.)(14.,15.)(0.,){/Straight}{-1}
\FALabel(10.,16.07)[b]{$q$}
\FAProp(6.,5.)(14.,5.)(0.,){/Sine}{0}
\FALabel(10.,3.93)[t]{$\gamma,\!\PZ$}
\FAProp(14.,15.)(14.,5.)(0.,){/Straight}{-1}
\FALabel(14.77,10.)[l]{$q$}
\FAVert(6.,15.){0}
\FAVert(6.,5.){0}
\FAVert(14.,15.){0}
\FAVert(14.,5.){0}
\end{feynartspicture}
\hspace{1em}
&
\begin{feynartspicture}(82,82)(1,1)
\FADiagram{}
\FAProp(0.,15.)(6.,15.)(0.,){/Sine}{0}
\FALabel(3.,16.07)[b]{$\gamma,\!\PZ$}
\FAProp(0.,5.)(6.,5.)(0.,){/Straight}{-1}
\FALabel(3.,3.93)[t]{$q$}
\FAProp(20.,15.)(14.,15.)(0.,){/Cycles}{0}
\FALabel(17.,16.07)[b]{$\Pg$}
\FAProp(20.,5.)(14.,5.)(0.,){/Straight}{1}
\FALabel(17.,4.18)[t]{$q$}
\FAProp(6.,15.)(6.,5.)(0.,){/Straight}{1}
\FALabel(6.77,10.)[l]{$q'$}
\FAProp(6.,15.)(14.,15.)(0.,){/Straight}{-1}
\FALabel(10.,16.07)[b]{$q'$}
\FAProp(6.,5.)(14.,5.)(0.,){/Sine}{0}
\FALabel(10.,3.93)[t]{$\PW$}
\FAProp(14.,15.)(14.,5.)(0.,){/Straight}{-1}
\FALabel(14.77,10.)[l]{$q'$}
\FAVert(6.,15.){0}
\FAVert(6.,5.){0}
\FAVert(14.,15.){0}
\FAVert(14.,5.){0}
\end{feynartspicture}
\hspace{1em}
&
\begin{feynartspicture}(82,82)(1,1)
\FADiagram{}
\FAProp(0.,15.)(6.,15.)(0.,){/Sine}{0}
\FALabel(3.,16.07)[b]{$\gamma,\!\PZ$}
\FAProp(0.,5.)(6.,5.)(0.,){/Straight}{-1}
\FALabel(3.,3.93)[t]{$q$}
\FAProp(20.,15.)(14.,15.)(0.,){/Straight}{1}
\FALabel(17.,16.07)[b]{$q$}
\FAProp(20.,5.)(14.,5.)(0.,){/Cycles}{0}
\FALabel(17.,3)[t]{$\Pg$}
\FAProp(6.,15.)(6.,5.)(0.,){/Sine}{0}
\FALabel(6.77,10.)[l]{$\PW$}
\FAProp(6.,15.)(14.,15.)(0.,){/Sine}{0}
\FALabel(10.,16.07)[b]{$\PW$}
\FAProp(6.,5.)(14.,5.)(0.,){/Straight}{-1}
\FALabel(10.,3.93)[t]{$q'$}
\FAProp(14.,15.)(14.,5.)(0.,){/Straight}{1}
\FALabel(14.77,10.)[l]{$q'$}
\FAVert(6.,15.){0}
\FAVert(6.,5.){0}
\FAVert(14.,15.){0}
\FAVert(14.,5.){0}
\end{feynartspicture}
&
\begin{feynartspicture}(82,82)(1,1)
\FADiagram{}
\FAProp(0.,15.)(6.,15.)(0.,){/Straight}{1}
\FALabel(3.,16.07)[b]{$\Pe$}
\FAProp(0.,5.)(6.,5.)(0.,){/Straight}{-1}
\FALabel(3.,3.93)[t]{$\Pe$}
\FAProp(20.,15.)(14.,15.)(0.,){/Straight}{1}
\FALabel(17.,16.07)[b]{$q$}
\FAProp(20.,5.)(14.,5.)(0.,){/Straight}{-1}
\FALabel(17.,4.18)[t]{$q$}
\FAProp(6.,15.)(6.,5.)(0.,){/Straight}{1}
\FALabel(6.77,10.)[l]{$\Pe$}
\FAProp(6.,15.)(14.,15.)(0.,){/Sine}{0}
\FALabel(10.,16.07)[b]{$\gamma,\!\PZ$}
\FAProp(6.,5.)(14.,5.)(0.,){/Sine}{0}
\FALabel(10.,3.93)[t]{$\gamma,\!\PZ$}
\FAProp(14.,15.)(14.,5.)(0.,){/Straight}{1}
\FALabel(14.77,10.)[l]{$q$}
\FAVert(6.,15.){0}
\FAVert(6.,5.){0}
\FAVert(14.,15.){0}
\FAVert(14.,5.){0}
\end{feynartspicture}
\hspace{1em}
\\[-1.5em]
\begin{feynartspicture}(82,82)(1,1)
\FADiagram{}
\FAProp(0.,15.)(6.,15.)(0.,){/Straight}{1}
\FALabel(3.,16.07)[b]{$\Pe$}
\FAProp(0.,5.)(6.,5.)(0.,){/Straight}{-1}
\FALabel(3.,3.93)[t]{$\Pe$}
\FAProp(20.,15.)(14.,15.)(0.,){/Straight}{1}
\FALabel(17.,16.07)[b]{$q$}
\FAProp(20.,5.)(14.,5.)(0.,){/Straight}{-1}
\FALabel(17.,4.18)[t]{$q$}
\FAProp(6.,15.)(6.,5.)(0.,){/Straight}{1}
\FALabel(6.77,10.)[l]{$\nu_\Pe$}
\FAProp(6.,15.)(14.,15.)(0.,){/Sine}{0}
\FALabel(10.,16.07)[b]{$\PW$}
\FAProp(6.,5.)(14.,5.)(0.,){/Sine}{0}
\FALabel(10.,3.93)[t]{$\PW$}
\FAProp(14.,15.)(14.,5.)(0.,){/Straight}{1}
\FALabel(14.77,10.)[l]{$q'$}
\FAVert(6.,15.){0}
\FAVert(6.,5.){0}
\FAVert(14.,15.){0}
\FAVert(14.,5.){0}
\end{feynartspicture}
&
\begin{feynartspicture}(82,82)(1,1)
\FADiagram{}
\FAProp(0.,15.)(6.,15.)(0.,){/Straight}{1}
\FALabel(3.,16.07)[b]{$\Pe$}
\FAProp(0.,5.)(6.,5.)(0.,){/Straight}{-1}
\FALabel(3.,3.93)[t]{$\Pe$}
\FAProp(20.,15.)(14.,15.)(0.,){/Straight}{-1}
\FALabel(17.,16.07)[b]{$q$}
\FAProp(20.,10.)(14.,10.)(0.,){/Cycles}{0}
\FALabel(17.,8.4)[t]{$\Pg$}
\FAProp(20.,5.)(14.,5.)(0.,){/Straight}{1}
\FALabel(17.,4.18)[t]{$q$}
\FAProp(6.,15.)(6.,5.)(0.,){/Straight}{1}
\FALabel(5.23,10.)[r]{$\Pe$}
\FAProp(6.,15.)(14.,15.)(0.,){/Sine}{0}
\FALabel(10.,16.07)[b]{$\gamma,\!\PZ$}
\FAProp(6.,5.)(14.,5.)(0.,){/Sine}{0}
\FALabel(10.,3.93)[t]{$\gamma,\!\PZ$}
\FAProp(14.,15.)(14.,10.)(0.,){/Straight}{-1}
\FALabel(13.23,12.5)[r]{$q$}
\FAProp(14.,10.)(14.,5.)(0.,){/Straight}{-1}
\FALabel(13.23,7.5)[r]{$q$}
\FAVert(6.,15.){0}
\FAVert(6.,5.){0}
\FAVert(14.,15.){0}
\FAVert(14.,10.){0}
\FAVert(14.,5.){0}
\end{feynartspicture}
\hspace{1em}
&
\begin{feynartspicture}(82,82)(1,1)
\FADiagram{}
\FAProp(0.,15.)(6.,15.)(0.,){/Straight}{1}
\FALabel(3.,16.07)[b]{$\Pe$}
\FAProp(0.,5.)(6.,5.)(0.,){/Straight}{-1}
\FALabel(3.,3.93)[t]{$\Pe$}
\FAProp(20.,15.)(14.,15.)(0.,){/Straight}{-1}
\FALabel(17.,16.07)[b]{$q$}
\FAProp(20.,10.)(14.,10.)(0.,){/Cycles}{0}
\FALabel(17.,8.4)[t]{$\Pg$}
\FAProp(20.,5.)(14.,5.)(0.,){/Straight}{1}
\FALabel(17.,4.18)[t]{$q$}
\FAProp(6.,15.)(6.,5.)(0.,){/Straight}{1}
\FALabel(5.23,10.)[r]{$\nu_\Pe$}
\FAProp(6.,15.)(14.,15.)(0.,){/Sine}{0}
\FALabel(10.,16.07)[b]{$\PW$}
\FAProp(6.,5.)(14.,5.)(0.,){/Sine}{0}
\FALabel(10.,3.93)[t]{$\PW$}
\FAProp(14.,15.)(14.,10.)(0.,){/Straight}{-1}
\FALabel(13.23,12.5)[r]{$q'$}
\FAProp(14.,10.)(14.,5.)(0.,){/Straight}{-1}
\FALabel(13.23,7.5)[r]{$q'$}
\FAVert(6.,15.){0}
\FAVert(6.,5.){0}
\FAVert(14.,15.){0}
\FAVert(14.,10.){0}
\FAVert(14.,5.){0}
\end{feynartspicture}
\end{tabular}
}
\vspace{-1.5em}
\caption{Diagrams for the $\gamma \Pg q\bar{q}$, $\PZ \Pg q\bar{q}$, $\Pe\Pe q\bar{q}$, and
$\Pe\Pe q\bar{q}\Pg$
vertex functions.} 
\label{fi:boxpenta} 
\end{figure}

The NLO EW virtual corrections to \refeq{born_qq} and
\refeq{born_qqg} receive contributions from self-energy,
vertex, box, and in the case with a gluon in the final state,
also pentagon diagrams.  The structural diagrams for the process with
gluon emission containing the generic contributions of all possible
vertex functions are shown in \reffig{fi:genericvertex}. The
structural diagrams for the process without gluon emission can be
obtained by taking the first four and the sixth diagrams of
\reffig{fi:genericvertex} and discarding the outgoing gluon.
The Feynman diagrams contributing to the 3-point vertex functions are
shown in \reffig{fi:three-point}, those contributing to the 4-point
and 5-point vertex functions in \reffig{fi:boxpenta}.  The diagrams
for the Z-boson, photon, and quark self-energies can be found for
example in \citere{Bohm:1986rj}.
The symbol $q$ stands for the quarks appearing in
\refeq{born_qq}--\refeq{born_qqa}, the symbols $q'$ for their
weak-isospin partners.  Since we neglect the masses of the external
fermions wherever possible, there are no contributions involving the
physical Higgs boson coupling to those particles.  For b quarks in the
final state, diagrams with W~bosons also have counterparts where the
W~bosons are replaced by would-be Goldstone bosons, which are included
in the calculation but not shown explicitly in the figures.  We also
do not depict diagrams that can be obtained by reversing the charge
flow of the external quark lines in the first six diagrams of
\reffig{fi:genericvertex}.

In total we have $\mathcal{O}(200)$ contributing diagrams in the 't
Hooft--Feynman gauge for the process with gluon emission and
$\mathcal{O}(80)$ for the process without gluon emission, counting
closed-fermion-loop diagrams for each family only once.

\begin{figure}
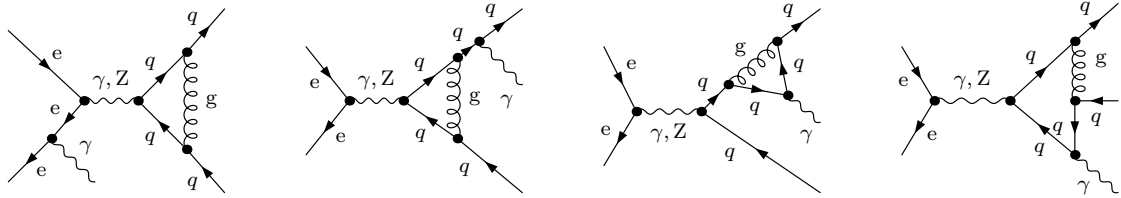
                                                 
\centerline{\footnotesize
\begin{feynartspicture}(90,90)(1,1)
\FADiagram{}
\FAProp(0.,16.5)(7.,10.)(0.,){/Straight}{1}
\FALabel(4.0747,13.9058)[bl]{$\Pe$}
\FAProp(0.,2.5)(4.,6.5)(0.,){/Straight}{-1}
\FALabel(2.61602,3.88398)[tl]{$\Pe$}
\FAProp(20.,18.5)(16.5,14.5)(0.,){/Straight}{-1}
\FALabel(17.5635,17.0407)[br]{$q$}
\FAProp(20.,1.5)(16.5,5.5)(0.,){/Straight}{1}
\FALabel(17.5635,2.95932)[tr]{$q$}
\FAProp(8.,2.5)(4.,6.5)(0.,){/Sine}{0}
\FALabel(6.61602,5.11602)[bl]{$\gamma$}
\FAProp(7.,10.)(4.,6.5)(0.,){/Straight}{1}
\FALabel(4.80315,8.77873)[br]{$\Pe$}
\FAProp(7.,10.)(12.,10.)(0.,){/Sine}{0}
\FALabel(9.5,11.07)[b]{$\gamma,\PZ$}
\FAProp(16.5,14.5)(16.5,5.5)(0.,){/Cycles}{0}
\FALabel(18.27,10.)[l]{$\Pg$}
\FAProp(16.5,14.5)(12.,10.)(0.,){/Straight}{-1}
\FALabel(13.634,12.866)[br]{$q$}
\FAProp(16.5,5.5)(12.,10.)(0.,){/Straight}{1}
\FALabel(13.634,7.13398)[tr]{$q$}
\FAVert(7.,10.){0}
\FAVert(4.,6.5){0}
\FAVert(16.5,14.5){0}
\FAVert(16.5,5.5){0}
\FAVert(12.,10.){0}
\end{feynartspicture}
\hspace{2em}
\begin{feynartspicture}(90,90)(1,1)
\FADiagram{}
\FAProp(0.,15.)(4.,10.)(0.,){/Straight}{1}
\FALabel(1.26965,12.0117)[tr]{$\Pe$}
\FAProp(0.,5.)(4.,10.)(0.,){/Straight}{-1}
\FALabel(2.73035,7.01172)[tl]{$\Pe$}
\FAProp(20.,18.5)(16.,15.5)(0.,){/Straight}{-1}
\FALabel(17.55,17.76)[br]{$q$}
\FAProp(20.,1.5)(14.,6.5)(0.,){/Straight}{1}
\FALabel(15.151,4.26558)[tr]{$q$}
\FAProp(20.,11.5)(16.,15.5)(0.,){/Sine}{0}
\FALabel(18,10)[l]{$\gamma$}
\FAProp(4.,10.)(9.,10.)(0.,){/Sine}{0}
\FALabel(6.5,11.07)[b]{$\gamma,\PZ$}
\FAProp(16.,15.5)(14.,14.)(0.,){/Straight}{-1}
\FALabel(14.55,15.51)[br]{$q$}
\FAProp(14.,6.5)(9.,10.)(0.,){/Straight}{1}
\FALabel(11.0911,7.46019)[tr]{$q$}
\FAProp(14.,6.5)(14.,14.)(0.,){/Cycles}{0}
\FALabel(15.07,10.25)[l]{$\Pg$}
\FAProp(9.,10.)(14.,14.)(0.,){/Straight}{1}
\FALabel(11.0117,12.7303)[br]{$q$}
\FAVert(4.,10.){0}
\FAVert(16.,15.5){0}
\FAVert(14.,6.5){0}
\FAVert(9.,10.){0}
\FAVert(14.,14.){0}
\end{feynartspicture}
\hspace{2em}
\begin{feynartspicture}(90,90)(1,1)
\FADiagram{}
\FAProp(0.,15.)(3.,9.)(0.,){/Straight}{1}
\FALabel(2.09361,12.9818)[bl]{$\Pe$}
\FAProp(0.,4.)(3.,9.)(0.,){/Straight}{-1}
\FALabel(0.650886,6.81747)[br]{$\Pe$}
\FAProp(20.,18.5)(16.,15.5)(0.,){/Straight}{-1}
\FALabel(17.55,17.76)[br]{$q$}
\FAProp(20.,1.5)(9.,9.)(0.,){/Straight}{1}
\FALabel(12.1553,5.65389)[tr]{$q$}
\FAProp(20.,8.)(17.,10.5)(0.,){/Sine}{0}
\FALabel(19.3409,7.34851)[tr]{$\gamma$}
\FAProp(3.,9.)(9.,9.)(0.,){/Sine}{0}
\FALabel(6.,7.93)[t]{$\gamma,\PZ$}
\FAProp(9.,9.)(11.5,11.5)(0.,){/Straight}{1}
\FALabel(9.63398,10.866)[br]{$q$}
\FAProp(16.,15.5)(17.,10.5)(0.,){/Straight}{-1}
\FALabel(17.5399,13.304)[l]{$q$}
\FAProp(16.,15.5)(11.5,11.5)(0.,){/Cycles}{0}
\FALabel(13.2002,14.1785)[br]{$\Pg$}
\FAProp(17.,10.5)(11.5,11.5)(0.,){/Straight}{-1}
\FALabel(13.9727,9.955)[t]{$q$}
\FAVert(3.,9.){0}
\FAVert(16.,15.5){0}
\FAVert(9.,9.){0}
\FAVert(17.,10.5){0}
\FAVert(11.5,11.5){0}
\end{feynartspicture}
\hspace{2em}
\begin{feynartspicture}(90,90)(1,1)
\FADiagram{}
\FAProp(0.,15.)(3.,10.)(0.,){/Straight}{1}
\FALabel(0.650886,12.1825)[tr]{$\Pe$}
\FAProp(0.,5.)(3.,10.)(0.,){/Straight}{-1}
\FALabel(2.34911,7.18253)[tl]{$\Pe$}
\FAProp(20.,19.)(16.,15.5)(0.,){/Straight}{-1}
\FALabel(17.4593,17.9365)[br]{$q$}
\FAProp(20.,10.)(16.,10.)(0.,){/Straight}{1}
\FALabel(18.,8.93)[t]{$q$}
\FAProp(20.,1.5)(16.,5.)(0.,){/Sine}{0}
\FALabel(17.4593,2.56351)[tr]{$\gamma$}
\FAProp(3.,10.)(10.,10.)(0.,){/Sine}{0}
\FALabel(6.5,11.07)[b]{$\gamma,\PZ$}
\FAProp(16.,15.5)(16.,10.)(0.,){/Cycles}{0}
\FALabel(17.92,13.7)[l]{$\Pg$}
\FAProp(16.,15.5)(10.,10.)(0.,){/Straight}{-1}
\FALabel(12.4326,13.4126)[br]{$q$}
\FAProp(16.,10.)(16.,5.)(0.,){/Straight}{1}
\FALabel(14.93,7.5)[r]{$q$}
\FAProp(16.,5.)(10.,10.)(0.,){/Straight}{1}
\FALabel(13.2559,6.14907)[tr]{$q$}
\FAVert(3.,10.){0}
\FAVert(16.,15.5){0}
\FAVert(16.,10.){0}
\FAVert(16.,5.){0}
\FAVert(10.,10.){0}
\end{feynartspicture}
}
\vspace{-1.5em}
\caption{Sample diagrams for virtual QCD corrections to the process 
$\Pe^+\Pe^-\rightarrow q\bar{q}\gamma$.} 
\label{fi:genericvertex_qqa} 
\end{figure}

The NLO QCD virtual corrections to \refeq{born_qqa} receive
contributions from self-energy, vertex, and box diagrams.  The
corresponding Feynman diagrams are shown in
\reffig{fi:genericvertex_qqa}, where we have omitted quark
self-energy contributions. We do not depict diagrams that can be
obtained by either reversing the charge flow of the external
lepton lines in the first diagram or of the external quark lines in
the last three diagrams of \reffig{fi:genericvertex_qqa}. In total we
have $\mathcal{O}(20)$ contributing diagrams in this case.

We treat the gauge-boson widths using the complex-mass scheme, which
has been worked out at the Born level in \citere{Denner:1999gp} and at
the one-loop level in \citere{Denner:2005fg}. In this framework the
masses of the Z and the W boson are complex quantities, defined at the
pole of the corresponding propagator in the complex plane. As a
consequence, derived quantities like the weak mixing angle also become
complex, and the renormalisation procedure has to be slightly
modified. Introducing complex masses everywhere in the Feynman rules
preserves all algebraic relations like Ward identities and therefore
also gauge invariance. Terms that break unitarity are beyond the
one-loop level (see \citere{Denner:2005fg}).

\subsubsection{Algebraic reduction of spinor chains}
We have performed two independent calculations of the virtual corrections.
In {\it version 1} of our calculation
we generated the amplitudes using {\sc FeynArts}~3.2
\cite{Hahn:2000kx} and employed {\sc FormCalc}~5
\cite{Hahn:1998yk} to algebraically manipulate the amplitudes, which
led to 150 different spinor structures. In order to reduce the number
of spinor structures, we applied the algorithm described in
\citere{Denner:2005fg} and extended it to the case with one external
gauge boson. In this way, we reduced all occurring spinor chains to
$\mathcal{O}(20)$ standard structures, the standard matrix elements
(SMEs), without creating coefficients that lead to numerical problems.

After the reduction of the spinor structures, we separate the matrix
elements into invariant coefficients $F_n$ and
SMEs $\hat{\mathcal{M}}$. The $F_n$ are linear
combinations of one-loop integrals with coefficients depending on
scalar kinematical variables, particle masses, and coupling factors,
the $\hat{\mathcal{M}}$ contain all spinorial objects and
the dependence on the helicities of the external particles (see
e.g.\ \citere{Denner:1991kt}):
\be
\mathcal{M}^{\sigma\sigma' \lambda}=\sum_{n} F_n^{\sigma\sigma' \lambda}\left(\{s,s_{ij},s_{ai}\}\right)
\hat{\mathcal{M}}^{\sigma\sigma' \lambda}_n \left(k_1,k_2,k_3,k_4,k_5\right).
\ee
All contributions to the matrix element involve the product of two
spinor chains corresponding to the incoming leptonic and the outgoing
hadronic current. These spinor chains can be contracted with one
another, with external momenta, or with the polarisation vector
of the outgoing photon or gluon.

In the following we describe the strategy to reduce all occurring
products of spinor chains and polarisation vectors to a few standard
structures. In this section we choose all particles incoming and
use the short-hand notation
\be
[A]^\pm_{ab}=\bar{v}_a(k_a)\,A\,\omega_\pm\,u_b(k_b)
\ee
for a spinor chain, where $\bar{v}_a(k_a)$ and $u_b(k_b)$ are spinors
for antifermions and fermions, respectively, with the chirality
projectors $\omega_\pm=(1\pm\gamma_5)/2$.  We denote the external
polarisation vector by $\varepsilon$.  Since we work with massless
external fermions, only odd numbers of Dirac matrices occur inside the
spinor chains.

The objects we want to simplify are of the form
\be
\bar{v}_1(k_1)\,A\,\omega_\rho\,u_2(k_2)\,\times\,\bar{v}_3(k_3)\,B\,
\omega_\tau\,u_4(k_4)=[A]^{\rho}_{12}[B]^{\tau}_{34}.
\ee
We make use of the Dirac algebra, the Dirac equation for the external
fermions, transversality of the polarisation vector, momentum
conservation, and relations resulting from the four-dimensionality of
space--time, which can be exploited after the cancellation of
UV divergences, which are dimensionally regularised in our work.%
\footnote{IR divergences, which may be regularised dimensionally, do
  not pose problems in this context, as shown explicitly in the
  appendix of \citere{Bredenstein:2008zb}. This fact is confirmed in
  our calculations where we alternatively used a four-dimensional mass
  regularisation scheme or dimensional regularisation, leading to the
  same IR-finite sum of virtual and real corrections.}  In four
dimensions one can relate a product of three Dirac matrices to a sum
where each term only consists of a single Dirac matrix multiplied by
either the metric tensor $g^{\mu\nu}$ or $\gamma_5$ and the totally
antisymmetric tensor $\epsilon^{\mu\nu\rho\sigma}$ through the
Chisholm identity as
\be
\gamma^\mu\gamma^\nu\gamma^\rho=g^{\mu\nu}\gamma^\rho - g^{\mu\rho}\gamma^\nu + g^{\nu\rho}\gamma^\mu
+\ri\epsilon^{\mu\nu\rho\sigma}\gamma_\sigma\gamma_5.
\ee
A further consequence of the four-dimensionality of space--time is
the fact that one can decompose the metric tensor $g^{\mu\nu}$ in
terms of four linearly independent orthonormal basis vectors $n_l$
(see e.g.\ \citeres{Denner:2003iy,Denner:2003zp})
\be
g^{\mu\nu}
=\sum_{k,l=1}^3 g_{kl}\, n_k^\mu n_l^\nu,
=n_0^\mu n_0^\nu-\sum_{l=1}^3n_l^\mu n_l^\nu,
\label{decomp-metric-tensor}
\ee
with $(n_k n_l)=g_{kl}$, where
$(g_{kl})=\mathrm{diag}(1,-1,-1,-1)$. A convenient choice of the four
vectors $n_l$ in terms of three linearly independent massless external
momenta $k_i,k_j,k_k$ is given by
\bea
n_0^\mu(k_i,k_j,k_k)&=&\frac{1}{\sqrt{2(k_ik_j)}}\left(k_i^\mu+k_j^\mu\right),\quad
n_1^\mu(k_i,k_j,k_k)=\frac{1}{\sqrt{2(k_ik_j)}}\left(k_i^\mu-k_j^\mu\right),\quad\nn\\
n_2^\mu(k_i,k_j,k_k)&=&-\frac{1}{\sqrt{2(k_ik_j)(k_ik_k)(k_jk_k)}}\Big\lbrack(k_jk_k)k_i^\mu
+(k_ik_k)k_j^\mu-(k_ik_j)k_k^\mu\Big\rbrack,\nn\\
n_3^\mu(k_i,k_j,k_k)&=&-\frac{1}{\sqrt{2(k_ik_j)(k_ik_k)(k_jk_k)}}
\epsilon^{\mu\nu\rho\sigma} k_{i,\nu}k_{j,\rho}k_{k,\sigma}.
\label{eq:nvec}
\eea
In particular, the construction of the forth independent momentum via
the totally antisymmetric tensor $\epsilon^{\mu\nu\rho\sigma}$
avoids the appearance of inverse Gram determinants.

%
For the reduction of spinor chains we use the following algorithm.  In
the first step we disconnect two spinor chains which are contracted
with each other using the decomposition \refeq{decomp-metric-tensor},
\be
\gamma_\mu \otimes\gamma^\mu=\dsl{n}_0\otimes\dsl{n}_0-\sum_{l=1}^3\dsl{n}_l\otimes\dsl{n}_l.
\ee
The choice of the external momenta in the above decomposition strongly
depends on the position of the contracted Dirac matrices inside the
spinor chain. It is advantageous to choose them in such a way that one
can make use of the Dirac equations
\mbox{$\bar{v}(k_i)\dsl{k}_i=0$}
and \mbox{$\dsl{k_i}u(k_i)=0$} and the
mass-shell condition $\dsl{k}_i^2=k_i^2=0$ in a very direct manner,
avoiding unnecessary anticommutations. We follow the algorithm
described in detail in \citere{Denner:2005fg}. After simplifying the
expressions using the identities above, there are remaining
contributions from the contraction of a basis vector $n_3$ with a
Dirac matrix inside the spinor chains, which can be eliminated by
employing the Chisholm identity as
\be
\dsl{n}_3(k_i,k_j,k_k)=-\frac{\ri\big[\dsl{k}_i\dsl{k}_j\dsl{k}_k-(k_ik_j)\dsl{k}_k+(k_ik_k)\dsl{k}_j
-(k_jk_k)\dsl{k}_i\big]\gamma_5}{\sqrt{2(k_ik_j)(k_ik_k)(k_jk_k)}}.
\label{N3slash}
\ee
In the calculation at hand, we have to deal with a maximum of three
contractions between the two spinor chains. After disconnecting them,
we are left with objects of the form
$\left[\dsl{p}\right]_{ab}^\pm\,$, where the vector $p$ can either
be an external momentum $k_j$, $j\neq a,b$, or the polarisation vector
$\varepsilon$ of the external gluon or photon.

In the next step, we reduce the spinor chains that do not involve
$\varepsilon$ using the relation
\bea
\dsl{k}_m&=&
 k_{m,\mu} g^{\mu\nu}\gamma_\nu
\stackrel{\mbox{\scriptsize{(\ref{decomp-metric-tensor})}}}{=}
k_{m,\mu}n_0^\mu\dsl{n}_0-\sum_{l=1}^3
k_{m,\mu}n_l^\mu\dsl{n}_l.
\label{loosep}
\eea
Choosing the indices $i,j,k$ in \refeq{eq:nvec} appropriately,
this allows to eliminate all external momenta in the spinor chains
apart from one for each chain via
\begin{eqnarray}
    \Bigl[ \dsl{k}_m \Bigr]_{ab}^{\pm}
   &\stackrel{\mbox{\scriptsize{(\ref{loosep})}}}{=} &
    k_{m\indexsep\mu} \,
    \sum_{i=0}^3 g^{i i} \,
                 n^{\mu}_i(k_a,k_b,k_n) \,
                 \Bigl[ \dsl{n}_i(k_a,k_b,k_n)
                 \Bigr]_{ab}^{\pm} \qquad \nn \\
  & \stackrel{\mbox{\scriptsize{(\ref{N3slash})}}}{=} &
      \frac{  (k_a k_n)\,(k_b k_m)
           -(k_a k_b)\,(k_n k_m)
           +(k_a k_m)\,(k_n k_b)
          \pm\ri\,\epsilon^{\mu\nu\rho\sigma} 
                  k_{a,\mu}k_{n,\nu}k_{b,\rho}k_{m,\sigma} }
       {2\,(k_a k_n)\,(k_b k_n)} \,
      \Bigl[ \dsl{k}_n \Bigr]_{ab}^{\pm} ,
   \label{Step2Case1}
\hspace{3em}
\end{eqnarray}
where $m\ne a,b,n$.
The described reduction allows us to express all occurring spinor
structures in terms of a linear combination of 20 SMEs 
\be
\left[\varepsilon\right]^\sigma_{12}\left[k_1\right]^\tau_{34},\,\,
\left[k_3\right]^\sigma_{12}\left[\varepsilon\right]^\tau_{34},\,\,
\mysp[\varepsilon,,k,1]\left[k_3\right]^\sigma_{12}\left[k_1\right]^\tau_{34},\,\,
\mysp[\varepsilon,,k,2]\left[k_3\right]^\sigma_{12}\left[k_1\right]^\tau_{34},\,\,
\mysp[\varepsilon,,k,3]\left[k_3\right]^\sigma_{12}\left[k_1\right]^\tau_{34}.
\label{SMEeps}
\ee
The reduction to this basis introduces at most two summands per spinor
chain.
Inserting the SMEs \refeq{SMEeps} into the amplitude
reduces its size by a factor of two. Since different reduction
strategies did not lead to more concise results we chose to use the
SMEs given in \refeq{SMEeps} in this calculation.

For the virtual corrections to $\sigma_{\mathrm{had}}$, 
where $k_5$ and $\varepsilon$ are absent, we use the four SMEs
\be
\left[k_3\right]^\sigma_{12}\left[k_1\right]^\tau_{34}.
\label{SMEsigmahad}
\ee

{\it Version 2} of our algebraic calculation starts from diagrammatic
expressions for the one-loop corrections generated by {\sc FeynArts}
1.0 \cite{Kublbeck:1990xc} and proceeds with the algebraic evaluation
of the loop amplitudes with an in-house {\sc Mathematica} program.
The algebraic manipulations do not make use of four-dimensional
identities for the Dirac chains for the $2\to3$ process, so that a
larger set of 64 SMEs had to be introduced. After rendering the
corresponding invariants $F_n$ UV~finite upon adding the counterterms
from the renormalisation and IR~finite upon including the ``endpoint
contributions'' from the subtraction function (see
\refsec{subtraction}), the SMEs are evaluated in four space--time
dimensions using the spinor formalism described in
\citere{Dittmaier:1998nn}.  For the $2\to2$ process the only Dirac
structure that involves divergences is
$[\gamma^\mu]^\sigma_{12}[\gamma_\mu]^\tau_{34}$, and all other SMEs
are proportional to these, as can be deduced from the identities given
in (3.10) of \citere{Denner:2005fg}.

\subsubsection{Evaluation of the loop integrals}
The tensor integrals are evaluated as in the calculation of
\citeres{Denner:2005es,Denner:2005fg}, i.e.\ they are numerically
reduced to scalar master integrals. The master integrals are computed
using complex masses according to
\citeres{'tHooft:1978xw,Beenakker:1988jr,Denner:1991qq}, using two
independent Fortran implementations which are in mutual agreement.
Results for different regularisation schemes are translated into each
other with the method of \citere{Dittmaier:2003bc}.  Tensor and scalar
5-point functions are directly expressed in terms of 4-point integrals
\cite{Denner:2002ii,Denner:2005nn}. Tensor 4-point and 3-point
integrals are reduced to scalar integrals with the Passarino--Veltman
algorithm \cite{Passarino:1978jh}. Although we already find sufficient
numerical stability with this procedure, we apply the dedicated
expansion methods of \citere{Denner:2005nn} in exceptional phase-space
regions where small Gram determinants appear.

UV divergences are regularised dimensionally. For the IR, i.e.\ soft
or collinear, divergences we either use pure dimensional
regularisation with massless gluons, photons, and fermions (except for
the top quark), or pure mass regularisation with infinitesimal photon,
gluon, and small fermion masses, which are only kept in the
mass-singular logarithms.  When using dimensional regularisation, the
rational terms of IR origin are treated as described in Appendix~A of
\citere{Bredenstein:2008zb}.

\subsection{Real Corrections}
\label{sec:real}

In this section we describe how we evaluate the last two terms in
\refeq{master_eq}. The real radiation corrections to the total
hadronic cross section [last term in \refeq{master_had}] are computed
along the same lines.

The processes we consider are given by
\begin{align}
&\Pe^+(k_1,\sigma_1)+\Pe^-(k_2,\sigma_2)\rightarrow 
q(k_3,\sigma_3)+\bar{q}(k_4,\sigma_4)
+\Pg(k_5,\lambda_1)+\gamma(k_6,\lambda_2),\quad q=\mathrm{u,d,c,s,b},
\label{process-brems} 
\\
&\Pe^+(k_1,\sigma_1)+\Pe^-(k_2,\sigma_2)\rightarrow 
q(k_3,\sigma_3)+\bar{q}(k_4,\sigma_4)
+q(k_5,\sigma_5)+\bar{q}(k_6,\sigma_6)
,\quad q=\mathrm{u,d,c,s,b}.
\label{process-4q}
\end{align}
The corresponding cross section is obtained as 
\bea
\int\rd\sigma_{\mathrm{real}} &=& \frac{1}{2s} \,
\fc\sum_{\sigma,\sigma'=\pm\frac{1}{2}}
\frac{1}{4}(1-2P_1\sigma)(1+2P_2\sigma)\,
\biggl\{
\sum_{\lambda_1,\lambda_2=\pm 1}
\int\mathrm{d}\Phi_4 \,
\vert\mathcal{M}_{\mathrm{real},q\bar q
  g\gamma}^{\sigma\sigma'\lambda_1\lambda_2}\vert^2 \,
\Theta_{\mathrm{cut}}\lrb \Phi_4 \rrb  \nonumber \\
& & {}+ \sum_{\sigma''=\pm\frac{1}{2}}
\int\mathrm{d}\Phi_4 \,
2\Re\left\{
\left(\mathcal{M}_{\mathrm{real},q\bar q q\bar q}^{\sigma\sigma'\sigma'',\mathrm{EW}}\right)^*
\mathcal{M}_{\mathrm{real},q\bar q q\bar q}^{\sigma\sigma'\sigma'',\mathrm{QCD}}
\right\}
\Theta_{\mathrm{cut}}\lrb \Phi_4 \rrb\biggr\} ,
\label{eq:sigma_real}
\eea
where we have used helicity conservation to simplify the helicity sum,
such that $\sigma$ denotes the helicity of the incoming electron, and
$\sigma'$ and $\sigma''=\sigma_5=-\sigma_6$ the helicities of the
outgoing quarks.  
The four-particle phase-space volume element reads
\be
\rd\Phi_4=\frac{1}{\left(2\pi\right)^{12}}\frac{\rd^3 \vec k_3}{2k_3^0}
\frac{\rd^3 \vec k_4}{2k_4^0}\frac{\rd^3 \vec k_5}{2k_5^0}
\frac{\rd^3 \vec k_6}{2k_6^0}
\left(2\pi\right)^{4}\delta^{(4)}\left(k_1+k_2-k_3-k_4-k_5-k_6\right),
\label{ps4}
\ee
and $\fc=4$ for both contributions. As in \refeq{eq:sigma0},
$\Theta_{\mathrm{cut}}\lrb \Phi_4 \rrb$ represents cuts used in the
event selection. 

\begin{figure}
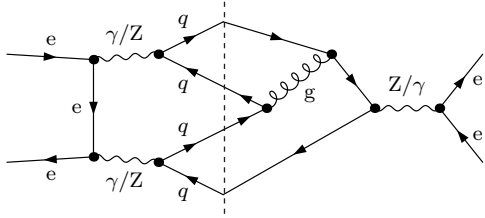
                                                 
\centerline{\footnotesize
\begin{feynartspicture}(200,90)(2,1)
\FADiagram{}
\FAProp(0.,15.)(8.,14.5)(0.,){/Straight}{1}
\FALabel(4.09669,15.817)[b]{$\Pe$}
\FAProp(0.,5.)(8.,5.5)(0.,){/Straight}{-1}
\FALabel(4.09669,4.18302)[t]{$\Pe$}
\FAProp(20.,18.)(14.,15.)(0.,){/Straight}{-1}
\FALabel(16.7868,17.4064)[br]{$q$}
\FAProp(20.,12.)(14.,15.)(0.,){/Straight}{1}
\FALabel(16.7868,12.5936)[tr]{$q$}
\FAProp(20.,8.)(14.,5.)(0.,){/Straight}{-1}
\FALabel(16.7868,7.40636)[br]{$q$}
\FAProp(20.,2.)(14.,5.)(0.,){/Straight}{1}
\FALabel(16.7868,2.59364)[tr]{$q$}
\FAProp(8.,14.5)(8.,5.5)(0.,){/Straight}{1}
\FALabel(6.93,10.)[r]{$\Pe$}
\FAProp(8.,14.5)(14.,15.)(0.,){/Sine}{0}
\FALabel(10.8713,15.8146)[b]{$\gamma/\PZ$}
\FAProp(8.,5.5)(14.,5.)(0.,){/Sine}{0}
\FALabel(10.8713,4.18535)[t]{$\gamma/\PZ$}
\FAVert(8.,14.5){0}
\FAVert(8.,5.5){0}
\FAVert(14.,15.){0}
\FAVert(14.,5.){0}
\FAProp(44.,15.)(40.,10.)(0.,){/Straight}{-1}
\FALabel(44.,13.0117)[tr]{$\Pe$}
\FAProp(44.,5.)(40.,10.)(0.,){/Straight}{1}
\FALabel(44.,7.91172)[tr]{$\Pe$}
\FAProp(20.,18.)(30.,15.)(0.,){/Straight}{1}
\FAProp(24.,10.)(30.,15.)(0.,){/Cycles}{0}
\FALabel(28.,10.6)[b]{$\Pg$}
\FAProp(40.,10.)(34.,10.)(0.,){/Sine}{0}
\FALabel(37.,11.07)[b]{$\PZ/\gamma$}
\FAProp(30.,15.)(34.,10.)(0.,){/Straight}{1}
\FAProp(20.,2.)(34.,10.)(0.,){/Straight}{-1}
\FAProp(24.,10.)(20.,12.)(0.,){/Straight}{1}
\FAProp(24.,10.)(20.,8.)(0.,){/Straight}{-1}
\FAVert(40.,10.){0}
\FAVert(30.,15.){0}
\FAVert(24.,10.){0}
\FAVert(34.,10.){0}
\DashLine(96,5)(96,85){2}
\end{feynartspicture}
}
\vspace{-1.0em}
\caption{Sample diagram for a non-trivial interference
between EW and QCD diagrams in
$\Pe^+\Pe^-\rightarrow q\bar{q}q\bar{q}$.} 
\label{fi:interference-eeqqqq} 
\end{figure}

We only consider the ${\cal O}(\alpha^3\alpha_{\mathrm{s}})$
corrections in our analysis. At this order,
the process (\ref{process-4q}) receives a contribution only from the
interference of EW with QCD amplitudes, as illustrated in
Fig.~\ref{fi:interference-eeqqqq}.  Owing to colour conservation, this
interference term is only non-zero for identical quark flavours
and for $\sigma'=\sigma''$. It is
non-singular over the full phase space defined by the event-selection
cuts.

The integral of the process (\ref{process-brems}) over the
four-particle phase space contains IR divergences due to the emission
of a soft or collinear photon or gluon.  Prior to numerical
implementation, one has to isolate these divergences and combine them
with the corresponding contributions from the virtual corrections. In
our implementation, we use three different methods for this task: two
variants of the phase-space slicing method
\cite{Baer:1988ux,Giele:1991vf,Giele:1993dj,Bohm:1993qx,Dittmaier:1993da,Wackeroth:1996hz,Baur:1998kt},
and the dipole subtraction method
\cite{Catani:1996vz,Dittmaier:1999mb,Dittmaier:2008md}.
In phase-space slicing, the phase space is split into singular and
non-singular regions. In the singular regions the integration over the
singular variables is performed analytically, while the non-singular
regions are integrated over fully numerically.  In the dipole
subtraction method, a subtraction function that mimics the singular
behaviour of the integrand is added and subtracted, leaving a finite
four-particle phase-space integration and a remainder, where the
integration that leads to singularities is carried out analytically.
Both methods rely on factorisation properties of the matrix elements
and of the phase space in the soft and collinear limits. The
singularities are treated analytically so that a numerical integration
does not pose any problems.

Both methods described above are valid for NLO calculations. We employ
them separately for the photon and the gluon in the calculation at
hand. However, they are not sufficient in the region where both the
photon and the gluon become soft or collinear at the same time.  This
region corresponds to two-jet production where the photon and the
gluon are unresolved.  In this region close to the kinematic endpoint
in the event-shape distributions fixed-order calculations are in any
case not appropriate. To prevent problems with numerical stability from
this region, we impose a lower cut-off on each event-shape variable in
the first bin of the distribution that corresponds to two-jet
production.

To be able to compare the results of our calculation to experimental
measurements and to improve the accuracy of the theoretical prediction
relevant for the $\alpha_{\mathrm{s}}$ determination, we have to
incorporate the kinematical cuts used in the specific experiment.  For
the event-selection cuts, $\Theta_{\mathrm{cut}}$, we apply the
procedure used by the ALEPH experiment, as described in \refsec{PI}.
Electromagnetic jets are removed by imposing an upper cut on the
photon energy in the jet. However, the cut on the photon energy also
removes events with a highly energetic photon collinear to a quark or
antiquark in the final state that lead to a configuration where the
photon and quark are in the same jet. The left-over collinear
singularity associated with this isolated-photon rejection is properly
accounted for by a contribution from the quark-to-photon fragmentation
function
(see next section).

Collinear photon emission in the initial state is regulated by the
mass of the electron and leads to large contributions of the form
$\alpha^n\ln^n(m_\Pe^2/s)$. The cut on the visible invariant mass
$M_{\mathrm{vis}}$ removes part of the hard photon emission collinear
to the incoming beam particles and therefore suppresses large
mass-singular logarithms.

\section{Treatment of soft and/or collinear singularities}
\label{sec:realcorr}
\setcounter{equation}{0}

In this section we describe the treatment of the soft and collinear
singularities for the process \refeq{process-brems}.

\subsection{Phase-space slicing}
\label{sec::softcoll}

In the phase-space slicing method, technical cut parameters are
introduced to
decompose the real radiation phase space into regions corresponding to
soft or collinear configurations (unresolved regions), and a resolved
region that is free from singularities. Consequently, the cross
section decomposes into a soft, a collinear, and a finite part. Since
we apply an isolation cut on the final-state photons (i.e.\ on a
specific, identified final-state particle), we must also include a
fragmentation contribution, so that the total real contribution
reads
\be
\rd\sigma_{\mathrm{real}}=
\rd\sigma_{\mathrm{soft}}+\rd\sigma_{\mathrm{coll}}
-\rd\sigma_{\mathrm{frag}}
+\rd\sigma_{\mathrm{finite}}.
\label{eq::realsum}
\ee In the following sections we first review the method for
collinear-safe observables, the necessary modifications for
non-collinear-safe event-selection cuts are described in
\refsecs{ncsslice} and \ref{quark-to-photonfragfac}.

\subsubsection{Collinear-safe observables}
\label{csslice}

Soft and collinear contributions contain IR divergences which are
evaluated analytically, while the finite contribution is evaluated
numerically.  Since no quark-mass singularities from final-state
radiation remain for collinear-safe observables, no fragmentation
contribution is necessary in this case, i.e.\ 
$\rd\sigma_{\mathrm{frag}}=0$ in \refeq{eq::realsum}.  We implemented
two different variants of phase-space slicing: the two-cut-off slicing
according to \citeres{Baur:1998kt,Denner:2000bj}, which uses mass
regularisation, and the one-cut-off slicing of
\citeres{Giele:1991vf,Giele:1993dj} within dimensional regularisation.
The application of both methods is described in detail below for
IR-divergent contributions due to unresolved photons. The unresolved
gluon contributions are obtained by an appropriate replacement of the
coupling constants.

In two-cut-off slicing, the splitting of the phase space into singular
and non-singular parts is achieved by introducing a cut $\delta_{\mathrm{s}}$ on
the photon energy $E_\gamma<\delta_{\mathrm{s}}\sqrt{s}/2=\Delta E$ in the CM
frame. The collinear region is defined by $E_\gamma>\Delta E$ and
$1>\cos\theta>1-\delta_{\mathrm{c}}$, where $\theta$ is the 
smallest angle between the photon and any charged fermion in the CM system.

In the soft region the squared matrix element
$\vert\mathcal{M}_{\mathrm{real}}\vert^2$ and the phase-space element
$\rd\Phi_4$ factorise so that we can apply the soft-photon
approximation, e.g.\ given in \citeres{Yennie:1961ad,Denner:1991kt},
and introduce an infinitesimal photon mass $m_\gamma$ as a regulator.
The resulting soft-photon contribution is
\bea
\rd\sigma_{\mathrm{soft}}&=&\rd\sigma_{\mathrm{Born}} \,
\frac{\alpha}{2\pi}\sum_{I=1}^{4}\sum_{J=I+1}^{4}(-1)^{I+J}Q_IQ_J
\left\{
2\ln\left(\frac{2\Delta E}{m_\gamma}\right)\left[
2-\ln\left(\frac{s_{IJ}^2}{m_I^2m_J^2}\right)
\right] \right.
\nn\\
&&{}
\left.
{}-2\ln\left(\frac{4E_IE_J}{m_Im_J}\right)
+\frac{1}{2}\ln^2\left(\frac{4E_I^2}{m_I^2}\right)+
\frac{1}{2}\ln^2\left(\frac{4E_J^2}{m_J^2}\right)+\frac{2\pi^2}{3}+
2\Li\left(1-\frac{4E_IE_J}{s_{IJ}}\right)
\right\},
\hspace{2em}
\eea
where we keep the masses of the fermions as regulators for the collinear 
singularities ($E_I\gg m_I$), 
i.e.\ we have $m_1=m_2=\Me$ and $m_3=m_4=m_q$,
and $\rd\sigma_{\mathrm{Born}}$ denotes the 
Born cross section for $\Pe^+\Pe^- \to q\bar q g$. 

Hard collinear contributions arise from the limit where the photon
becomes collinear either with one of the incoming beams or with the
outgoing (anti-)quark. These contributions contain mass-singular
logarithms, which are regularised by the fermion masses. Their
integrated forms are again proportional to the Born cross section for
$\Pe^+\Pe^- \to q\bar q g$. 
We split the collinear photon
contributions into those from initial-state radiation and those from
final-state radiation:
\be \rd\sigma_{\mathrm{coll}}
=\rd\sigma^{{\rm initial}}_{\mathrm{coll}} 
+ \rd\sigma^{{\rm final}}_{\mathrm{coll}}. 
\ee
In the case of initial-state photon emission, the available
$\Pe^+\Pe^-$ CM energy is reduced, and the Born process is probed at
this reduced energy.  Moreover, the mass regularisation introduces a
spin-flip term. We indicate these two features in the argument of the
Born cross section, where $k_i$ and $P_i$ being are the momenta and
the degrees of polarisation of the respective beams.  The integrated
initial-state and final-state collinear contributions thus read 
\bea
\rd\sigma_{\mathrm{coll}}^{\mathrm{initial}} &=&\sum_{a=1}^2
\frac{\alpha}{2\pi}Q_a^2\int_0^{1-\delta_{\mathrm{s}}} \rd
z\biggl\{\rd\sigma_{\mathrm{Born}}\left(k_a\to zk_a\right)
P_{ff}(z)\left[\ln\left(\frac{{s}}{m_a^2}\frac{\delta_{\mathrm{c}}}{2}\frac{1}{z}\right)
  -1\right]
\nn\\
&&\hspace*{33.5mm}{}+\rd\sigma_{\mathrm{Born}} \left(k_a\to
  zk_a,P_a\to -P_a\right)(1-z)\biggr\}
\label{eq.:collin}
\eea
and
\bea
\rd \sigma_{\mathrm{coll}}^{\mathrm{final}}&=&
\sum_{i=3}^4
\frac{\alpha}{2\pi}Q_i^2\, \rd\sigma_{\mathrm{Born}}
\left\{
\left[
\frac{3}{2}+2\ln\left(\frac{\Delta E}{E_i}\right)
\right]
\left[
1-\ln\left(\frac{2E_i^2\delta_{\mathrm{c}}}{m_i^2}\right)
\right]
+3-\frac{2\pi^2}{3}\right\},
\label{slicing_final}
\eea
where 
\be 
P_{ff}(z)=\frac{1+z^2}{1-z} 
\ee 
denotes the $f\to f\gamma$ splitting function.  The parameters
$\delta_{\mathrm{s}}$ and $\delta_{\mathrm{c}}$ govern the splitting
of the phase space in the different regions, but the final result does
not depend on them. They have to be chosen small enough to guarantee
that the applied approximations are valid. Therefore, varying these
parameters in a certain range and showing independence of the results
serves as an important check of the calculation.

In the one-cut-off slicing, soft and collinear regions are defined by a
cut $y_{{\rm min}}$ on two-particle invariants $s_{IJ} = (k_I+k_J)^2$
normalised to the $\Pe^+\Pe^-$ 
CM energy squared,
$y_{IJ} = s_{IJ}/s$. The soft region of parton $I$ is defined by 
$|y_{IJ}| < y_{{\rm min}}$ for all $J$, 
while the collinear region of partons $I$ and $J$ 
is defined by $|y_{IJ}| < y_{{\rm min}}$, while all other invariants $|y_{IK}| \ge
 y_{{\rm min}}$ for $K\neq J$. 

Following \citere{Giele:1993dj}, the soft and collinear divergent 
contributions are first calculated in the unphysical kinematical situation 
of all particles outgoing, and later on continued to the physical kinematics 
by means of a crossing function, 
 such that 
\be 
\rd\sigma_{\mathrm{coll}}
=\rd\sigma^{{\rm out}}_{\mathrm{coll}} 
+ \rd\sigma^{{\rm cross}}_{\mathrm{coll}} 
.\ee
The combination of the first term with the soft contribution 
yields \cite{Giele:1991vf}
\bea
\rd\sigma_{\mathrm{soft}}
+\rd\sigma^{{\rm out}}_{\mathrm{coll}}&=&\rd\sigma_{\mathrm{Born}}
\frac{\alpha}{2\pi} \,   \frac{(4\pi)^\e}{\Gamma(1-\e)}
\sum_{I,J=1 \atop I\ne J}^{4}(-1)^{I+J}Q_IQ_J\nn\\
&&\times
\left[\frac{1}{\e^2} \left(\frac{\mu^2}{|s_{IJ}|}\right)^\e + \frac{3}{2\e}
\left(\frac{\mu^2}{|s_{IJ}|}\right)^\e
-\ln^2 \left( \frac{|y_{IJ}|}{y_{{\rm min}}}\right)
+ \frac{3}{2} \ln \left( \frac{|y_{IJ}|}{y_{{\rm min}}}\right)
+ \frac{7}{2} - \frac{\pi^2}{3} 
\right]\,,\quad
\label{eq:sliceff}
\eea
where we used dimensional regularisation in $d=4-2\e$ dimensions with
mass parameter $\mu$ required to maintain a dimensionless coupling.

Kinematical crossing of $\Pe^+\Pe^-$ to the initial state introduces 
crossing functions, which were derived for QCD using factorisation 
of parton distributions in the $\overline{{\rm MS}}$ scheme in
\citere{Giele:1993dj}. For photon radiation off incoming electrons,
this $\overline{{\rm MS}}$ expression is converted using
\citere{Baur:1998kt} to a mass-regularised form, involving the
physical electron mass.  This results in the following crossing term,
\bea
\label{eq:onecutcoll}
\rd \sigma_{\mathrm{coll}}^{\mathrm{cross}}
&=&\sum_{a=1,2}\frac{\alpha}{2\pi}Q_a^2\int_0^{1}
\rd z \, \rd\sigma_{\mathrm{Born}}\left(k_a\to zk_a\right) \, \Biggl\{ 
\left[ \frac{\pi^2}{3} - \frac{5}{4} \right] \delta(1-z) 
\nonumber \\
&& + \left[ P_{ff}(z) \left( \ln (y_{{\rm min}}) +
    \ln\left(\frac{s}{m_a^2(1-z)}\right) -1\right) +1-z\right]_+ \Biggr\},
\eea
where we use the usual $\lsb\dots\rsb_+$ prescription
\be
\int_0^1\rd x \lsb f(x)\rsb_+g(x)=\int_0^1\rd x f(x)\lsb g(x)-g(1)\rsb.
\ee
The results \refeq{eq:sliceff} and \refeq{eq:onecutcoll} apply to
unpolarized cross sections, but the polarization effects of the incoming
particles can be easily restored
by the spin-flip terms of \refeq{eq.:collin},
\be
\rd \sigma_{\mathrm{coll}}^{\mathrm{pol}} =
\rd \sigma_{\mathrm{coll}} 
+\sum_{a=1,2}\frac{\alpha}{2\pi}Q_a^2\int_0^{1} \rd z \, 
\Bigl[\rd\sigma_{\mathrm{Born}}\left(k_a\to zk_a,P_a\to -P_a\right)
-\rd\sigma_{\mathrm{Born}}\left(k_a\to zk_a\right) \Bigr] (1-z).
\label{eq:onecutcoll_pol}
\ee

\subsubsection{Non-collinear-safe observables}
\label{ncsslice}
The upper cut on the photon energy inside jets affects the slicing
procedure. Imposing this cut in the soft and finite parts of 
\refeq{eq::realsum} is trivial. Since the cut acts
outside the soft region, this part is not affected at all.
Only the collinear singular part needs some non-trivial
adjustments.
More precisely, only collinear final-state radiation needs to be
considered, since photons collinear to the incoming electrons or
positrons can never appear inside jets within detector coverage.
In \refsec{csslice} we parametrised the collinear region in
terms of the energy fraction $z$ carried by the (anti-)quark that
results from the splitting. The experimental cut, however, is a cut on
the energy fraction $1-z$ of the photon after the splitting, i.e.\ 
the cut on the energy fraction of the photon reads
$1-z< z_{\mathrm{cut}}$.  

In the two-cut-off slicing approach, imposing this cut
generalises \refeq{slicing_final} to
\be
\mathrm{d}\sigma_{\mathrm{coll}}^{\mathrm{final}}(z_{\mathrm{cut}})
=\sum_{i=3}^4
\frac{\alpha}{2\pi}Q_i^2
\rd\sigma_{\mathrm{Born}} \, 
\int_{1-z_{\mathrm{cut}}}^{1-\Delta E/E_i}\mathrm{d}z \Bigg\{
P_{ff}(z)\left[\mathrm{ln}\left(\frac{2E_i^2\delta_{\mathrm{c}}}{m_i^2}z^2\right)-1\right]
+ (1-z)\Bigg\}.
\ee
Performing the integration yields
\begin{align}
\label{cutint}
\mathrm{d}\sigma_{\mathrm{coll}}^{\mathrm{final}}(z_{\mathrm{cut}})
=&\sum_{i=3}^4\frac{\alpha}{2\pi}Q_i^2 \,
\rd\sigma_{\mathrm{Born}}\left\{
\left[
-2 z_{\mathrm{cut}}  + \frac{z_{\mathrm{cut}}^2}{2} - 2\ln\left(\frac{\Delta E/E_i}{z_{\mathrm{cut}}}\right) 
\right]\ln\left(\frac{2E_i^2\delta_{\mathrm{c}}}{m_i^2}\right)
+2\ln\left(\frac{\Delta E/E_i}{z_{\mathrm{cut}}}\right)
\right. \nn
\\
&\left.{}
-4\Li\left(z_{\mathrm{cut}}\right)
+(1-z_{\mathrm{cut}})(3-z_{\mathrm{cut}}) \ln(1-z_{\mathrm{cut}})\right.
+5z_{\mathrm{cut}}-\frac{z_{\mathrm{cut}}^2}{2}
\Bigg\}.
\end{align}
In \refeq{cutint} we have the original
dependence on the slicing parameters and the mass regulators, already
present in \refeq{slicing_final}, plus an additional term that depends
on the slicing parameters, the mass regulators, and the cut on the
photon energy. It is exactly this term that gives rise to left-over
singularities. 

Within the one-cut-off approach, a cut on the 
final-state photon energy in jets  can be conveniently taken into
account by including 
a collinear photon contribution in $\rd\sigma_{\mathrm{coll}}$,
\be \rd\sigma_{\mathrm{coll}}
=\rd\sigma^{{\rm out}}_{\mathrm{coll}} 
+ \rd\sigma^{{\rm cross}}_{\mathrm{coll}} 
- \rd\sigma^{{\gamma}}_{\mathrm{coll}}(z_{\mathrm{cut}}),
\ee
which subtracts the contributions of collinear photons with energies
above $z_{\mathrm{cut}}$. 
This contribution reads
\begin{eqnarray}
\mathrm{d}\sigma_{\mathrm{coll}}^{\gamma}(z_{\mathrm{cut}})
&=& \sum_{i=3}^4
\frac{\alpha}{2\pi}Q_i^2
\rd\sigma_{\mathrm{Born}} \, 
\int^{1-z_{\mathrm{cut}}}_{0}\mathrm{d}z \,
\Biggl\{\frac{(4\pi\mu^2)^\epsilon}{\Gamma(1-\epsilon)}\,
\frac{P_{ff}^{(\e)}(z)}{\left[z(1-z)\right]^\epsilon} 
\int_0^{s_{\mathrm{min}}}
\rd s_{q\gamma}\frac{1}{s_{q\gamma}^{1+\epsilon}}
\Biggr\}\nonumber \\
&=& -\sum_{i=3}^4
\frac{\alpha}{2\pi}Q_i^2
\rd\sigma_{\mathrm{Born}} \, 
\int^{1-z_{\mathrm{cut}}}_{0}\mathrm{d}z \,\Biggl\{ 
\frac{1}{\e} \,\left(\frac{4\pi\mu^2}{s_{\mathrm{min}}}\right)^\epsilon\frac{1}{\Gamma(1-\epsilon)}
\frac{P^{(\epsilon)}_{ff}(z)}{\left[z(1-z)\right]^\epsilon} \Biggr\}\,,
\label{eq:sliceint}
\end{eqnarray}
where 
the $\epsilon$-dependent splitting function $P^{(\epsilon)}_{ff}$ is given by 
\be 
P^{(\epsilon)}_{ff}(z)=\frac{1+z^2-\epsilon\,(1-z)^2}{1-z}.  
\ee
The derivation of this collinear unresolved photon factor is described in 
detail in~\citeres{Glover:1993xc,Poulsen:2006}. 

In both slicing approaches, we observe that the introduction of the cut on the 
final-state photon energy results in uncompensated collinear singularities 
in (\ref{cutint}) and (\ref{eq:sliceint}). These are properly accounted 
for by factorisation of the quark-to-photon fragmentation function, 
explained in detail in 
Section~\ref{quark-to-photonfragfac} below.

\subsection{Dipole subtraction}
\label{subtraction}

The basic idea of any subtraction method is to subtract an auxiliary
function from the integrand that features the same singular behaviour
in the soft and collinear limits. The partially integrated auxiliary
function is then added to the virtual corrections (and counterterms
from the factorisation of parton distributions and fragmentation
functions) resulting in analytic cancellation of IR singularities.  We
use the dipole subtraction method, first introduced in
\citere{Catani:1996vz} for massless QCD and later generalised to
massive fermions for collinear-safe and non-collinear-safe observables
in \citere{Dittmaier:1999mb} and \citere{Dittmaier:2008md}, respectively. 
Since we regulate IR
divergences with particle masses, we follow the description of
\citeres{Denner:2000bj,Dittmaier:1999mb,Dittmaier:2008md}. 
Again we first describe the method
for the case of an unresolved photon; the gluon case is obtained by
a suitable replacement of couplings. In this section we suppress the
weighted sum over initial-state polarisations.

\subsubsection{Collinear-safe observables}
\label{cssub}

Suppressing flux and colour factors, the
$\Theta$-function related to the phase-space cuts, 
dipole subtraction is based on the general formula
\be
\int\rd\Phi_4\sum_\lambda\vert\mathcal{M}_{\mathrm{real}}\vert^2=
\int\rd\Phi_4\lrb\sum_\lambda\vert\mathcal{M}_{\mathrm{real}}\vert^2-
\vert\mathcal{M}_{\mathrm{sub}}\vert^2\rrb
+\int\rd\Phi_4\,\vert\mathcal{M}_{\mathrm{sub}}\vert^2,
\label{master_sub}
\ee 
where $\lambda$ labels the photon polarisation.
The subtraction function $\mathcal{M}_{\mathrm{sub}}$ is
constructed from ordered pairs $IJ$ of charged fermions, where fermion
$I$ is called the emitter and fermion $J$ the spectator. Only the
kinematics of the emitter leads to singularities. The spectator fermion
is used to balance energy-momentum conservation when combining the
momentum of the photon with the momentum of the emitter. Making the
dependence of the subtraction function on the emitter--spectator pair
explicit, we can write
\be
\vert\mathcal{M}_{IJ,\mathrm{sub}}\lrb\Phi_4\rrb\vert^2=
-(-1)^{I+J}Q_IQ_Je^2\sum_{\tau=\pm}g_{IJ,\tau}^{(\mathrm{sub})}
\lrb k_I,k_J,k\rrb\vert\mathcal{M}_{\mathrm{Born}}(\tilde{\Phi}_{3,IJ},
\tau\sigma_I)\vert^2,
\label{def_subtractionij}
\ee where $\sigma_I$ is the helicity of the emitter and $k=k_6$ the
photon momentum.  
The sum over $\tau=\pm$ takes care of a possible spin-flip resulting from
collinear photon emission.
The subtraction function
$\mathcal{M}_{IJ,\mathrm{sub}}$ depends on the whole four-particle
phase space $\Phi_4$, whereas the Born matrix elements depend only on
three-particle phase spaces $\tilde{\Phi}_{3,IJ}$.  The mappings of
the four-particle on the three-particle phase spaces, which are
different for each emitter--spectator pair $IJ$ and explicitly given
in \citere{Dittmaier:1999mb}, ensure proper factorisation in each
singular limit.

In the massless case, the dipole factors are explicitly given by
\bea
g_{ij,+}^{(\mathrm{sub})}(k_i,k_j,k)&=&\frac{1}{(k_ik)\lrb1-y_{ij}\rrb}
\lsb\frac{2}{1-z_{ij}\lrb1-y_{ij}\rrb}-1-z_{ij}\rsb,\nn\\
g_{ia,+}^{(\mathrm{sub})}(k_i,k_a,k)&=&\frac{1}{(k_ik)x_{ia}}
\lsb\frac{2}{2-x_{ia}-z_{ia}}-1-z_{ia}\rsb,\nn\\
g_{ai,+}^{(\mathrm{sub})}(k_a,k_i,k)&=&\frac{1}{(k_ak)x_{ia}}
\lsb\frac{2}{2-x_{ia}-z_{ia}}-1-x_{ia}\rsb,\nn\\
g_{ab,+}^{(\mathrm{sub})}(k_a,k_b,k)&=&\frac{1}{(k_ak)x_{ab}}
\lsb\frac{2}{1-x_{ab}}-1-x_{ab}\rsb,\nn\\
g_{ij,-}^{(\mathrm{sub})}(k_i,k_j,k)&=&
g_{ia,-}^{(\mathrm{sub})}(k_i,k_a,k)=
g_{ai,-}^{(\mathrm{sub})}(k_a,k_i,k)=
g_{ab,-}^{(\mathrm{sub})}(k_a,k_b,k)=0,
\eea
where we denote final-state particles with the letters $i,j$ and
initial-state particles with the letters $a,b$ and use the definitions
\bea
x_{ab}&=&\frac{k_ak_b-k_ak-k_bk}{k_ak_b}, \qquad
x_{ia}=\frac{k_ak_i+k_ak-k_ik}{k_ak_i+k_ak}, \qquad
y_{ij}=\frac{k_ik}{k_ik_j+k_ik+k_jk},
\nn\\
z_{ia}&=&\frac{k_ak_i}{k_ak_i+k_ak}, \qquad
z_{ij}=\frac{k_ik_j}{k_ik_j+k_jk}.
\label{sub_invariants}
\eea
The finite part of the real corrections thus reads
\be
\int\rd\sigma^{\mathrm{finite}}_{\mathrm{real}}=
\frac{1}{2s}\int\rd\Phi_4\lsb
\sum_\lambda\vert\mathcal{M}_{\mathrm{real}}\vert^2 \,
\Theta_{\mathrm{cut}}(\Phi_4) -
\sum_{I,J=1 \atop I\neq J}^4\vert\mathcal{M}_{IJ,\mathrm{sub}}\vert^2 \,
\Theta_{\mathrm{cut}}(\tilde{\Phi}_{3,IJ})
\rsb,
\label{eq:sigmasubfinite}
\ee
where from now on we suppress the spin sums of the fermions in the notation.
The cuts on the momenta of the final-state particles are included in
terms of the functions $\Theta_{\mathrm{cut}}(\Phi_4)$ and
$\Theta_{\mathrm{cut}}(\tilde{\Phi}_{3,IJ})$, where the arguments
signal which momenta are subject to the cuts.  Note that already at
this point collinear safety is assumed, because the emitter momentum
in $\tilde{\Phi}_{3,IJ}$ tends to the sum of emitter and photon
momenta in the collinear limit by construction, i.e.\ a recombination
of these two particles in the collinear limit is understood. This has
to be changed in the treatment of non-collinear-safe observables
considered below.
Note that the vanishing of the spin-flip parts $g_{IJ,-}$ only holds
in the difference \refeq{eq:sigmasubfinite}, where the collinear singularities cancel;
the spin-flip parts contribute when the collinear singular region in 
$\vert\mathcal{M}_{IJ,\mathrm{sub}}\vert^2$ is integrated over, as will become
apparent in the following.

We turn now to the treatment of the singular contributions. Splitting
the four-particle phase space into a three-particle phase space and
the photonic part,
\be
\int\rd\Phi_4=\int_0^1\rd
x\int\rd\tilde{\Phi}_{3,IJ}(x)\int\rd\Phi_{\gamma,IJ},
\ee
where the photonic part of the phase space, $\rd\Phi_{\gamma,IJ}$,
depends on the mass regulators $m_I$ and $m_\gamma$ of the fermions
and of the photon, we have to compensate for the reduction of the CM
energy in the case of ISR radiation, which is indicated by the
$x$-dependence of the three-particle phase space.  The integration
over $x$, thus, becomes process dependent and cannot be carried out
analytically in general. Instead, it is possible to split off
the singular contribution resulting from the endpoint of this
integration at $x\to1$ upon introducing a $[\dots]_+$ distribution.
Leaving details of the analytic integrations over the singular phase
spaces $\rd\tilde{\Phi}_{3,IJ}(x)$ to \citere{Dittmaier:1999mb}, the
result for the integrated singular part of the real corrections reads
\bea \int\rd\sigma^{\mathrm{sing}}_{\mathrm{real}}
&=&-\frac{\alpha}{2\pi}\sum_{{I,J=1} \atop {I\neq
    J}}^{4}\sum_{\tau=\pm}(-1)^{I+J}Q_IQ_J
\nn\\
&& \times\biggl\{ \int_0^1\frac{\rd x}{2sx}\int\rd
\tilde{\Phi}_{3,IJ}(x)\lsb\mathcal{G}_{IJ,\tau}^{(\mathrm{sub})}(\tilde{s}_{IJ},x)\rsb_+
\Bigl\vert\mathcal{M}_{\mathrm{Born}}
\Bigl(\tilde{\Phi}_{3,IJ}(x),\tau\sigma_I\Bigr)\Bigr\vert^2 \,
\Theta_{\mathrm{cut}}\Bigl(\tilde{\Phi}_{3,IJ}(x)\Bigr)
\nn\\
&&{} +\frac{1}{2s} \int\rd\Phi_3
\,G_{IJ,\tau}^{(\mathrm{sub})}(s_{IJ})\bigl\vert\mathcal{M}_{\mathrm{Born}}(\Phi_3,\tau\sigma_I)\bigr\vert^2
\, \Theta_{\mathrm{cut}}(\Phi_3) \biggr\}, \eea where the functions
$\mathcal{G}_{IJ,\tau}^{(\mathrm{sub})}(\tilde{s}_{IJ},x)$ and
$G_{IJ,\tau}^{(\mathrm{sub})}(s_{IJ})$ result from the analytic
integration over the photonic part of the phase space.
For the final--final emitter--spectator case, there is no convolution 
part $\mathcal{G}_{ij,\tau}^{(\mathrm{sub})}$, i.e.\
\be
\mathcal{G}_{ij,\tau}^{(\mathrm{sub})}(s_{ij},x)=0, \qquad
G_{ij,\tau}^{(\mathrm{sub})}(s_{ij}) = 
8\pi^2\int\rd\Phi_{\gamma,ij} \, 
g_{ij,\tau}^{(\mathrm{sub})}\lrb k_i,k_j,k\rrb,
\ee
in contrast to the other emitter--spectator cases, where
\be
\mathcal{G}_{IJ,\tau}^{(\mathrm{sub})}(s_{IJ},x)=
8\pi^2\, x \int\rd\Phi_{\gamma,IJ} \, 
 g_{IJ,\tau}^{(\mathrm{sub})}\lrb k_I,k_J,k\rrb,
\qquad
G_{IJ,\tau}^{(\mathrm{sub})}(s_{IJ}) = \int_0^1\rd x\,
\mathcal{G}_{IJ,\tau}^{(\mathrm{sub})}(s_{IJ},x).
\ee
It should be noted that the invariant $\tilde{s}_{IJ}$ in the
integration over
$\lsb\mathcal{G}_{IJ,\tau}^{(\mathrm{sub})}(\tilde{s}_{IJ},x)\rsb_+$
consistently takes the values
$\tilde{s}_{IJ}=2\tilde{k}_I\tilde{k}_J$, i.e.\ in the ``endpoint'' at
$x=1$ this variable corresponds to the invariant $s_{IJ}=2k_Ik_J$ of
the three-particle phase space $\Phi_3=\tilde{\Phi}_{3,IJ}(x=1)$
corresponding to the phase space without photon.  The explicit results
for the functions $\mathcal{G}$ and $G$ read
\bea
\mathcal{G}_{ia,+}^{(\mathrm{sub})}(\tilde{s}_{ia},x)&=&\frac{1}{1-x}
\lsb2\ln\lrb\frac{2-x}{1-x}\rrb-\frac{3}{2}\rsb,
\nn\\
\mathcal{G}_{ai,+}^{(\mathrm{sub})}(\tilde{s}_{ai},x)&=&P_{ff}(x)
\lsb\ln\lrb\frac{\vert \tilde{s}_{ai}\vert}{m_a^2 x}\rrb-1\rsb
-\frac{2}{1-x}\ln(2-x)+(1+x)\ln(1-x), 
\nn\\
\mathcal{G}_{ab,+}^{(\mathrm{sub})}(\tilde{s}_{ab},x)&=&P_{ff}(x)
\lsb\ln\lrb\frac{s}{m_a^2}\rrb-1\rsb,
\nn\\
\mathcal{G}_{ij,\pm}^{(\mathrm{sub})}(\tilde{s}_{ij},x)&=&
\mathcal{G}_{ia,-}^{(\mathrm{sub})}(\tilde{s}_{ia},x)=0, \quad
\mathcal{G}_{ab,-}^{(\mathrm{sub})}(\tilde{s}_{ab},x)=
\mathcal{G}_{ai,-}^{(\mathrm{sub})}(\tilde{s}_{ai},x)=1-x
\label{mathcalG}
\eea
and
\bea
G_{ij,+}^{(\mathrm{sub})}({s}_{ij})&=&
\mathcal{L}\lrb{s}_{ij},m_i^2\rrb-\frac{\pi^2}{3}+1, \qquad
G_{ia,+}^{(\mathrm{sub})}({s}_{ia})=
\mathcal{L}\lrb\vert{s}_{ia}\vert,m_i^2\rrb-\frac{\pi^2}{2}+1,
\nn\\
G_{ai,+}^{(\mathrm{sub})}({s}_{ai})&=&
\mathcal{L}\lrb\vert{s}_{ai}\vert,m_a^2\rrb+\frac{\pi^2}{6}-\frac{3}{2},
\qquad
G_{ab,+}^{(\mathrm{sub})}({s}_{ab})=
\mathcal{L}\lrb s_{ab},m_a^2\rrb-\frac{\pi^2}{3}+\frac{3}{2},
\nn\\
G_{ij,-}^{(\mathrm{sub})}({s}_{ij})&=&
G_{ia,-}^{(\mathrm{sub})}({s}_{ia})=
G_{ai,-}^{(\mathrm{sub})}({s}_{ai})=
G_{ab,-}^{(\mathrm{sub})}({s}_{ab})=\frac{1}{2},
\label{subG}
\eea
where the auxiliary function 
\be
\mathcal{L}\lrb{s},m_i^2\rrb=\ln\lrb\frac{m_\gamma^2}{{s}}\rrb
\lsb\ln\lrb\frac{m_i^2}{{s}}\rrb+1\rsb
-\frac{1}{2}\ln^2\lrb\frac{m_i^2}{{s}}\rrb
+\frac{1}{2}\ln\lrb\frac{m_i^2}{{s}}\rrb
\ee
contains the soft and collinear singularities of the endpoint parts.
These $G$ terms, which have the same kinematics as the lowest-order 
contribution, exactly cancel
the corresponding singular contribution from the virtual
corrections.
\subsubsection{Non-collinear-safe observables}
\label{ncssub}
In \citere{Dittmaier:2008md} the dipole subtraction method for
collinear-safe photon radiation~\cite{Dittmaier:1999mb}, as briefly
described above, has been generalised to non-collinear-safe radiation.
There, in the collinear photon radiation cone around a charged
final-state particle, the fraction $z$ of the charged particle's
energy is not fully integrated over, but the cut procedure affects the
range of the $z$~integration.  In the auxiliary three-particle phase
spaces $\tilde{\Phi}_{3,ij}(x)$ introduced in
\refsec{cssub} the role of $z$ is played by the variables
$z_{ij}$ and $z_{ia}$ defined in \refeq{sub_invariants}.  The
transition from the collinear-safe to the non-collinear-safe case
requires both a modification in the subtraction procedure and a
non-trivial change in the analytical integration of the subtracted
parts that will be re-added again.  We briefly describe these
generalisations in the following and refer to the original formulation
in \citere{Dittmaier:2008md} for more details.

In the subtraction procedure, as given in \refeq{eq:sigmasubfinite},
the cut prescription $\Theta_{\mathrm{cut}}(\tilde{\Phi}_{3,iJ})$
is modified in such a way that the auxiliary momentum
$\tilde k_{iJ}$ for the emitter--photon system is replaced
by the two collinear momenta $\tilde k_i = z_{iJ}\tilde k_{iJ}$
and $\tilde k=(1-z_{iJ})\tilde k_{iJ}$ for the emitter and the
photon, respectively. This procedure concerns only contributions
of final-state emitters.

On the side of the re-added subtraction function this implies that
the cut on $z_{iJ}$ has to be respected as well. In practice this
means that in most cases at least the non-singular parts of 
this integration have to be performed numerically.
The detailed procedure is described in the following separately
for the two cases of 
final--final and final--initial emitter--spectator pairs. 

\paragraph{Final-state emitter and final-state spectator}
\mbox{}\\[.3em]
The integration of the radiator functions for a final-state emitter 
$i$ and a final-state spectator $j$ is of the form
\be
G_{ij,\tau}^{(\mathrm{sub})}(\tilde{s}_{ij}) = \frac{\tilde{s}_{ij}}{2}
\int\rd y_{ij}\,(1-y_{ij}) \int\rd z_{ij} \,
g_{ij,\tau}^{(\mathrm{sub})}(k_i,k_j,k),
\label{ff_G_coll}
\ee
where $y_{ij}$ and $z_{ij}$ are given in \refeq{sub_invariants}. 
The limits of integration, which can be explicitly found in
\citere{Dittmaier:1999mb},
depend on the regulator masses and Lorentz invariants of the
emitter, spectator, and photon system.
The explicit integration leads to the results given in \refeq{subG}
in the massless limit. In the non-collinear-safe case
we want to use information on the photon momentum in the collinear cone, 
which is controlled by the variable $z_{ij}$, i.e.\
we have to interchange the order of the integrations in
\refeq{ff_G_coll} and leave the integration over $z_{ij}$ open. 
To this end, we consider
\be
\bar{\mathcal{G}}_{ij,\tau}^{(\mathrm{sub})}(\tilde{s}_{ij},z_{ij})
= \frac{\tilde{s}_{ij}}{2}
\int_{y_1(z_{ij})}^{y_2(z_{ij})}\rd y_{ij} \, (1-y_{ij}) \,
g_{ij,\tau}^{(\mathrm{sub})}(k_i,k_j,k),
\label{ff_mathcalG_int}
\ee
where the limits $y_{1,2}(z_{ij})$ of the $y_{ij}$ depend on the
mass regulators (see \citere{Dittmaier:2008md} for details).
The soft singularity contained in \refeq{ff_mathcalG_int} can be split off
by employing a $[\dots]_+$ distribution,
\be
\bar{\mathcal{G}}_{ij,\tau}^{(\mathrm{sub})}(\tilde{s}_{ij},z)=
G_{ij,\tau}^{(\mathrm{sub})}(\tilde{s}_{ij})\delta(1-z)+\lsb\bar{\mathcal{G}}_{ij,\tau}^{(\mathrm{sub})}(\tilde{s}_{ij},z)\rsb_+,
\ee
so that this singularity appears only in the quantity
$G_{ij,\tau}^{(\mathrm{sub})}(\tilde{s}_{ij})$ already given in
\refeq{subG}.  In the limit of small fermion masses, the integral in
\refeq{ff_mathcalG_int} can be carried out explicitly, resulting in
\be
\bar{\mathcal{G}}_{ij,+}^{(\mathrm{sub})}(\tilde{s}_{ij},z)=P_{ff}(z)\lsb
\ln\lrb\frac{\tilde{s}_{ij}z}{m_i^2}\rrb-1\rsb+(1+z)\ln(1-z),
\qquad
\bar{\mathcal{G}}_{ij,-}^{(\mathrm{sub})}(\tilde{s}_{ij},z)=1-z.
\ee
The explicit form of the $ij$ contribution
$|\mathcal{M}_{\mathrm{sub},ij}\lrb\Phi_1\rrb|^2$ then reads
\bea\label{final-final-gen}
\int \rd \Phi_1 \,
|\mathcal{M}_{\mathrm{sub},ij}\lrb\Phi_1;\sigma_i\rrb|^2&=&
-\frac{\alpha}{2\pi}\sum_{\tau=\pm}(-1)^{i+j}Q_i Q_j \int \rd
\tilde{\Phi}_{0,ij}\int_0^1\rd z
\nn\\
&&{}\times\lcb G_{ij,\tau}^{(\mathrm{sub})}\lrb \tilde{s}_{ij}\rrb\delta(1-z)
+\lsb\bar{\mathcal{G}}_{ij,\tau}^{(\mathrm{sub})}\lrb
\tilde{s}_{ij},z\rrb\rsb_+\rcb
\nn\\
&&{}\times \left|\mathcal{M}_0\lrb\tilde{k}_i,\tilde{k}_j;\tau\sigma_i\rrb\right|^2 \,
\Theta_{\mathrm{cut}}\lrb k_i=z\tilde{k}_i,
k=(1\!-\!z)\tilde{k}_i,\tilde{k}_j,\lcb k_n\rcb\rrb.
\hspace{2em}
\eea
The term in curly brackets in \refeq{final-final-gen} consists of a
term proportional to a $\delta$-function in $z$, which is the usual
endpoint contribution, and a $\lsb\dots\rsb_+$ prescription, acting
only on $\Theta_{\mathrm{cut}}$. In our case $\Theta_{\mathrm{cut}}$
just provides a lower cut-off on the $z$-integration,
$\Theta_{\mathrm{cut}}=\theta(z-1+z_{\mathrm{cut}})$, and we find
\bea
\int \rd \Phi_1 |\mathcal{M}_{\mathrm{sub},ij}\lrb\Phi_1;\sigma_i\rrb|^2
&=&-\frac{\alpha}{2\pi}\sum_{\tau=\pm}(-1)^{i+j}Q_i Q_j \int \rd
\tilde{\Phi}_{0,ij} \,
\left|\mathcal{M}_0\lrb\tilde{k}_i,\tilde{k}_j;\tau\sigma_i\rrb\right|^2
\nn\\
&& {}
\times
\lcb G_{ij,\tau}^{(\mathrm{sub})}\lrb \tilde{s}_{ij}\rrb
-\int_0^{1-z_{\mathrm{cut}}}\rd z \,\,\bar{\mathcal{G}}_{ij,\tau}^{(\mathrm{sub})}
\lrb\tilde{s}_{ij},z\rrb\rcb.
\label{eq:ffint}
\eea
The $z$-integration in the second term of \refeq{eq:ffint} can be carried out explicitly,
and the sum over $\tau$ can be performed, because we consider only unpolarized final states.
In this way we obtain
\begin{align}
\sum_{\tau=\pm}\int_0^{1-z_{\mathrm{cut}}}\rd z \,
\bar{\mathcal{G}}_{ij,\tau}^{(\mathrm{sub})}\lrb
\tilde{s}_{ij},z\rrb
=&-\frac{\pi^2}{3}+ \lsb \frac{1}{2} 
-2\ln\lrb \frac{\tilde{s}_{ij}}{m_i^2} \rrb\rsb \ln\lrb z_{\mathrm{cut}}\rrb
\nn\\
&{}+\frac{1}{2}(1-z_{\mathrm{cut}})\lsb 3
-(3-z_{\mathrm{cut}})\ln\lrb \frac{\tilde{s}_{ij}(1-z_{\mathrm{cut}})}{m_i^2 z_{\mathrm{cut}}} 
\rrb\rsb
+ 2\Li\lrb z_{\mathrm{cut}}\rrb.
\label{eq:Gijbarint}
\end{align}

\paragraph{Final-state emitter and initial-state spectator}
\mbox{}\\[.3em]
In the case of a final-state emitter $i$ and an initial-state spectator $a$, 
the integration of
$\vert\mathcal{M}_{ia,\mathrm{sub}}\vert^2$ over
$x=x_{ia}$ is performed using a $\lsb\dots\rsb_+$ prescription,
\be
\frac{-\tilde{s}_{ia}}{2}
\int_{0}^{x_1}\rd x_{ia}\int\rd z_{ia} \,
g_{ia,\tau}^{(\mathrm{sub})}(k_i,k_a,k)\dots
=\int_{0}^{1}\rd x\lcb
G_{ia,\tau}^{(\mathrm{sub})}(\tilde{s}_{ia})\delta(1-x)+\lsb\mathcal{G}_{ia,\tau}^{(\mathrm{sub})}(\tilde{s}_{ia},x)\rsb_+
\rcb\dots,
\ee
where the ellipses stand for $x$-dependent, process-specific functions
like the Born matrix element squared or flux factors.  Here we used
the fact that the upper limit $x_1$ of the $x$-integration is equal to
one up to mass terms that are only relevant for the regularisation of
the singularities which eventually appear in
$G_{ia,\tau}^{(\mathrm{sub})}$.  The integration over $x$ is usually
done numerically.  Details of the derivation of the functions
$G_{ia,\tau}^{(\mathrm{sub})}$ and
$\mathcal{G}_{ia,\tau}^{(\mathrm{sub})}$, which are given in
\refeq{subG} and \refeq{mathcalG}, can be found in
\citere{Dittmaier:1999mb}.

In the non-collinear-safe case (see again \citere{Dittmaier:2008md}
for details), the ellipses also involve terms like the cut function
that depend on $z_{ia}$.  Therefore the whole integration has to be
done in general numerically. To this end, a procedure is employed that
isolates the occurring singularities in the endpoint. The basic idea
in this procedure is the use of a generalised $\lsb\dots\rsb_+$
prescription that acts on multiple variables.  Denoting the usual
$\lsb\dots\rsb_+$ prescription in an $n$-dimensional integral over the
variables $r_i$, $i=1,\ldots,n$, by
\be
\int\rd^n {\bf{r}}\lsb g({\bf{r}})\rsb_{+,(a)}^{(r_i)}f({\bf{r}})\equiv\int\rd^n {\bf{r}}\,g({\bf{r}})
\lrb f({\bf{r}})-f({\bf{r}})\vert_{r_i=a}\rrb,
\ee
the natural generalisation to two variables reads
\bea
\int\rd^n {\bf{r}}\lsb g({\bf{r}})\rsb_{+,(a,b)}^{(r_i,r_j)}f({\bf{r}})&\equiv&
\int\rd^n {\bf{r}}\lsb\lsb g({\bf{r}})\rsb_{+,(a)}^{(r_i)}\rsb_{+,(b)}^{(r_j)}f({\bf{r}})
\nn\\
&=&\int\rd^n {\bf{r}}\,g({\bf{r}})
\lrb f({\bf{r}})-f({\bf{r}})\vert_{r_i=a}-f({\bf{r}})\vert_{r_j=b}+
f({\bf{r}})\vert_{{r_i=a} \atop {r_j=b}}\rrb.
\eea
To recover the usual notation, we drop the subscripts $a$ and/or $b$ 
if they are equal to one.
The generic form of the integral we want to perform is
\be
I[f]\equiv \frac{-\tilde{s}_{ia}}{2}
\int_{0}^{x_1}\rd x\int_{z_1(x)}^{z_2(x)}\rd z \,
g_{ia,\tau}^{(\mathrm{sub})}(k_i,k_a,k) f(x,z),
\ee
with $f(x,z)$ denoting an integrable function of $x=x_{ia}$ 
and $z=z_{ia}$. After introducing the multiple $\lsb\dots\rsb_+$
distribution, all soft and collinear singularities are integrated
out, and the result contains only regular integrations
over $x$ and $z$ within unit boundaries,
\bea
I[f]&=&
\int_0^1\rd x\int_0^1\rd z \lsb \bar{g}_{ia,\tau}^{(\mathrm{sub})}(x,z)\rsb_+^{(x,z)}f(x,z)+
\int_0^1\rd x f(x,1)\lsb \mathcal{G}_{ia,\tau}^{(\mathrm{sub})}(\tilde{s}_{ia},x)\rsb_+\nn\\
&&{}+
\int_0^1\rd z f(1,z)\lsb \bar{\mathcal{G}}_{ia,\tau}^{(\mathrm{sub})}(\tilde{s}_{ia},z)\rsb_+ +
f(1,1)G_{ia,\tau}^{(\mathrm{sub})}(\tilde{s}_{ia}),
\eea
with the additional integration kernels
\bea
\bar{g}_{ia,+}^{(\mathrm{sub})}(x,z)&=&\frac{1}{1-x}\lrb\frac{2}{2-x-z}-1-z\rrb,
\qquad
\bar{g}_{ia,-}^{(\mathrm{sub})}(x,z)=0,
\nonumber\\
\bar{\mathcal{G}}_{ia,+}^{(\mathrm{sub})}(\tilde{s}_{ia},z)&=&P_{ff}(z)\lsb
\ln\lrb\frac{\vert\tilde{s}_{ia}\vert z}{m_i^2}\rrb-1
\rsb-\frac{2\ln(2-z)}{1-z}+(1+z)\ln(1-z),
\nonumber\\
\bar{\mathcal{G}}_{ia,-}^{(\mathrm{sub})}(\tilde{s}_{ia},z)&=&1-z.
\eea
Using this result, the explicit form of the $ia$ contribution
$|\mathcal{M}_{\mathrm{sub},ia}\lrb\Phi_1\rrb|^2$ reads
\bea
\lefteqn{
\int \rd \Phi_1 \,
|\mathcal{M}_{\mathrm{sub},ia}\lrb\Phi_1;\sigma_i\rrb|^2
=
-\frac{\alpha}{2\pi}\sum_{\tau=\pm}
(-1)^{a+i}Q_a Q_i\int_0^1 \rd x  \int \rd
\tilde{\Phi}_{0,ia}\lrb \tilde{s}_{ia},x\rrb\int_0^1\rd z
} &&
\nn\\
&&{}\times\frac{1}{x}\lcb G_{ia,\tau}^{(\mathrm{sub})}\lrb \tilde{s}_{ia}\rrb\delta(1-x)\delta(1-z)
+\lsb\mathcal{G}_{ia,\tau}^{(\mathrm{sub})}\lrb
\tilde{s}_{ia},x\rrb\rsb_+\delta(1-z)\right.
\nn\\
&&\hspace*{10mm}{}\left.+
\lsb\bar{\mathcal{G}}_{ia,\tau}^{(\mathrm{sub})}\lrb
\tilde{s}_{ia},z\rrb\rsb_+\delta(1-x)+
\lsb\bar{g}_{ia,\tau}^{(\mathrm{sub})}\lrb
x,z\rrb\rsb_+^{x,z}\rcb
\nn\\
&&{}\times
\left|\mathcal{M}_0\lrb\tilde{k}_i(x),\tilde{k}_a(x);\tau\sigma_i\rrb\right|^2 \,
\Theta_{\mathrm{cut}}\lrb k_i=z\tilde{k}_i(x),
k=(1\!-\!z)\tilde{k}_i(x),\lcb k_n(x)\rcb\rrb.
\label{eq:figen}
\eea
Inserting the explicit form of the cut function,
$\Theta_{\mathrm{cut}}=\theta(z-1+z_{\mathrm{cut}})$, this yields
\bea
\lefteqn{
\int \rd \Phi_1 \,
|\mathcal{M}_{\mathrm{sub},ia}\lrb\Phi_1;\sigma_i\rrb|^2
=-\frac{\alpha}{2\pi}\sum_{\tau=\pm}(-1)^{a+i}
Q_a Q_i} &&
\nn\\
&& {}
\times\biggl\{
\int \rd
\tilde{\Phi}_{0,ia}\lrb \tilde{s}_{ia}\rrb
\lsb G_{ia,\tau}^{(\mathrm{sub})}\lrb
\tilde{s}_{ia}\rrb  - \int_0^{1-z_{\mathrm{cut}}}\rd z\,
\bar{\mathcal{G}}_{ia,\tau}^{(\mathrm{sub})}\lrb \tilde{s}_{ia},z\rrb  
\rsb\left|\mathcal{M}_0\lrb\tilde{k}_i,\tilde{k}_a;\tau\sigma_i\rrb\right|^2
\\
&&{}
+\int_0^1 \rd x  \int \rd\tilde{\Phi}_{0,ia}\lrb \tilde{s}_{ia},x\rrb
\frac{1}{x}\lsb\mathcal{G}_{ia,\tau}^{(\mathrm{sub})}\lrb
\tilde{s}_{ia},x\rrb
-\int_0^{1-z_{\mathrm{cut}}}\rd z\,\bar{g}_{ia,\tau}^{(\mathrm{sub})}\lrb
x,z\rrb
\rsb_+
\left|\mathcal{M}_0\lrb\tilde{k}_i(x),\tilde{k}_a(x);\tau\sigma_i\rrb\right|^2\biggr\}.
\nn
\label{eq:fiint}
\eea
The $z$-integrations and the sum over $\tau$ can again be performed, 
resulting in
\bea
\lefteqn{
\sum_{\tau=\pm}\int_0^{1-z_{\mathrm{cut}}}\rd z \,
\bar{\mathcal{G}}_{ia,\tau}^{(\mathrm{sub})}\lrb
\tilde{s}_{ia},z\rrb
=
-\frac{\pi^2}{2}+ \lsb \frac{1}{2} 
-2\ln\lrb \frac{\vert\tilde{s}_{ia}\vert}{m_i^2} \rrb\rsb 
\ln\lrb z_{\mathrm{cut}}\rrb
} &&
\nn\\
&&{}+\frac{1}{2}(1-z_{\mathrm{cut}})\lsb 3
-(3-z_{\mathrm{cut}})\ln\lrb \frac{\vert \tilde{s}_{ia}\vert(1-z_{\mathrm{cut}})}
{m_i^2z_{\mathrm{cut}}}  \rrb\rsb
+ 2\Li\lrb z_{\mathrm{cut}}\rrb - 2\Li\lrb -z_{\mathrm{cut}}\rrb 
\label{eq:Giabarint}
\eea
and
\be
\sum_{\tau=\pm}\int_0^{1-z_{\mathrm{cut}}}\rd z \,
\bar{g}_{ia,\tau}^{(\mathrm{sub})}\lrb x,z\rrb
=\frac{1}{1-x} \lsb 
-\frac{1}{2}(1-z_{\mathrm{cut}})(3-z_{\mathrm{cut}})
+2\ln\lrb\frac{2-x}{1-x+z_{\mathrm{cut}}}\rrb \rsb.
\label{eq:giaint}
\ee
\subsection{The quark-to-photon fragmentation function}
\label{quark-to-photonfragfac}
In the previous sections we described how we can deal with identified
particles in the final state that lead to non-IR-safe
observables using phase-space slicing and the subtraction method. To
restore IR safety one factorises the resulting left-over
singularities into an experimentally determined fragmentation
function, in our case the quark-to-photon fragmentation function.

In \citere{Glover:1993xc} a method has been proposed in how to measure
the quark-to-photon fragmentation function, i.e.~the probability of a
quark splitting into a quark and a photon, using the
$\Pe^+\Pe^-\rightarrow n\,\mathrm{jet}+\mathrm{photon}$ cross section.
The method has been extended to next-to-leading order in QCD 
in \citere{GehrmannDeRidder:1997wx}.
The key feature of the proposed method is the democratic clustering of
both hadrons and photons into jets, where one keeps track of the
fraction of photonic energy in the jet. This treatment of the photon
in the jet enhances the non-perturbative part of the quark-to-photon
fragmentation function \cite{Koller:1978kq,Glover:1993xc}, which in
turn can be measured in $\Pe^+\Pe^-$ annihilation.

In \citere{Glover:1993xc} the quark-to-photon fragmentation function
was theoretically defined using dimensional regularisation and
one-cut-off slicing. To be able to use the results obtained in this
way in our calculation, we need to translate this definition to mass
regularisation. We first summarize the results in dimensional
regularisation.

Since fragmentation is a long-distance process, it cannot be
calculated entirely in perturbation theory. The fragmentation 
function $D_{q\rightarrow\gamma}(z_\gamma)$ describes the probability 
of a quark fragmenting into a jet containing a photon carrying $z_\gamma$ of 
the jet energy. Photons inside hadronic jets can have two origins: (a) The 
perturbatively calculable radiation of a photon off a quark, which 
contains a collinear divergence, described by a 
perturbative contribution $\d\sigma_{{\rm coll}}^\gamma$, 
dependent on the method used to regulate the collinear singularity; 
(b) the non-perturbative production of 
a photon in the hadronisation process of the quark into a hadronic jet, 
which is described by a bare non-perturbative fragmentation function  
$D^{{\rm bare}}_{q\rightarrow\gamma}(z_\gamma)$. 
Both processes refer to the infrared dynamics inside the quark jet, and 
can a priori not be separated from each other. 

Within dimensional regularisation and one-cutoff-slicing, the photon
fragmentation contribution was studied in detail in
\citeres{Glover:1993xc,Poulsen:2006}. Exploiting the universal
factorisation of matrix elements and phase space in the collinear
limit, one obtains for the cross section for the emission of one
collinear photon with energy fraction $z_\gamma$ above
$z_{\mathrm{cut}}$ off a quark $q$ and invariant mass of the
photon--quark pair below $s_{\mathrm{min}}$,
\bea
\d\sigma_{{\rm coll}}^\gamma(z_{\mathrm{cut}})
&=& 
\frac{\alpha Q_q^2}{2\pi}  \,\d \sigma_0 
\int_{z_{\mathrm{cut}}}^1 \d z_\gamma                      
\frac{\left({4\pi\mu^2}\right)^\epsilon}{\Gamma(1-\epsilon)}
\frac{P^{(\epsilon)}_{ff}(1-z_\gamma)}{\left[z_\gamma(1-z_\gamma)\right]^\epsilon}
\int_0^{s_{\mathrm{min}}}
\rd s_{q\gamma}\frac{1}{s_{q\gamma}^{1+\epsilon}}
\nonumber\\
&=&- 
\frac{\alpha Q_q^2}{2\pi}  \,\d \sigma_0                      
\int_{z_{\mathrm{cut}}}^1 \d z_\gamma                      
\frac{1}{\epsilon}\left(
\frac{4\pi\mu^2}{s_{\mathrm{min}}}\right)^\epsilon\frac{1}{\Gamma(1-\epsilon)}
\frac{P^{(\epsilon)}_{ff}(1-z_\gamma)}{\left[z_\gamma(1-z_\gamma)\right]^\epsilon}
\nonumber\\
&=& 
- \frac{\alpha Q_q^2}{2\pi} \,\d \sigma_0 
\int_{z_{\mathrm{cut}}}^1 \d z_\gamma
\nonumber\\&&{}                      
 \times\left[                       
\frac{1}{\epsilon}\left(
\frac{4\pi\mu^2}{s_{\mathrm{min}}}\right)^\epsilon\frac{1}{\Gamma(1-\epsilon)}
{P_{ff}(1-z_\gamma)}
 - P_{ff}(1-z_\gamma) \ln\left(z_\gamma(1-z_\gamma)\right) - z_\gamma \right]
\,,\qquad
\label{frag_p_eps}
\eea
where $\d \sigma_0$ is a  reduced cross section with the quark--photon 
system replaced by its parent quark. The same expression was obtained in 
(\ref{eq:sliceint}) above for the hard-photon cut contribution,
where $z=1-z_\gamma$.

The infrared singularity present in this perturbative contribution
must be balanced by a divergent piece in the bare fragmentation
function, which contributes to the photon-emission cross section as
\begin{equation}
\d \sigma_{{\rm frag}}(z_{\mathrm{cut}}) = 
\d \sigma_0 
\int_{z_{\mathrm{cut}}}^1 \d z_\gamma                      
D^{{\rm bare}}_{q\rightarrow\gamma}(z_\gamma)
\end{equation}
To make this cancellation explicit, one introduces a 
factorisation scale $\muF$ into the bare fragmentation function, which
then reads in dimensional regularisation
\bea
D_{q\rightarrow\gamma}^{\mathrm{bare,DR}}(z_\gamma)&=&D_{q\rightarrow\gamma}(z_\gamma,\muF)+
\frac{\alpha Q_q^2}{2\pi}                    
\frac{1}{\epsilon}\left(
\frac{4\pi\mu^2}{\muF^2}\right)^\epsilon\frac{1}{\Gamma(1-\epsilon)}
P_{ff}(1-z_\gamma)\,.
\label{eq:massfactDR}
\eea
The factorised non-perturbative fragmentation function 
$D_{q\rightarrow\gamma}(z_\gamma,\muF)$ is then finite, and can be 
determined from experimental data. Its variation with the scale $\muF$ 
is described by the Altarelli--Parisi evolution equation, which reads to 
leading order 
\begin{equation}
\frac{{\rm d}D_{q \to \gamma}(z_\gamma,\muF)}{{\rm d}\ln \muF^2}
=\frac{\alpha Q_{q}^2}{2 \pi}P_{ff}(1-z_\gamma).
\label{evolution}
\end{equation} 

The fixed-order exact solution at ${\cal O}(\alpha)$ is given by
\begin{equation}
 D_{q \to \gamma}(z_\gamma,\muF)=\frac{\alpha Q_{q}^2}{2 \pi} 
\,P_{ff}(1-z_\gamma)
\ln \left(\frac{\muF^2}{\mu_{0}^2}\right) + D_{q \to \gamma}
(z_\gamma,\mu_{0}),
\label{eq:npFF}
\end{equation}
where $D_{q \to \gamma}(z_\gamma,\mu_{0})$ is the quark-to-photon
fragmentation function at some initial scale $\mu_{0}$. This function
and the initial scale $\mu_0$ cannot be calculated and have to be
determined from experimental data.  The first determination of $D_{q
  \to \gamma}(z,\mu_{0})$ was performed by the ALEPH
collaboration~\cite{aleph} using the ansatz
\be
D_{q\rightarrow\gamma}^{\mathrm{ALEPH}}(z_\gamma,\muF)=
\frac{\alpha Q_q^2}{2\pi}                        
\left[
P_{ff}(1-z_\gamma)\ln\left(\frac{\muF^2}{\mu_0^2}
\frac{1}{(1-z_\gamma)^2}\right)+C
\right]
,
\label{ff_ALEPH}
\ee
with fitting parameters $C$ and $\mu_0$. The fit to the
photon-plus-one-jet rate~\cite{aleph} yielded
 \begin{equation}
\mu_0 = 0.22~\mbox{GeV};\qquad C=-12.1\,.
\end{equation}

Through the factorisation formula (\ref{eq:massfactDR}), a
relation between bare fragmentation function and the collinear photon
contribution is established. As a result, the sum of both
contributions is finite, but depends on the slicing parameter
$s_{\mathrm{min}}$:
\bea
\lefteqn{\d\sigma_{{\rm coll}}^\gamma(z_{\mathrm{cut}}) 
+ \d\sigma_{{\rm frag}}(z_{\mathrm{cut}}) 
= 
\d \sigma_0 
\int_{z_{\mathrm{cut}}}^1 \d z_\gamma  }\qquad
\nonumber\\&&{}   \times                  
\left( D_{q\rightarrow\gamma}(z_\gamma,\muF)
+
\frac{\alpha Q_q^2}{2\pi}                        
\left[
P_{ff}(1-z_\gamma)\ln\lrb\frac{s_{\mathrm{min}}}{\muF^2}(1-
z_\gamma)z_\gamma\rrb+z_\gamma
\right]\right) \,
.
\label{frag_MSbar}
\eea 
Inserting the solution
(\ref{eq:npFF}) into (\ref{frag_MSbar}), $\d\sigma_{{\rm coll}}^\gamma
+ \d\sigma_{{\rm frag}}$ becomes independent of the factorisation
scale $\muF$.

We can now determine the contribution of the quark-to-photon
fragmentation function to 3-jet production. Since photons above
$z_{{\rm cut}}$ are vetoed we have to subtract the
photon-fragmentation contribution from the real corrections, with the
hard-photon cut procedure included, and the virtual corrections,
resulting in the IR-safe cross section
\be
\int \rd\sigma^{\mbox{\scriptsize IR-safe}}=\int\rd\sigma_{\mathrm{virt}}+
\int\rd\sigma_{\mathrm{real}}(z_{\mathrm{cut}})
-\int\rd\sigma_{\mathrm{frag}}(z_{\mathrm{cut}}).
\label{master_real_frag}
\ee
Taking into account all quarks (and antiquarks) in the final state, the
photon-fragmentation contribution reads:
\be
\rd\sigma_{\mathrm{frag}}(z_{\mathrm{cut}})=\sum_{i=3}^4
\rd\sigma_{\mathrm{Born}}\int_{z_{\mathrm{cut}}}^1\rd z_\gamma \,
D^{\mathrm{bare}}_{q_i\rightarrow\gamma}(z_\gamma).
\label{eq:dsigfinalzcut}
\ee
Physically we can motivate this approach as follows.  The hard photon cut removes 
collinear quark-photon pairs above a predefined photon energy fraction. This 
results in an incomplete cancellation of collinear singularities between the real 
photon radiation and the virtual
QED-type corrections (where the hard photon cut is not acting, since the virtual photon is not observed). By
subtracting $\rd\sigma_{\mathrm{frag}}$ from the
$\mathcal{O}\lrb\alpha\rrb$ corrections, we correct for the effect of
the hard-photon cut and compensate for excess terms related to
collinear photon emission. In this way we can define an IR-safe
quantity even in the presence of the hard-photon cuts. 

In order to use the fragmentation function in our calculation with
mass regularization we have to translate \refeq{eq:massfactDR} to mass
regularization. For consistency with previous sections, we use from
now on the quark energy $z=1-z_\gamma$ as collinear variable.

Using results of \citere{Baur:1998kt}, the collinear photon
contribution in mass regularisation and one-cutoff slicing is obtained
as 
\bea \d\sigma_{{\rm coll}}^\gamma(z_{\mathrm{cut}}) &=& \frac{\alpha Q_q^2}{2\pi}
\,\d \sigma_0                      
\int_0^{1-z_{\mathrm{cut}}} \d z
\int_{m_q^2/z}^{s_{\mathrm{min}}} \frac{\rd s_{q\gamma}}{s_{q\gamma}-m_q^2}
\left[P_{ff}(z)-\frac{2m_q^2}{s_{q\gamma}-m_q^2}\right] \nn\\&=&
\frac{\alpha Q_q^2}{2\pi}\,\d \sigma_0                      
\int_0^{1-z_{\mathrm{cut}}} \d z
\left[P_{ff}(z)\ln\left(\frac{s_{\mathrm{min}}}{m_q^2}\frac{z}{1-z}\right)-\frac{2z}{1-z}\right]\,.
\label{ff_m}
\eea
Using this result and the independence of \refeq{frag_MSbar} on the
regularisation scheme we find the bare fragmentation function
in mass regularization
\bea
D_{q\rightarrow\gamma}^{\mathrm{bare,MR}}(1-z)&=&
D_{q\rightarrow\gamma}(1-z,\muF)+
\frac{\alpha Q_q^2}{2\pi}                    
P_{ff}(z)\left[\ln\left(
\frac{m_q^2}{\muF^2}(1-z)^2
\right)+1
\right],
\label{bareff_mass}
\eea
where the finite terms ensure that the $\overline{{\rm MS}}$ scheme
factorised fragmentation function
$D_{q\rightarrow\gamma}(z_\gamma,\muF)$ is identical in the
dimensionally regularised and the mass-regularised expressions.  After
inserting the ALEPH ansatz \refeq{ff_ALEPH} for $D_{q \to
  \gamma}(z_\gamma,\muF)$, we obtain
\be
D_{q\rightarrow\gamma}^{\mathrm{bare,MR}}(1-z)=
\frac{\alpha Q_q^2}{2\pi}                        
\left[
P_{ff}(z)\left[\ln\left(\frac{m_q^2}{\mu_0^2}\frac{(1-z)^2}{z^2}\right)+1\right]+C
\right]\,
.
\label{bareff_ALEPH}
\ee
Integrating \refeq{bareff_ALEPH} over $z$ results in
\bea
\int_0^{1-z_{\mathrm{cut}}}\rd z \,D_{q\rightarrow\gamma}^{\mathrm{bare,MR}}(1-z)&=&
\frac{\alpha Q_q^2}{2\pi}                        
\biggl[
C(1-z_{\mathrm{cut}}) - \frac{(1-z_{\mathrm{cut}})^2}{2} + \ln\left(z_{\mathrm{cut}}\right)+4 \Li\left(1-z_{\mathrm{cut}}\right)\nn\\
&&{}-2\ln\left(z_{\mathrm{cut}}\right) \ln\left(\frac{m_q^2}{\mu_0^2} 
\frac{z_{\mathrm{cut}}^2}{(1 -z_{\mathrm{cut}})^2}\right)
+2\ln^2\lrb z_{\mathrm{cut}}\rrb\nn\\
&&{}-
\frac{1}{2}(1-z_{\mathrm{cut}}) (3-z_{\mathrm{cut}}) 
\ln\left(\frac{m_q^2}{\mu_0^2} 
\frac{z_{\mathrm{cut}}^2}{(1-z_{\mathrm{cut}})^2}\right)
\biggr]
.
\label{bareff_ALEPH_z_int}
\eea
In the case of the two-cut-off phase-space-slicing method,
subtracting $\rd\sigma_{\mathrm{frag}}(z_{\mathrm{cut}})$, i.e.\ 
\refeq{eq:dsigfinalzcut} with \refeq{bareff_ALEPH_z_int} from
  $\rd\sigma_{\mathrm{coll}}^{\mathrm{final}}(z_{\mathrm{cut}})$,
  \refeq{cutint}, leads to
\begin{align}
\rd\sigma_{\mathrm{coll}}^{\mathrm{final}}(z_{\mathrm{cut}})
&{}-\mathrm{d}\sigma_{\mathrm{frag}}(z_{\mathrm{cut}})
=\sum_{i=3}^4
\frac{\alpha}{2\pi}Q_i^2\rd\sigma_{\mathrm{Born}}\biggl[
\frac{1}{2} - \frac{2\pi^2}{3} + 4 z_{\mathrm{cut}} - C (1-z_{\mathrm{cut}})\nn\\
&{}-\left(\frac{3}{2}+2\ln\left(\frac{\Delta E}{E_i}\right)\right)\ln\left(\frac{2E_i^2\delta_{\mathrm{c}}}{m_i^2}\right)
+2\ln\left(\frac{\Delta E}{E_i}\right)
\nn\\&{}
+\left[\ln\left(\frac{2E_i^2\delta_{\mathrm{c}}}{\mu_0^2}z_{\mathrm{cut}}\right)
-\frac{3}{2}\right]\ln\left(z_{\mathrm{cut}}^2\right)
+\frac{1}{2}(1-z_{\mathrm{cut}}) (3-z_{\mathrm{cut}}) 
\ln\left(\frac{2E_i^2\delta_{\mathrm{c}}}{\mu_0^2}z_{\mathrm{cut}}^2\right)
\biggr]\nn\\
=&\,\,{}\rd\sigma_{\mathrm{coll}}^{\mathrm{final}}-\sum_{i=3}^4
\frac{\alpha}{2\pi}Q_i^2\rd\sigma_{\mathrm{Born}}
\biggl\{(4+C)(1-z_{\mathrm{cut}})\nn\\
&{}
-\lsb\ln\lrb \frac{2E_i^2\delta_{\mathrm{c}}}{\mu_0^2} z_{\mathrm{cut}} 
- \frac{3}{2}\rrb\rsb
\ln\lrb z_{\mathrm{cut}}^2\rrb
-\frac{1}{2}(1-z_{\mathrm{cut}}) (3-z_{\mathrm{cut}}) 
\ln\lrb \frac{2E_i^2\delta_{\mathrm{c}}}{\mu_0^2}
z_{\mathrm{cut}}^2\rrb
\biggr\}
\label{dsigfinalzcut_aleph}
\end{align}
with $\rd\sigma_{\mathrm{coll}}^{\mathrm{final}}$ from \refeq{slicing_final}.
This result consists of the original collinear contribution that
cancels against the virtual corrections, and an additional, finite
contribution, depending on the cut on the quark energy
$z_{\mathrm{cut}}$. All collinear singularities have cancelled.

In the case of the subtraction method, we can use charge conservation
\be
Q_i^2=-\sum_{J=1\atop J\ne i}^4 (-1)^{(i+J)}Q_iQ_J
=-\sum_{a=1}^2 (-1)^{(i+a)}Q_iQ_a
 -\sum_{j=3\atop j\ne i}^4 (-1)^{(i+j)}Q_iQ_j
\ee
in \refeq{eq:dsigfinalzcut} to split
$\rd\sigma_{\mathrm{frag}}(z_{\mathrm{cut}})$ into a contribution from
final-state emitter and final-state spectator and one from final-state
emitter and initial-state spectator
\be
\rd\sigma_{\mathrm{frag}}(z_{\mathrm{cut}})=-\sum_{i=3}^4
\Biggl( \sum_{a=1}^2 (-1)^{i+a}\frac{Q_iQ_a}{Q_i^2} 
      +\sum_{{j=3}\atop {j\ne i}}^4 (-1)^{i+j}\frac{Q_iQ_j}{Q_i^2} \Biggr)
\rd\sigma_{\mathrm{Born}}\int_0^{1-z_{\mathrm{cut}}}\rd z\, 
D_{q_i\rightarrow\gamma}^{\mathrm{bare,MR}}(1-z).\;
\label{frag_split}
\ee
In the case of a final-state emitter and final-state spectator, we can
subtract the second part of \refeq{frag_split} from the contribution
of \refeq{eq:Gijbarint} to find
\begin{align}
\frac{1}{2 s}\sum_{i=3}^4\sum_{{j=3}\atop {j\ne i}}^4
&(-1)^{i+j}\frac{Q_iQ_j}{Q_i^2}
\int\rd\tilde{\Phi}_{0,ij}\left|\mathcal{M}_0\lrb\tilde{k}_i,\tilde{k}_j\rrb\right|^2
\nonumber\\&{}\times
\int_0^{1-z_{\mathrm{cut}}}\rd z \lcb
\frac{\alpha}{2\pi}Q_i^2 \sum_{\tau^\pm}\bar{\mathcal{G}}_{ij,\tau}^{(\mathrm{sub})}\lrb
\tilde{s}_{ij},z\rrb + D_{q_i\rightarrow\gamma}^{\mathrm{bare,MR}}(1-z)\rcb \nn\\
=\frac{\alpha}{4\pi s}&\sum_{i=3}^4\sum_{{j=3}\atop {j\ne i}}^4
(-1)^{i+j}\frac{Q_iQ_j}{Q_i^2}
\int\rd\tilde{\Phi}_{0,ij}\left|\mathcal{M}_0\lrb\tilde{k}_i,\tilde{k}_j\rrb\right|^2
\,\biggl\{
\lrb1+C+\frac{z_{\mathrm{cut}}}{2}\rrb
(1-z_{\mathrm{cut}}) 
\nn\\&\quad{}
-\lsb
\frac{1}{2}(1-z_{\mathrm{cut}})(3-z_{\mathrm{cut}})
+2\ln\lrb z_{\mathrm{cut}}\rrb\rsb
\ln\lrb\frac{\tilde{s}_{ij}}{\mu_0^2}\frac{z_{\mathrm{cut}}}{1-z_{\mathrm{cut}}}\rrb
\nn\\&\quad{}
+2\Li\lrb 1-z_{\mathrm{cut}}\rrb
+\frac{3}{2}\ln\lrb z_{\mathrm{cut}}\rrb
\biggr\}.
\label{ncssubplusfrag_ij}
\end{align}
Analogously, in the case of a final-state emitter and initial-state
spectator using \refeq{eq:Giabarint}, we are left with
\begin{align}
\frac{1}{2 s}\sum_{i=3}^4\sum_{a=1,2}
&(-1)^{i+a}\frac{Q_iQ_a}{Q_i^2}
\int\rd\tilde{\Phi}_{0,ia}\left|\mathcal{M}_0\lrb\tilde{k}_i,\tilde{k}_a\rrb\right|^2
\nonumber\\&\times
\int_0^{1-z_{\mathrm{cut}}}\rd z 
\lcb\frac{\alpha}{2\pi}Q_i^2\sum_{\tau^\pm}\bar{\mathcal{G}}_{ia,\tau}^{(\mathrm{sub})}\lrb
\tilde{s}_{ia},z\rrb + D_{q_i\rightarrow\gamma}^{\mathrm{bare,MR}}(1-z)\rcb=\nn\\
=\frac{\alpha}{4\pi s}&\sum_{i=3}^4\sum_{a=1,2}
(-1)^{i+a}Q_iQ_a
\int\rd\tilde{\Phi}_{0,ia}\left|\mathcal{M}_0\lrb\tilde{k}_i,\tilde{k}_a\rrb\right|^2
\biggl\{
-\frac{\pi^2}{6}+\lrb 1+C+\frac{z_{\mathrm{cut}}}{2}\rrb
(1-z_{\mathrm{cut}})\nn\\
&\quad{}-\lsb 
\frac{1}{2}(1-z_{\mathrm{cut}})(3-z_{\mathrm{cut}})
+2\ln\lrb z_{\mathrm{cut}}\rrb\rsb
\ln\lrb\frac{|\tilde{s}_{ia}|}{\mu_0^2}\frac{z_{\mathrm{cut}}}{1-z_{\mathrm{cut}}}\rrb
\nn\\&\quad{}
+2\Li\lrb 1-z_{\mathrm{cut}}\rrb-2\Li\lrb -z_{\mathrm{cut}}\rrb
+\frac{3}{2}\ln\lrb z_{\mathrm{cut}}\rrb 
\biggr\}.
\label{ncssubplusfrag_ia}
\end{align}
Both \refeq{ncssubplusfrag_ij} and \refeq{ncssubplusfrag_ia} are
finite and only depend on the value of $z_{\mathrm{cut}}$, but not on
the mass regulators.
\subsection{Higher-order initial-state radiation}
\label{hoisr}
In order to achieve an accuracy at the per-mille level, we also
include effects stemming from higher-order ISR
using the structure-function approach as described in
\citeres{Altarelli:1996gh,Denner:2000bj}. The factorisation
theorem states that the leading-logarithmic (LL) initial-state QED
correction can be written as a convolution of the lowest-order cross
section with structure functions according to
\be
\int\rd\sigma^{\mathrm{LL}}=\int_0^1\rd x_1\int_0^1\rd x_2 \, 
\Gamma_{\Pe\Pe}^{\mathrm{LL}}(x_1,Q^2)\Gamma_{\Pe\Pe}^{\mathrm{LL}}(x_2,Q^2)
\int\rd \sigma_{\mathrm{Born}}(x_1k_1,x_2k_2),
\label{LL_cross}
\ee
where $x_1$ and $x_2$ denote the fractions of the incoming momenta just before the hard scattering, $Q^2$
is the typical scale at which the scattering occurs, and the structure functions up to
$\mathcal{O}(\alpha^3)$ are given by \cite{Altarelli:1996gh}
\bea
\Gamma^{\mathrm{LL}}_{\Pe\Pe}&=&\frac{\exp\lrb-\frac{1}{2}\beta_{\Pe}\gamma_{\mathrm{E}}
+\frac{3}{8}\beta_{\Pe}\rrb}{\Gamma\lrb1+\frac{1}{2}\beta_{\Pe}\rrb}\frac{\beta_{\Pe}}{2}(1-x)
^{\frac{\beta_{\Pe}}{2}-1}-\frac{\beta_{\Pe}}{4}(1+x)\nn\\
&&{}-\frac{\beta_{\Pe}^2}{32}\lcb\frac{1+3x^2}{1-x}\ln(x)+4(1+x)\ln(1-x)+5+x\rcb\nn\\
&&{}-\frac{\beta_{\Pe}^3}{384}\biggl\{(1+x)\lsb6\Li(x)+12\ln^2(1-x)-3\pi^2\rsb\nn\\
&&{}+\frac{1}{1-x}\biggr[\frac{3}{2}(1+8x+3x^2)\ln(x)+6(x+5)(1-x)\ln(1-x)\nn\\
&&{}+12(1+x^2)\ln(x)\ln(1-x)-\frac{1}{2}(1+7x^2)\ln^2(x)
+\frac{1}{4}(39-24x-15x^2)\biggr]\biggr\},
\label{structure_function}
\eea
where
\be
\beta_{\Pe}=\frac{2\alpha}{\pi}\left[\ln\lrb\frac{Q^2}{m_\Pe^2}\rrb-1\right],
\ee
%
$\Gamma$ is the Gamma function, and $\gamma_{\mathrm{E}}$ the
Euler--Mascheroni constant.  In the calculation at hand we use
$Q^2=s$.

When we add \refeq{LL_cross} to the one-loop result, we have to
subtract the lowest-order and one-loop contributions
$\rd\sigma^{\mathrm{LL},1}$ already contained in
\refeq{structure_function} to avoid double counting.  They read
\bea
\int\rd\sigma^{\mathrm{LL},1}&=&\int_0^1\rd x_1\int_0^1\rd x_2\lsb
\delta(1-x_1)\delta(1-x_2)+\Gamma_{\Pe\Pe}^{\mathrm{LL},1}(x_1,Q^2)\delta(1-x_2)\right.\nn\\
&&\left.{}+\Gamma_{\Pe\Pe}^{\mathrm{LL},1}(x_2,Q^2)\delta(1-x_1)\rsb\int\rd
\sigma_{\mathrm{Born}}(x_1k_1,x_2k_2),
\eea
where the one-loop structure functions are given by
\be
\Gamma_{\Pe\Pe}^{\mathrm{LL},1}=\frac{\beta_{\Pe}}{4}\lrb\frac{1+x^2}{1-x}\rrb_+.
\ee

\section{Implementation}
\label{sec:num}
\setcounter{equation}{0}

The real and virtual corrections described in the two previous sections are
implemented in a parton-level event generator. This program generates
final states with two and three particles for the hadronic cross sections, and
with three and four particles for event-shape distributions. It allows for
arbitrary, infrared-safe cuts on the final-state particles to be applied.

\subsection{Event selection}
\label{sec:es}

The infrared-safety requirement has one important implication that was
up to now not necessarily realized in the experimental studies. In ISR
events, where a photon is radiated close to the beam-pipe and not
observed, the event-shape variables must be computed in the CM system
of the observed hadrons. Only this transformation to the hadronic CM
system ensures that two-jet-like configurations are correctly
identified with the kinematic limit $y\to 0$ of the event-shape
distributions, as can be seen on the example of thrust: a partonic
final state with a quark--antiquark pair and an unobserved photon
yields two jets which are not back-to-back in the $\Pe^+\Pe^-$ CM
frame. The reconstructed thrust axis in this frame is in the direction
of the difference vector of the two jet momenta, resulting in $T<1$,
even for an ideal two-particle final state.

Unfortunately, most experimental studies of event-shape distributions
at LEP computed the shape variables in the $\Pe^+\Pe^-$ CM frame.
Modelling the ISR photonic corrections using standard parton-shower
programs, these distributions were then corrected (by often large
correction factors, shifting the two-jet peak from within the
distribution back to the kinematical edge) bin-by-bin. It is therefore
very difficult in practise to compare our results with data from LEP.
Despite its importance this task goes beyond the present paper.

In our implementation, we proceed as follows:
\begin{enumerate}
\item For all final-state particles, an acceptance cut is applied on
  the polar angle $\theta_i$ of the particle $i$ with respect to the
  beam direction: $|\cos\theta_i|\leq \cos\theta_{\mathrm{cut}}$.
  Particles that do not pass this cut are unobserved and thus
  discarded, i.e.\ their momenta are set to zero.
\item For the remaining final-state particles, the observed
  final-state invariant mass squared $s^\prime$ is computed. The event
  is rejected if $s' < s_{{\rm cut}}$.
\item If the event is accepted, it is boosted to the CM frame of
the observed final-state particles.
\item To identify isolated-photon events, all observed final-state
  particles (including the photon) are clustered into jets using the
  Durham algorithm with $E$ recombination and $y_{{\rm cut}} =
  0.002$. After the clustering, the photon is inside one of the jets
  (or makes up one jet), carrying a fraction $z$ of the jet energy. If
  $z>z_{{\rm cut}}$, the photon is considered isolated and the event
  is rejected.
\item All remaining events are accepted.
\end{enumerate}

Once an event passes the above set of cuts, we proceed with the
calculation of the event-shape variables, defined in \refsec{jetrate},
in the CM frame of the observed final-state particles.  We impose an
additional cut individually for each histogram such that the
singularity in the two-jet region is avoided, typically as a lower
cut-off for the respective observables (see \refsec{se:input}).  
The cut affects only the first bin of the histogram and does not cause
a distortion of the shape of the distribution.

\subsection{Input parameters}
\label{se:input}

For the numerical calculation, we
use the following set of input parameters \cite{Amsler:2008zzb}:
\bea
\hspace*{-5mm}
\begin{array}[b]{r@{\,}l@{\qquad }r@{\,}l@{\qquad }r@{\,}l@{\qquad}}
G_{\mu} &= 1.16637\times 10^{-5}\GeV^{-2}, &
\alpha(0) &= 1/137.03599911,&\alpha_{G_\mu} &= 1/132.49395637 \\
\alpha_{\mathrm{s}}(\MZ) &= 0.1176,\\
\MW^{\LEP} &= 80.403\GeV, & \GW^{\LEP} &= 2.141\GeV, && \\  
\MZ^{\LEP} &= 91.1876\GeV,& \GZ^{\LEP} &= 2.4952\GeV, && \\
\Me &= 0.51099892 \MeV, & \Mt &= 171.0\GeV,
&\MH &= 120\GeV.
\end{array}\hspace*{-10mm}
\eea

We employ the complex-mass scheme \cite{Denner:2005es}, where a fixed
width enters the resonant W- and Z-boson propagators, in contrast to the
approach used at LEP to fit the W~and Z~resonances, where running
widths are taken.  Therefore, we have to convert the ``on-shell''
values of $M_V^{\LEP}$ and $\Ga_V^{\LEP}$ ($V=\PW,\PZ$), resulting
from LEP, to the ``pole values'' denoted by $M_V$ and $\Ga_V$. The
relation between the two sets is given by
\cite{Bardin:1988xt}
\be\label{eq:m_ga_pole}
M_V = M_V^{\LEP}/
\sqrt{1+(\Ga_V^{\LEP}/M_V^{\LEP})^2},
\qquad
\Ga_V = \Ga_V^{\LEP}/
\sqrt{1+(\Ga_V^{\LEP}/M_V^{\LEP})^2},
\ee
leading to
\bea
\begin{array}[b]{r@{\,}l@{\qquad}r@{\,}l}
\MW &= 80.375\ldots\GeV, & \GW &= 2.140\ldots\GeV, \\
\MZ &= 91.1535\ldots\GeV,& \GZ &= 2.4943\ldots\GeV.
\label{eq:m_ga_pole_num}
\end{array}
\eea
The scale dependence of $\alphas$ is determined according to the
two-loop running. The number of active flavours at $\MZ$ is $n_\mathrm{F}=5$, 
resulting in  $\Lambda_5=0.221$~GeV. The scale dependence is matched 
to two-loop order at the top threshold~\cite{Chetyrkin:2000yt}.

We neglect effects due to quark mixing and set the CKM matrix to
unity. Throughout this work, we parametrise the couplings appearing in
LO in the $G_\mu$ scheme, i.e.\ we use $\alpha_{G_\mu}$, whereas we
fix the electromagnetic coupling appearing in the relative corrections
by $\alpha=\alpha(0)$.  If not stated otherwise, we use the parameters
\be
\cos\theta_{\mathrm{cut}}=0.965,\quad s_{\mathrm{cut}}=0.81s,\quad z_{\mathrm{cut}}=0.9,\quad y_{\mathrm{cut}}=0.002
\label{esparas}
\ee
in accordance with the event-selection criteria used in the ALEPH
analysis \cite{Barate:1996fi} and employ the Durham jet algorithm
together with the $E$ recombination scheme for the reconstruction of
isolated photons (see \refsec{jetrate}).

As mentioned in \refsec{sec:es}, we implement a cut such that the
singularity in the two-jet region is avoided. This cut requires the
variables $T$, $\rho$, $B_{\mathrm{T}}$, $B_{\mathrm{W}}$, and $C$ to
be greater than $0.005$, whereas $Y_3$ and $y_\mathrm{cut}$ for
$\sigma_{\mbox{\scriptsize3-jet}}$ are required to be greater than $0.00005$.
%

\subsection{Checks of the implementation}

The reliability of our results is ensured as follows:
\begin{itemize}
\item \emph{UV finiteness} is checked by varying the reference scale $\mu$ of dimensional regularisation
and finding that our results are independent of this variation.
\item \emph{IR finiteness} is verified through varying the infinitesimal photon mass $m_\gamma$ 
in mass regularization and observing
that the sum of the virtual corrections and the soft-photonic corrections in both the slicing and subtraction
approach is invariant.
In dimensional regularisation the independence of $\mu$ is checked as for UV divergences.
\item \emph{Mass singularities} related to collinear photon emission or exchange are shown to cancel between
the virtual and the subtraction endpoint contributions by varying the small
masses of the external fermions.
\item \emph{Two completely independent} calculations have been
  performed within our collaboration. We find complete agreement of
  the results for $\sigma_{\mathrm{had}}$, jet rates, and event-shape
  distributions at the level of the Monte Carlo integration error
  which typically is at the per-mille level.
\end{itemize}

\begin{figure}
\begin{center}
\epsfig{file=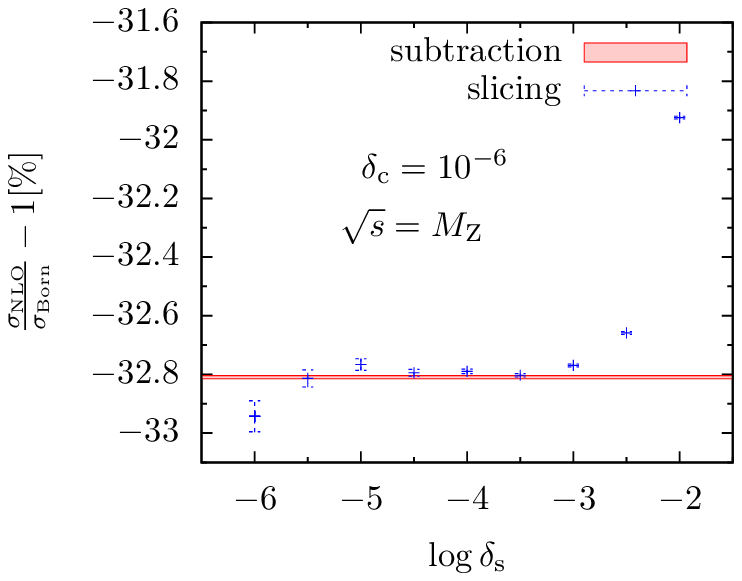,width=6cm}
\quad\epsfig{file=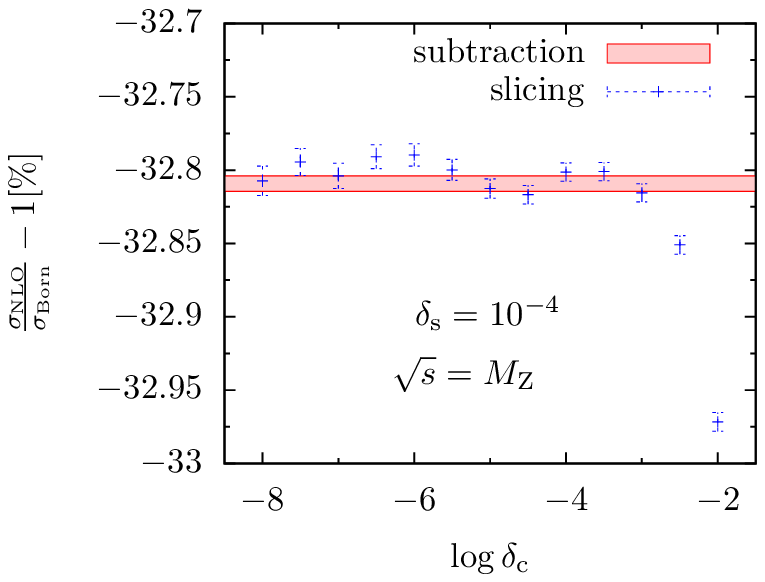,width=6cm}
\end{center}
\vspace*{-2em}
\caption{Slicing cut dependence of
  $\sigma_{\mathrm{had}}\left(M_\PZ\right)$.
For comparison also the results of the subtraction method are shown.}
\label{fig:sighad_slpara}
\end{figure}

\begin{figure}
\begin{center}
\epsfig{file=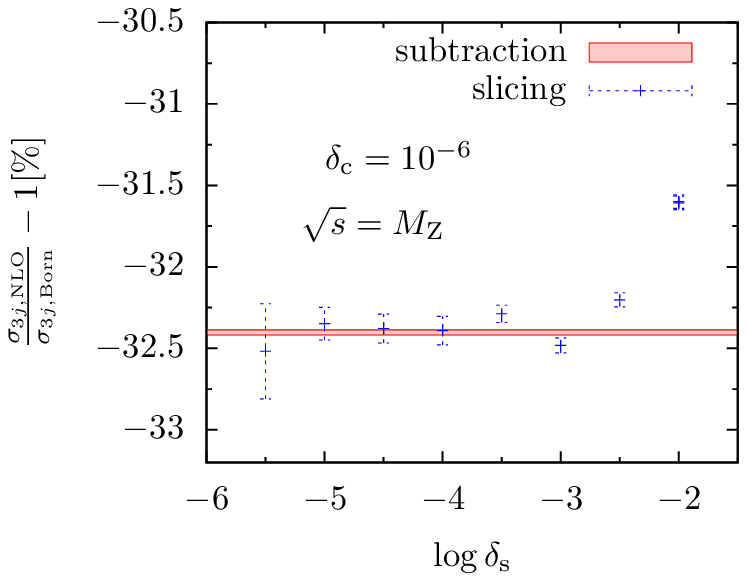,width=6cm}
\quad\epsfig{file=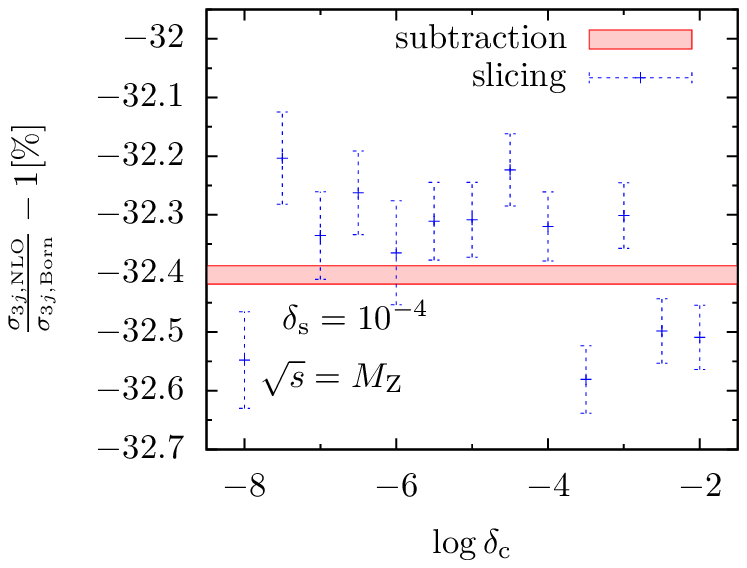,width=6cm}
\end{center}
\vspace*{-1em}
\caption{Dependence of the three-jet rate on the slicing parameters at
  $\sqrt{s}=M_\PZ$ for $y_{\mathrm{cut}}=0.0006$. For comparison also the results of the subtraction method are shown.}
\label{fig:sig3j_slpara}
\end{figure}

\begin{figure}
\begin{center}
\epsfig{file=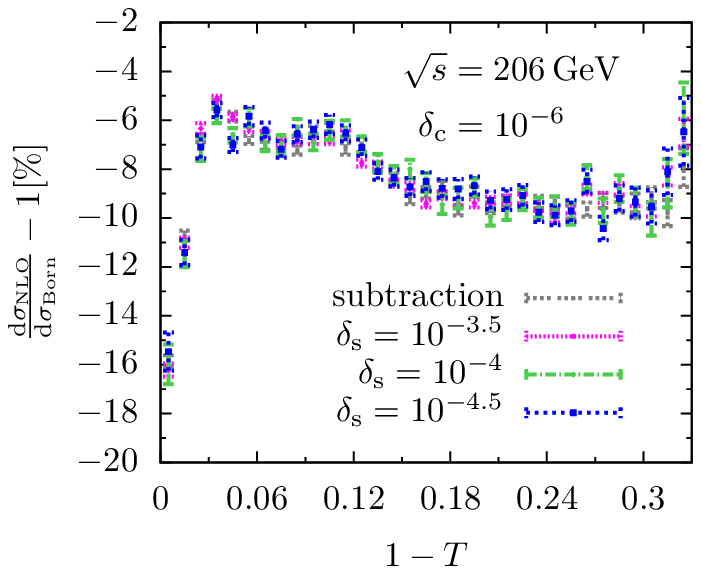,width=6cm}
\quad
\epsfig{file=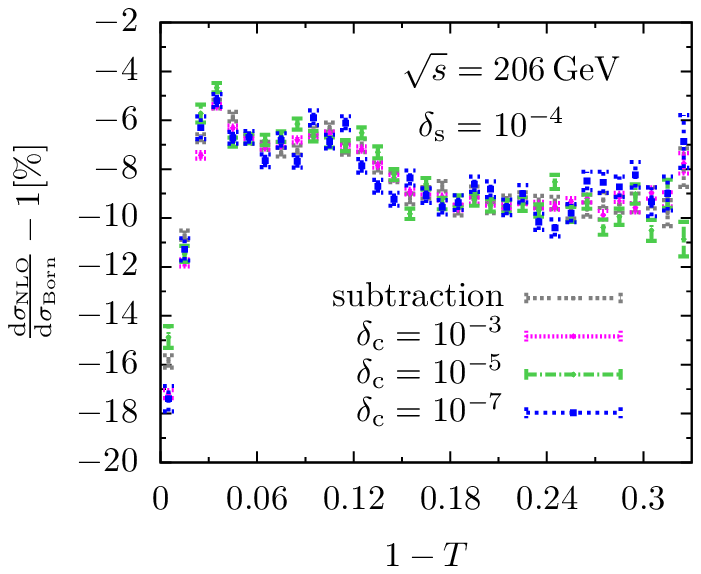,width=6cm}
\end{center}
\vspace*{-1em}
\caption{Dependence of the differential thrust distribution
on the slicing parameters at $\sqrt{s}=206\GeV$. For comparison also the results of the subtraction method are shown.}
\label{fig:T_slpara}
\end{figure}

Furthermore, we compare the results obtained with phase-space slicing
and the subtraction method, which are completely independent
techniques. In \reffig{fig:sighad_slpara} we show the mutual agreement
of both techniques for the NLO EW results for $\sigma_{\mathrm{had}}$,
in \reffig{fig:sig3j_slpara} for the full
$\mathcal{O}{\lrb\alpha\rrb}$ results for the three-jet rate with
$y_\mathrm{cut}=0.0006$ at $\sqrt{s}=\MZ$, and in
\reffig{fig:T_slpara} for the full $\mathcal{O}{{\lrb\alpha\rrb}}$
results for the thrust distribution at $\sqrt{s}=206\GeV$.  Note that
\reffig{fig:sighad_slpara} and \reffig{fig:sig3j_slpara} refer to the
corrections to $\Pe^+\Pe^-\to q\bar{q}$ and
$\Pe^+\Pe^-\to q\bar{q}\Pg$, respectively, i.e.\ to two independent
implementations.
We find that within integration errors the slicing results become
independent of the cut-offs for $\delta_s\lesssim 10^{-3}$ and
$\delta_c\lesssim 10^{-4}$ and fully agree with the results obtained
using the subtraction method. For the sake of clarity, we show only
curves for values of the slicing parameters that lie on the plateau in
\reffig{fig:T_slpara}. For values of the slicing parameters outside
the plateau, the behaviour follows the same pattern as in
Figs.~\ref{fig:sighad_slpara} and \ref{fig:sig3j_slpara}.

It turns out that the subtraction method is more efficient in terms of
run-time compared to phase-space slicing. To obtain results of comparable statistical quality, the
 phase-space slicing method required about one order of magnitude more events than the subtraction 
 method.  We therefore used the
subtraction method to obtain the results presented in the following
sections, both for $\sigma_{\mathrm{had}}$ and jet rates and
event-shape distributions.

\section{Numerical results}
\label{sec:results}
\setcounter{equation}{0}

The numerical parton-level event generator described above can be used
to compute the ${\cal O}(\alpha)$ electroweak corrections to the total
hadronic cross section, to event-shape distributions, and to the
three-jet rate.

\subsection{Results for the total hadronic cross section}
\label{se:results_sighad}

\begin{table}
\begin{tabular*}{\textwidth}{@{\extracolsep{\fill}}rcccc}
\toprule
&$\sqrt{s}=M_\PZ$&$\frac{\rd\sigma_i-\rd\sigma_{\mathrm{Born}}}{\rd\sigma_{\mathrm{Born}}}[\%]$&$\sqrt{s}=133\GeV$&$\frac{\rd\sigma_i-\rd\sigma_{\mathrm{Born}}}{\rd\sigma_{\mathrm{Born}}}[\%]$\\
\midrule
$\sigma_{\mathrm{had}}^{\mathrm{Born}}/\nb$&$38.2845(15)$&&$0.068858(2)$&\\
$\sigma_{\mathrm{had}}^{\mathrm{weak}}/\nb$&$37.8541(2)$&$~{-1.1}$&$0.068348(2)$&$-0.7$\\
$\sigma_{\mathrm{had}}^{\mathrm{NLO}}/\nb$&$25.729(3)$&$-32.8$&$0.06269(2)$&$-9.0$\\
$\sigma_{\mathrm{had}}^{\mathrm{NLO+h.o.LL}}/\nb$&$27.341(3)$&$-28.6$&$0.06208(2)$&$-9.8$\\
\bottomrule
\end{tabular*}
\\[0.7cm]
\begin{tabular*}{\textwidth}{@{\extracolsep{\fill}}rcccc}
\toprule
&$\sqrt{s}=172\GeV$&$\frac{\rd\sigma_i-\rd\sigma_{\mathrm{Born}}}{\rd\sigma_{\mathrm{Born}}}[\%]$&$\sqrt{s}=206\GeV$&$\frac{\rd\sigma_i-\rd\sigma_{\mathrm{Born}}}{\rd\sigma_{\mathrm{Born}}}[\%]$\\
\midrule
$\sigma_{\mathrm{had}}^{\mathrm{Born}}/\nb$&$0.0276993(8)$&&$0.0170486(5)$&\\
$\sigma_{\mathrm{had}}^{\mathrm{weak}}/\nb$&$0.0276780(8) $&$~{-0.1}$&$0.0170626(5)$&$~0.1$\\
$\sigma_{\mathrm{had}}^{\mathrm{NLO}}/\nb$&$0.024770(7) $&$-10.6$&$0.015197(4)$&$-10.9$\\
$\sigma_{\mathrm{had}}^{\mathrm{NLO+h.o.LL}}/\nb$&$0.024633(7) $&$-11.1$&$0.015127(4)$&$-11.3$\\
\bottomrule
\end{tabular*}
\\[0.7cm]
\begin{tabular*}{\textwidth}{@{\extracolsep{\fill}}rcccc}
\toprule
&$\sqrt{s}=500\GeV$&$\frac{\rd\sigma_i-\rd\sigma_{\mathrm{Born}}}{\rd\sigma_{\mathrm{Born}}}[\%]$&$\sqrt{s}=1000\GeV$&$\frac{\rd\sigma_i-\rd\sigma_{\mathrm{Born}}}{\rd\sigma_{\mathrm{Born}}}[\%]$\\
\midrule
$\sigma_{\mathrm{had}}^{\mathrm{Born}}/\nb$&$0.00241881(7)$&&$0.00059139(2)$&\\
$\sigma_{\mathrm{had}}^{\mathrm{weak}}/\nb$&$0.00238722(7)$&$~{-1.3}$&$0.00056838(2)$&$~{-3.9}$\\
$\sigma_{\mathrm{had}}^{\mathrm{NLO}}/\nb$&$0.0020665(7)$&$-14.6$&$0.0004856(2)$&$-17.9$\\
$\sigma_{\mathrm{had}}^{\mathrm{NLO+h.o.LL}}/\nb$&$0.0020585(7)$&$-14.9$&$0.0004836(2)$&$-18.2$\\
\bottomrule
\end{tabular*}
\caption{Total hadronic cross section $\sigma_{\mathrm{had}}\left(\sqrt{s}\right)$ for LEP1 and LEP2 energies, and for $\sqrt{s}=500\GeV$ and $\sqrt{s}=1\TeV$.}
\label{tab:sighad}
\end{table}
The total hadronic cross section as a function of the CM energy and
the corresponding NLO electroweak corrections have been shown in
\citere{Denner:2009gx}. Here we list some numbers that have been used
to extract $\delta_{\sigma,1}$ and $\delta_{\sigma,\ge2,\mathrm{LL}}$, as
defined in \refeq{sig0NLO} and \refeq{sig0LL}, which enter the
calculation of normalised event-shape distributions and jet rates.
We use the event selection as described in \refsec{sec:es} with the cut
parameters given in \refeq{esparas}.  In the following, `weak' refers to
the electroweak NLO corrections without purely photonic corrections.

\reftab{tab:sighad} shows the Born contribution to
$\sigma_{\mathrm{had}}$ in the first row, the weak
$\mathcal{O}(\alpha)$ contribution in the second row, the full
$\mathcal{O}(\alpha)$ contribution in the third row, and the full
$\mathcal{O}(\alpha)+$ h.o. LL contribution in the fourth row for LEP1
and LEP2 energies, as well as for $\sqrt{s}=500\GeV$ and
$\sqrt{s}=1000\GeV$. We show the absolute results in nanobarn in the
second and fourth columns and the relative corrections in per cent in
the third and fifths columns.  The numbers in parentheses give the
uncertainties from Monte Carlo integration in the last digits of the
predictions.
For most energies, the electroweak corrections are sizable and
negative, ranging between $-30\%$ at the $\PZ$ peak and about $-9\%$
at energies above. The numerically largest contribution is
always due to ISR. 
Above $120\GeV$ the magnitude of the corrections is increased due to
LL resummation of ISR, whereas it is decreased in the resonance
region.  The virtual one-loop weak corrections (from fermionic and
massive bosonic loops) are moderate of a few per cent and
always negative for $\MZ<\sqrt{s}<1\TeV$.

\subsection{Results for the event-shape distributions and jet rates}
\label{se:results_distri}
In the following we present the results of our calculation for the
three-jet rate as well as for event-shape distributions as described
in \refsec{jetrate}. We show our findings for $\sqrt{s}=\MZ$ as used
at LEP1 and the selected LEP2 energies $172\GeV$ and $206\GeV$. To
stress the relevance of our work for future linear colliders, we also
show results for $\sqrt{s}=500\GeV$.

The precise size and shape of the corrections depend on the observable
$y$ in question. However, they share the common feature that $q\bar
q\gamma$ final states contribute only in the two-jet region, typically
for small values of $y$.

\begin{figure}
\begin{center}
\epsfig{file=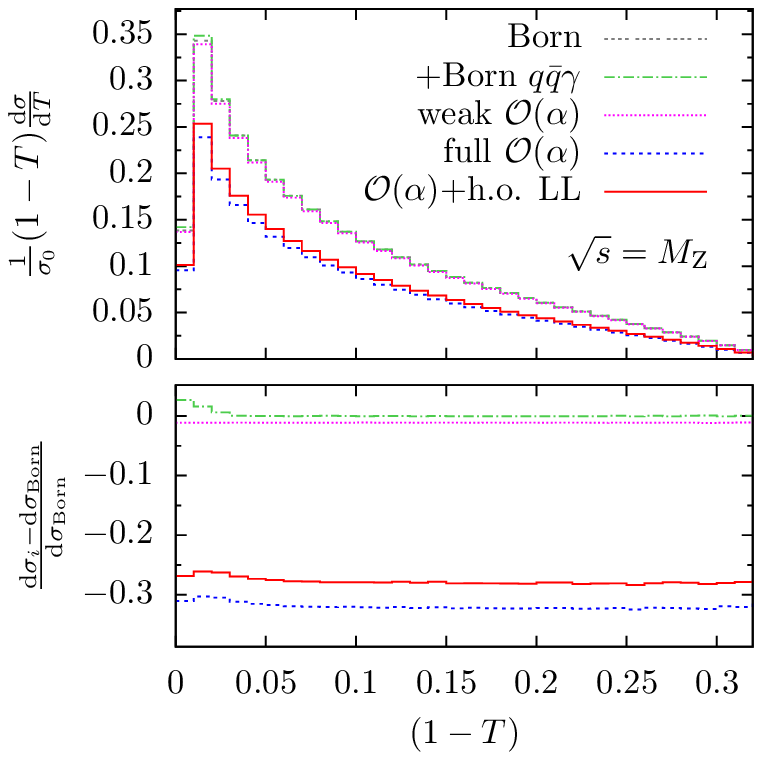,width=6.3cm}
\quad
\epsfig{file=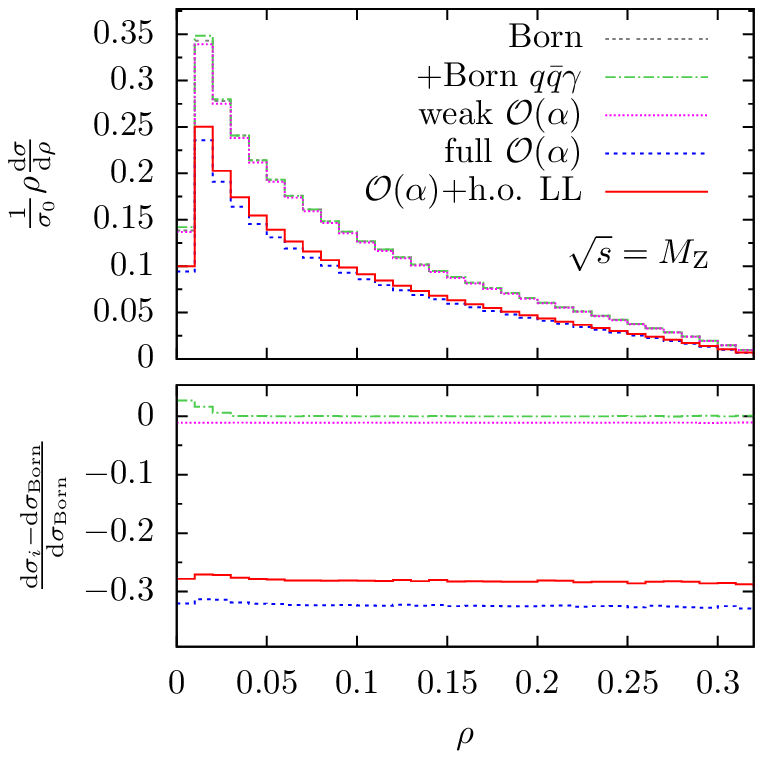,width=6.3cm}\\[2mm]
\epsfig{file=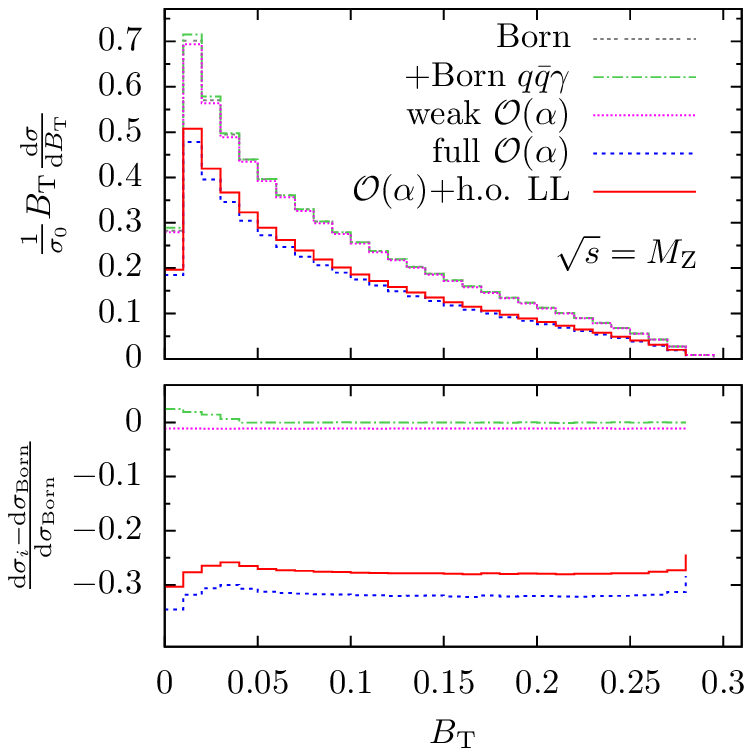,width=6.3cm}
\quad
\epsfig{file=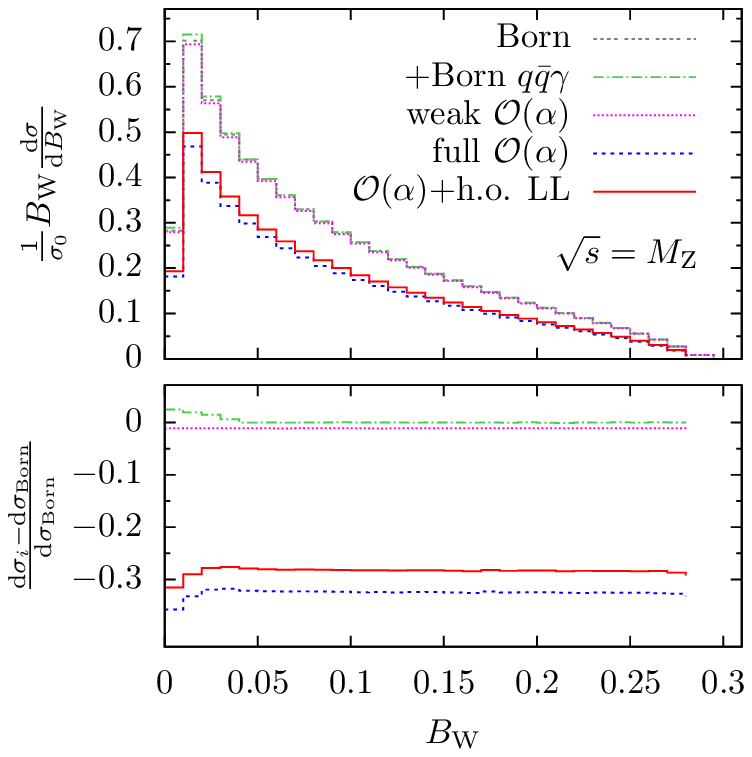,width=6.3cm}\\[2mm]
\epsfig{file=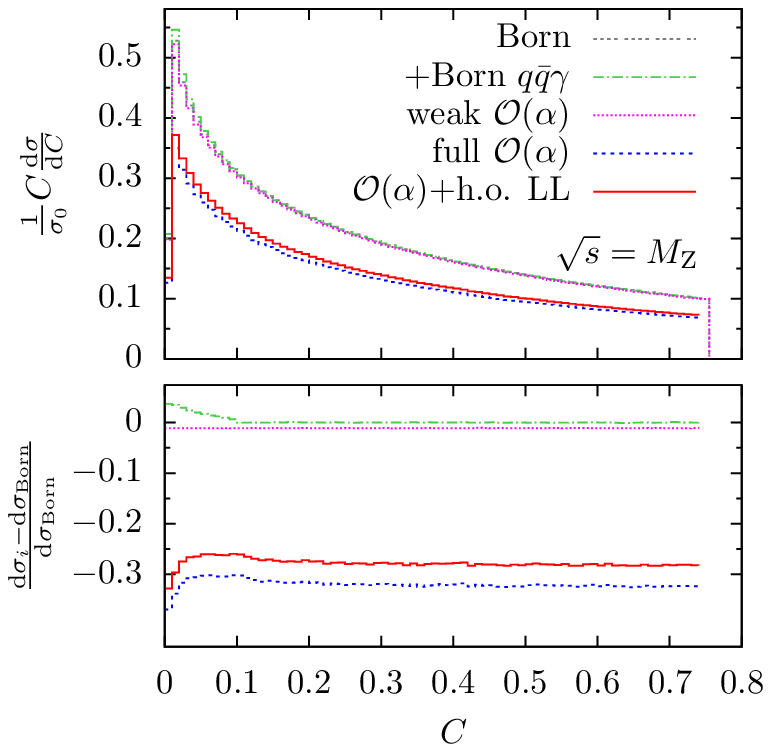,width=6.3cm}
\quad
\epsfig{file=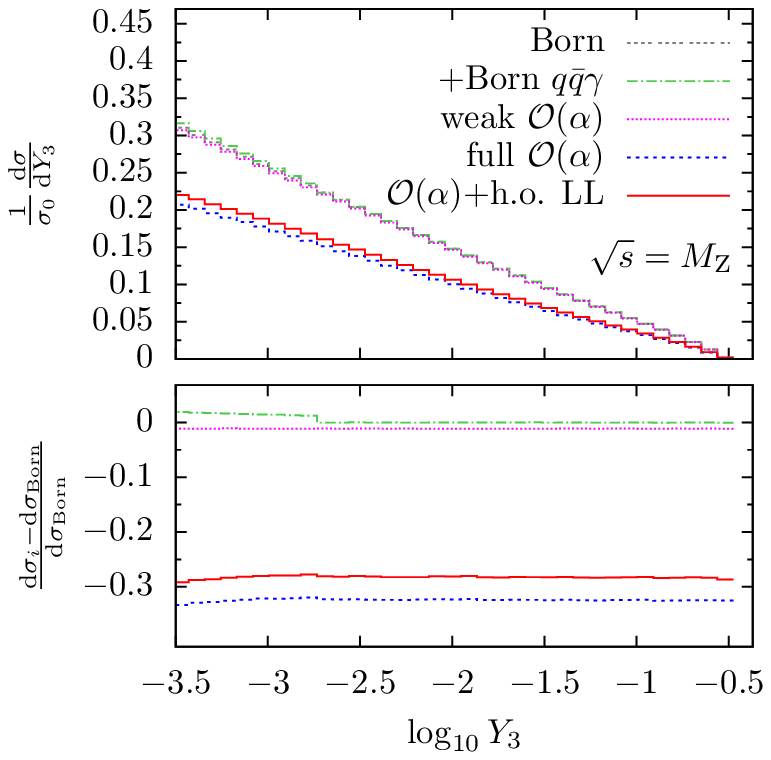,width=6.3cm}
\end{center}
\vspace*{-2em}
\caption{The event-shape distributions normalised to $\sigma_0$ at $\sqrt{s}=M_\PZ$.}
\label{fig:distri_MZ_nonorm}
\end{figure}

In a first step, we show our results for the distributions normalised
to $\sigma_0$ for $\sqrt{s}=\MZ$ in \reffig{fig:distri_MZ_nonorm}.
The Born contribution is given by the $A$ term of \refeq{dsdy_EW},
while the full ${\cal O}(\alpha)$ corrections contain the tree-level
$q\bar q \gamma$ contribution $\delta_\gamma$ and the NLO electroweak
contribution $\delta_A$ of \refeq{dsdy_EW}. The $T$, $\rho$,
$B_{\mathrm{T}}$, $B_{\mathrm{W}}$, and $C$ distributions are weighted
by the respective variable $y$, evaluated at each bin centre. The
relative corrections in the lower boxes are obtained by dividing the
respective contributions to the corrections by the Born distribution
given by the $A$ term.  We observe large negative corrections due to
ISR, and moderate weak corrections in all distributions.  The
corrections are mainly constant for large $y$ (note that we plot $1-T$
instead of $T$), where the isolated-photon veto rejects all
contributions from $q\bar q\gamma$ final states. Near the two-jet
limit, the contribution from $q\bar q\gamma$ final states dominates
the relative corrections. Moreover, it turns out that the
electromagnetic corrections depend non-trivially on the
event-selection cuts (see \refsec{se:results_paradep} for a more
detailed discussion).  We observe a significant decrease from the
second bin to the first bin in all distributions, caused by the lower
cut-off that we impose individually for all distributions. Since the
cut-off acts both in the Born and the NLO contribution, we find a
meaningful result for the relative corrections in the first bin.  In
the $Y_3$ distribution we clearly see the onset of the $q\bar q\gamma$
final states for $Y_3=0.002$.  Since we always cluster photons with
$y<y_\mathrm{cut}=0.002$ in the event selection (see \refsec{sec:es}),
the contribution from $q\bar q\gamma$ final states is removed if
$Y_3>0.002$ and only plays a role for $Y_3< 0.002$.
\begin{figure}
\begin{center}
\epsfig{file=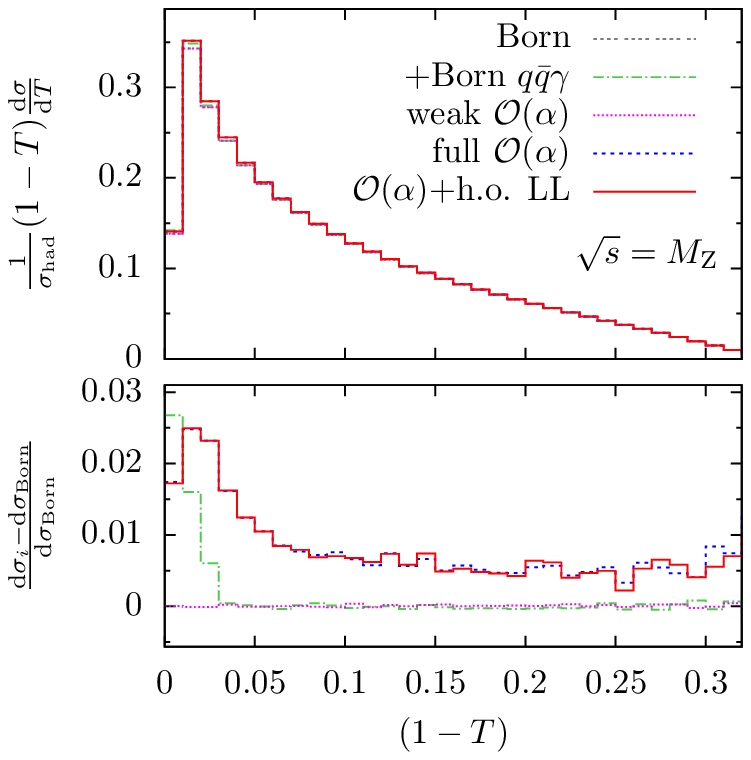,width=6.3cm}
\quad
\epsfig{file=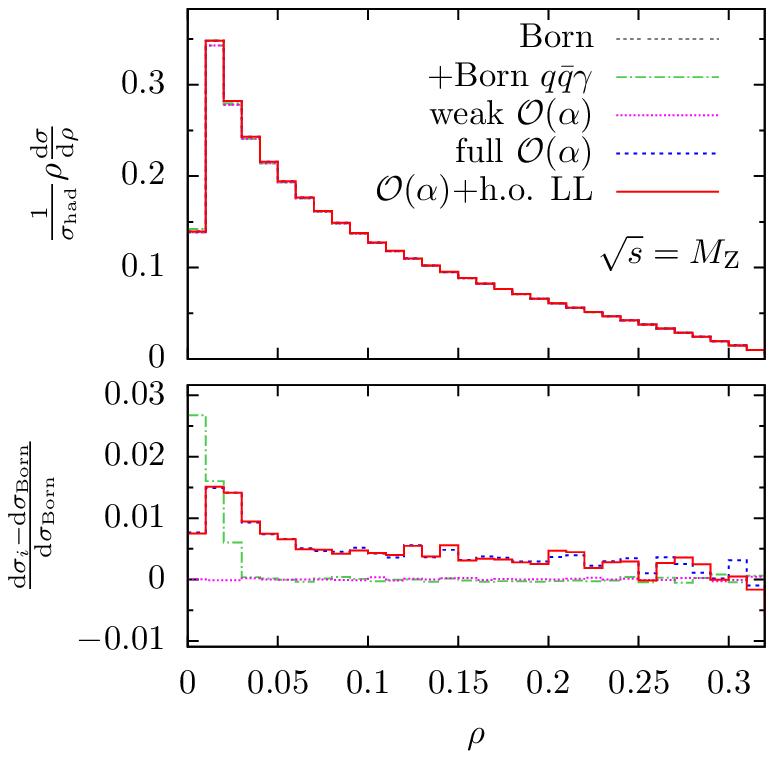,width=6.3cm}\\[2mm]
\epsfig{file=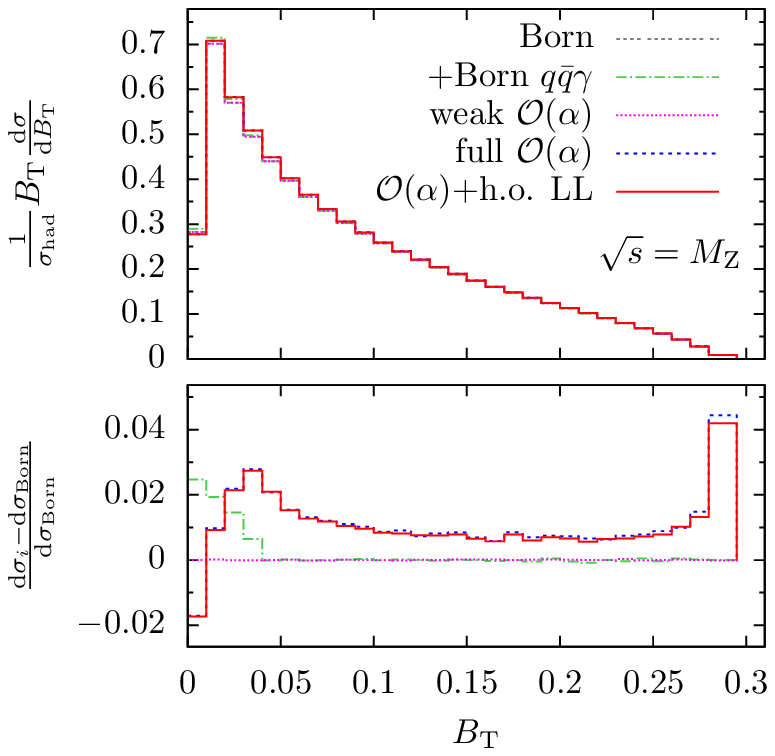,width=6.3cm}
\quad
\epsfig{file=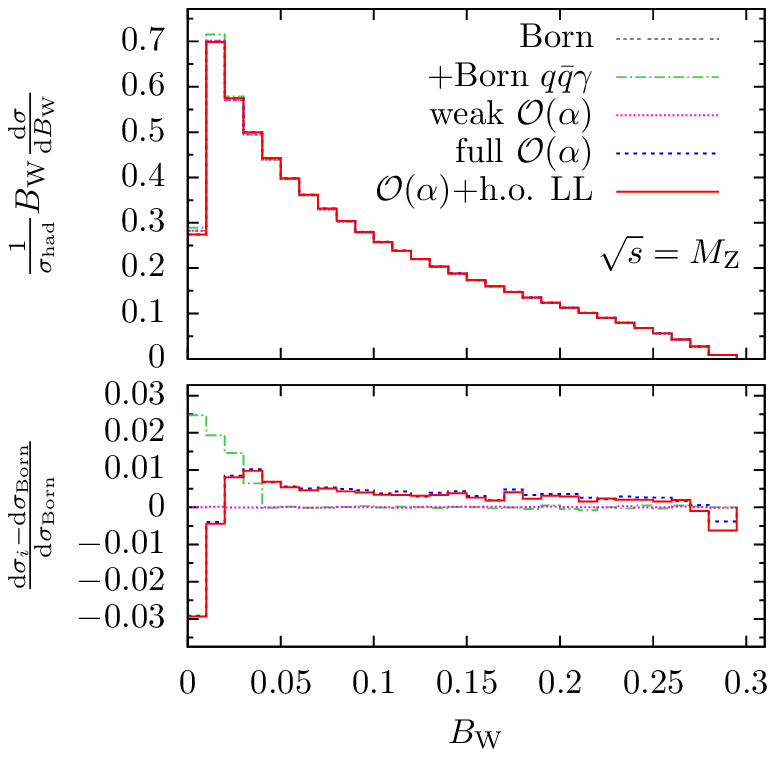,width=6.3cm}\\[2mm]
\epsfig{file=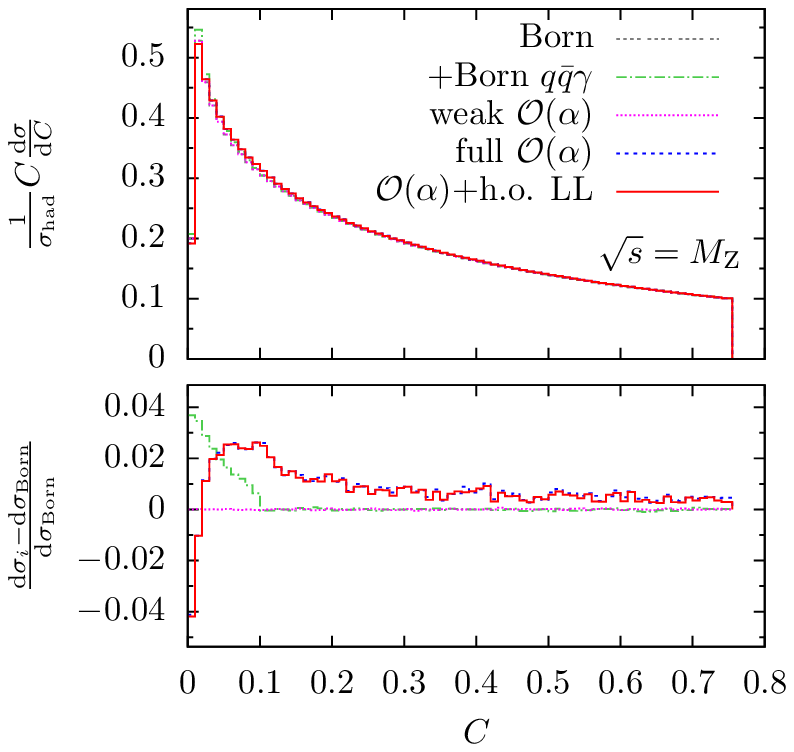,width=6.3cm}
\quad
\epsfig{file=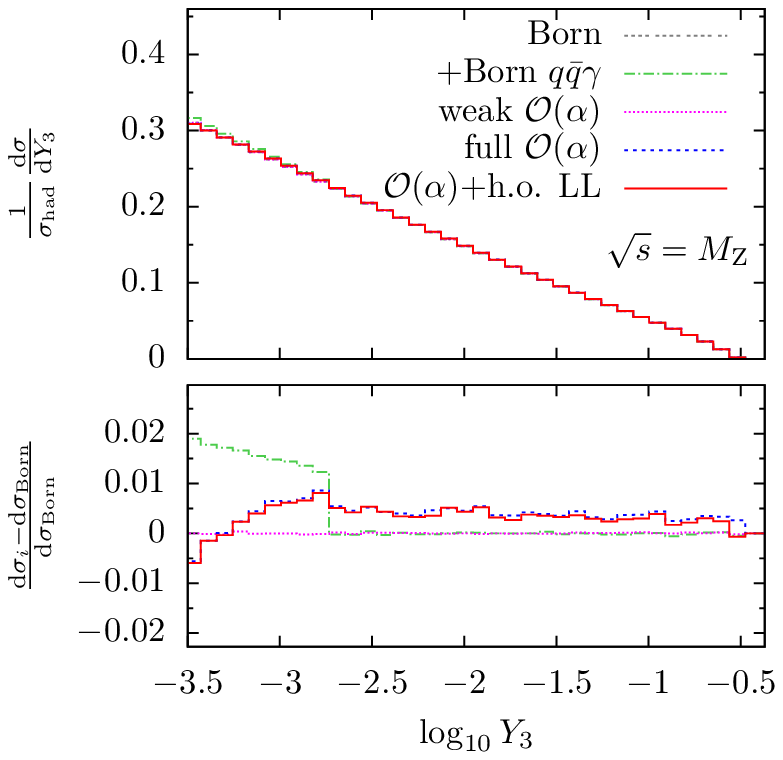,width=6.3cm}
\end{center}
\vspace*{-2em}
\caption{The event-shape distributions normalised 
to $\sigma_{\mathrm{had}}$ at $\sqrt{s}=M_\PZ$.}
\label{fig:distri_MZ_1}
\end{figure}
\begin{figure}
\begin{center}
\epsfig{file=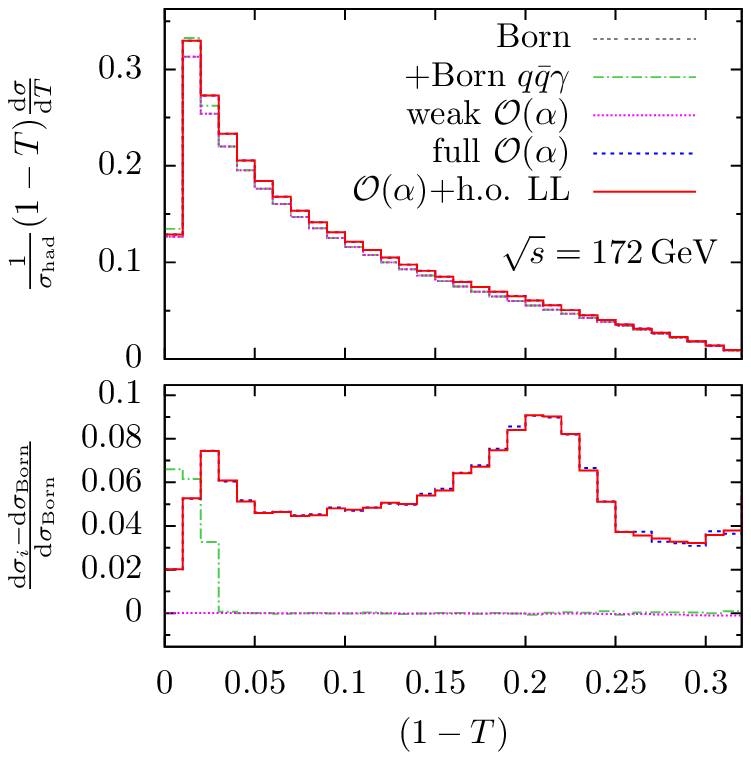,width=6.3cm}
\quad
\epsfig{file=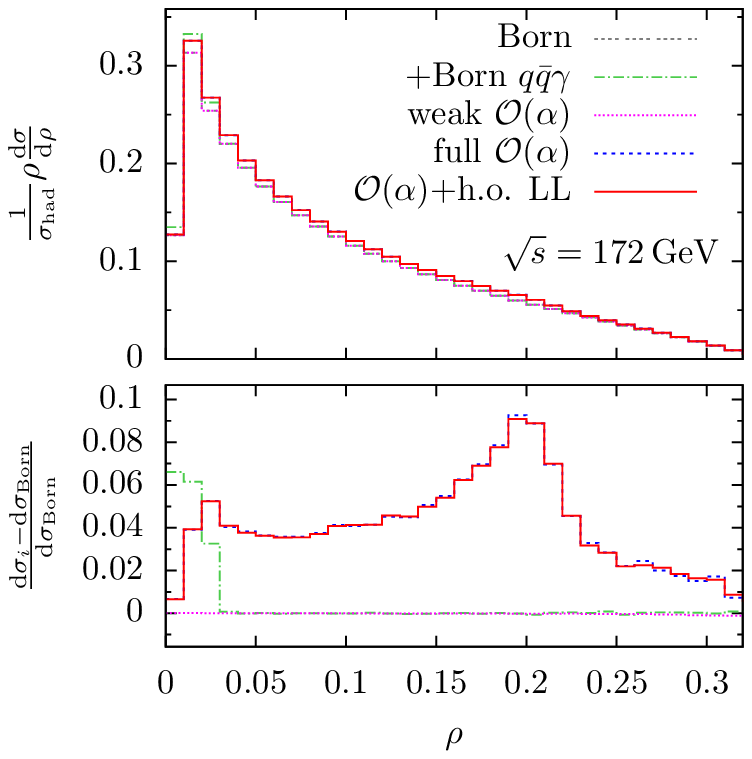,width=6.3cm}\\[2mm]
\epsfig{file=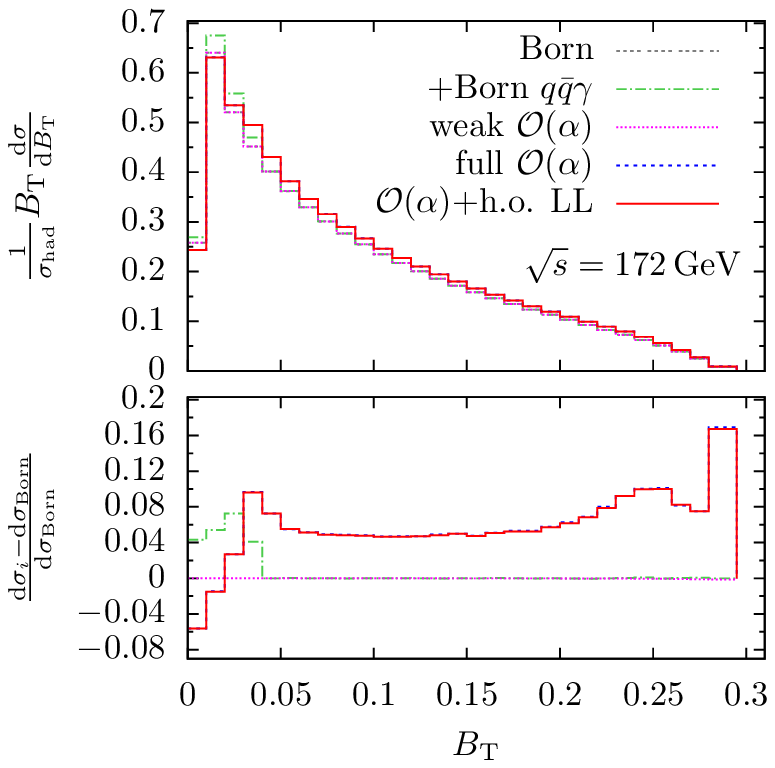,width=6.3cm}
\quad
\epsfig{file=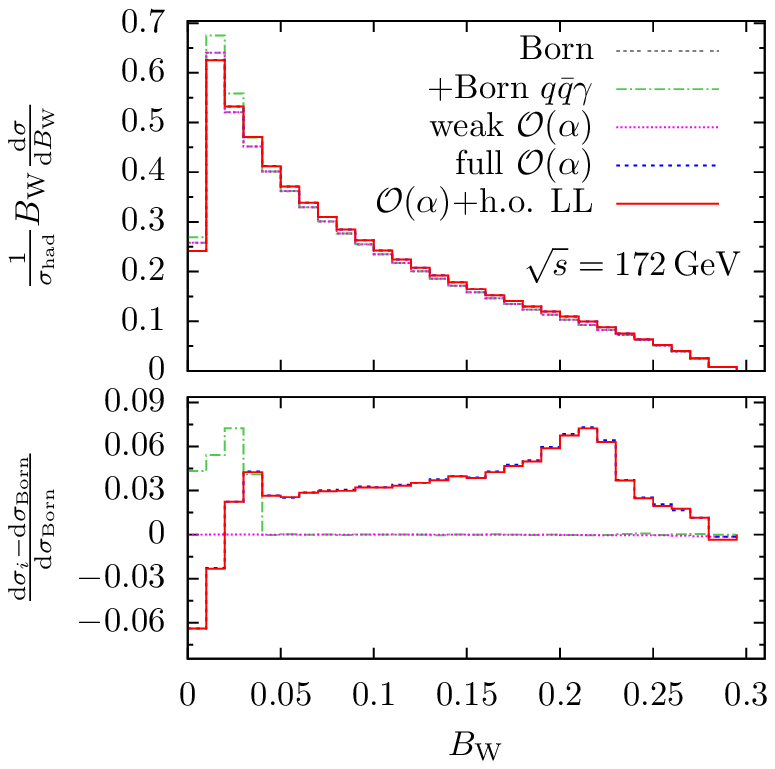,width=6.3cm}\\[2mm]
\epsfig{file=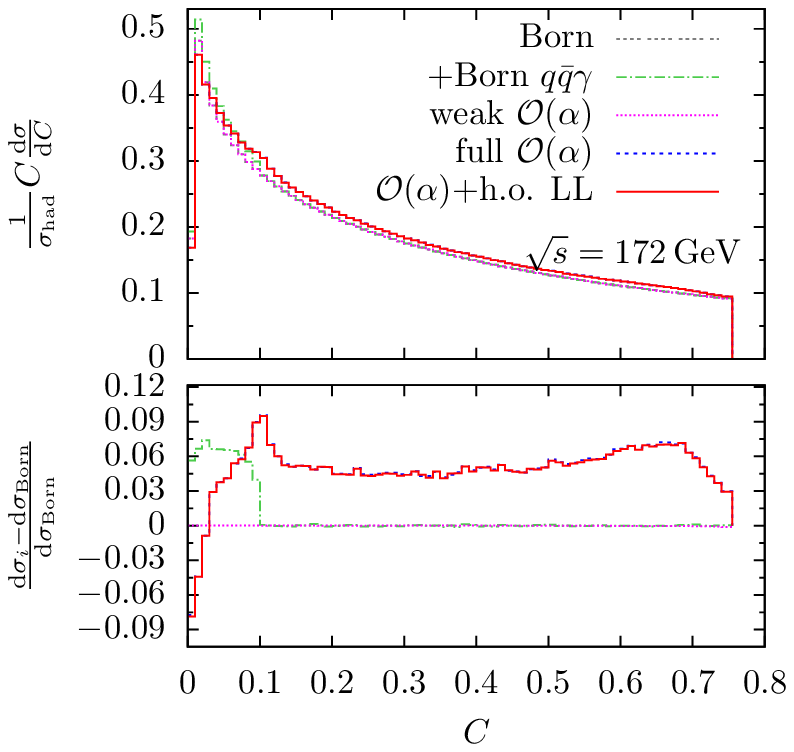,width=6.3cm}
\quad
\epsfig{file=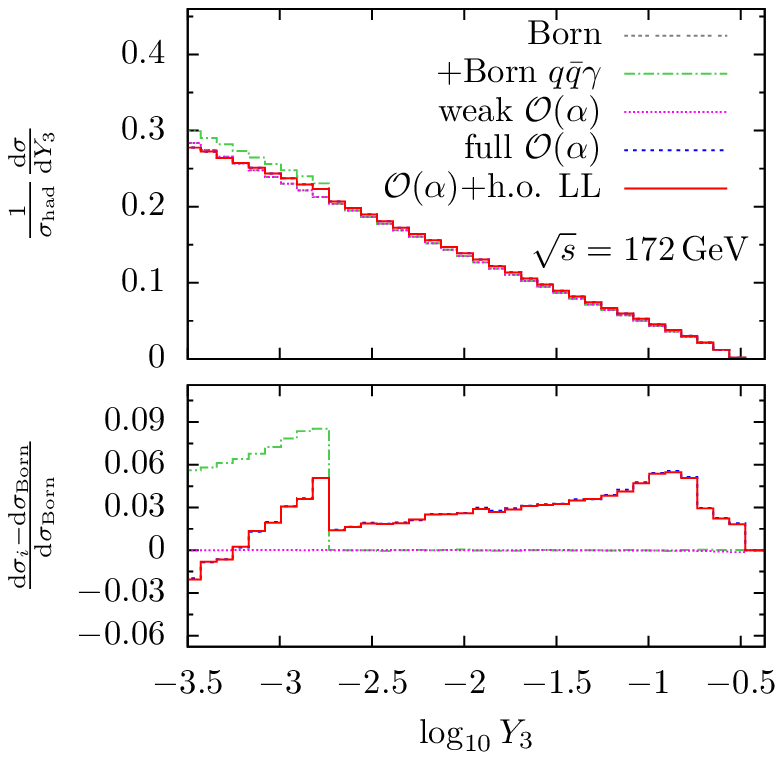,width=6.3cm}
\end{center}
\vspace*{-2em}
\caption{The event-shape distributions normalised 
to $\sigma_{\mathrm{had}}$ at $\sqrt{s}=172\GeV$.}
\label{fig:distri_172_1}
\end{figure}
\begin{figure}
\begin{center}
\epsfig{file=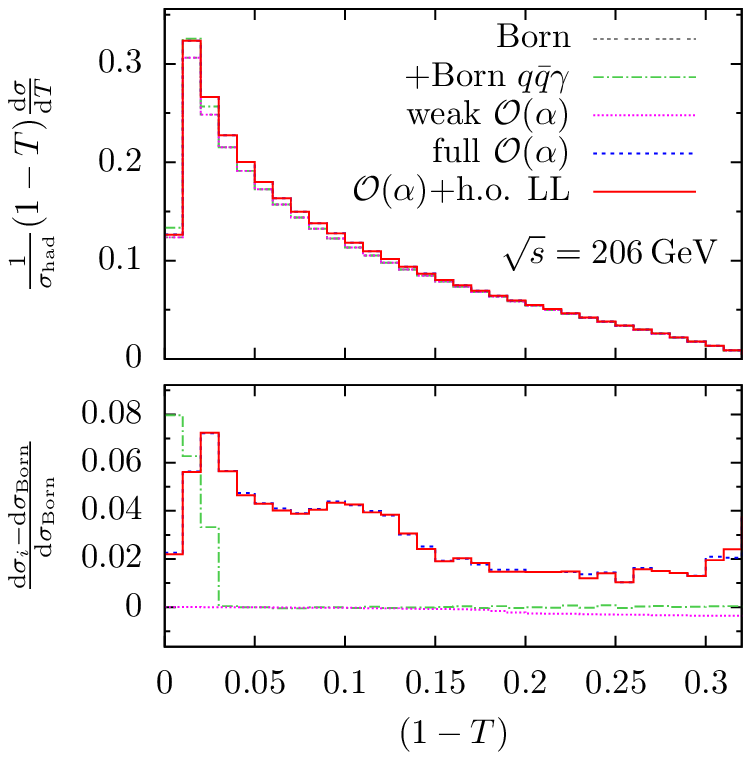,width=6.3cm}
\quad
\epsfig{file=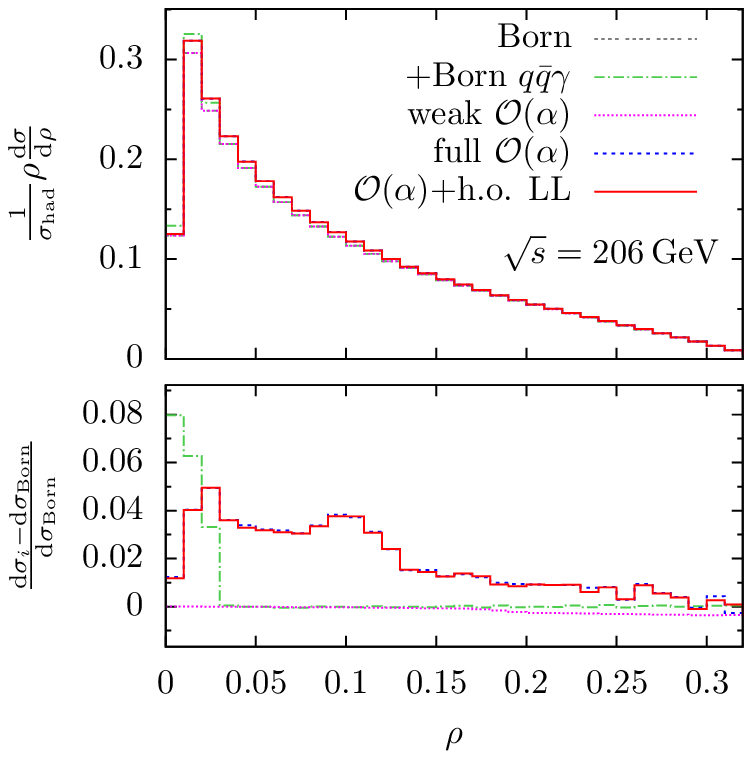,width=6.3cm}\\[2mm]
\epsfig{file=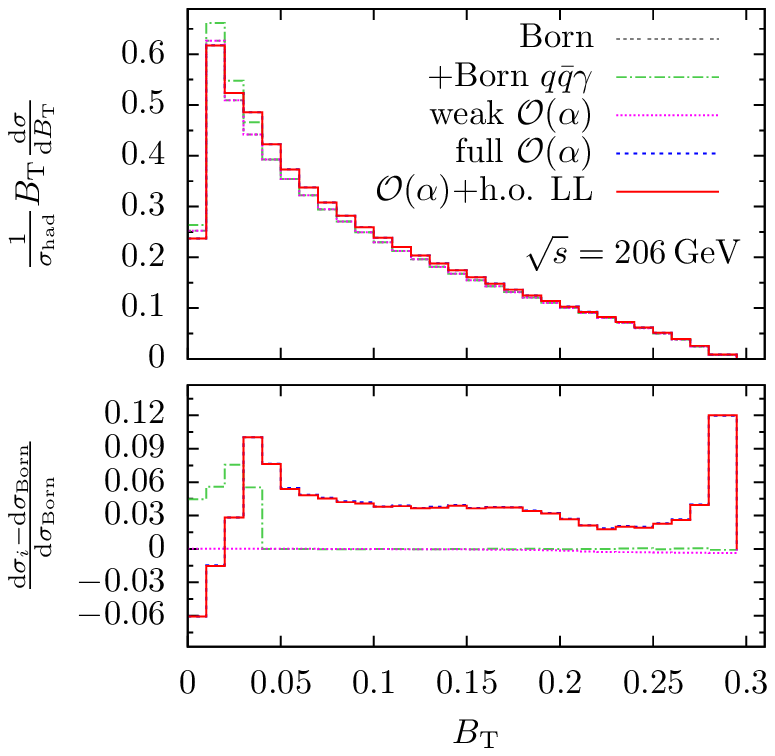,width=6.3cm}
\quad
\epsfig{file=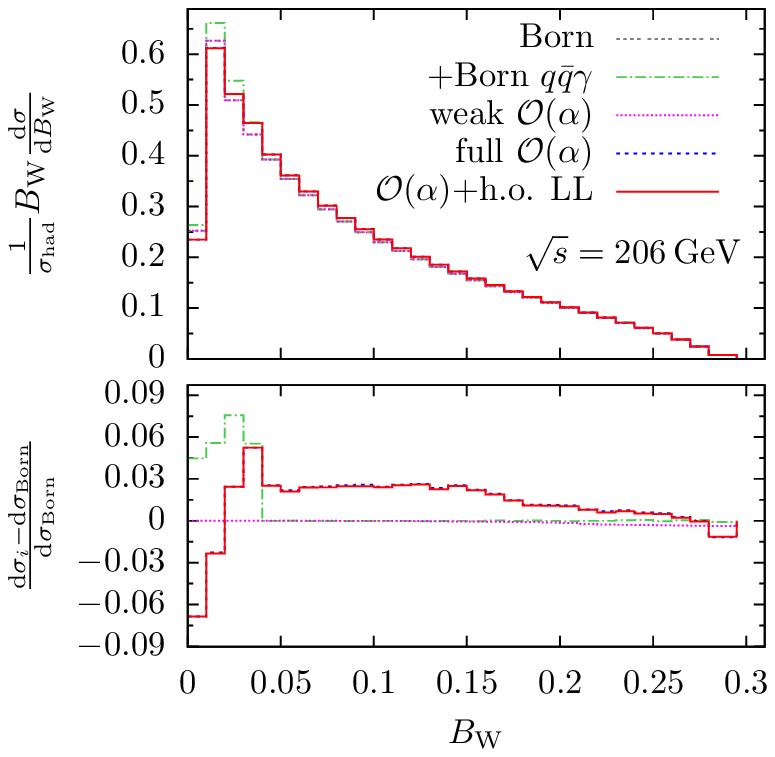,width=6.3cm}\\[2mm]
\epsfig{file=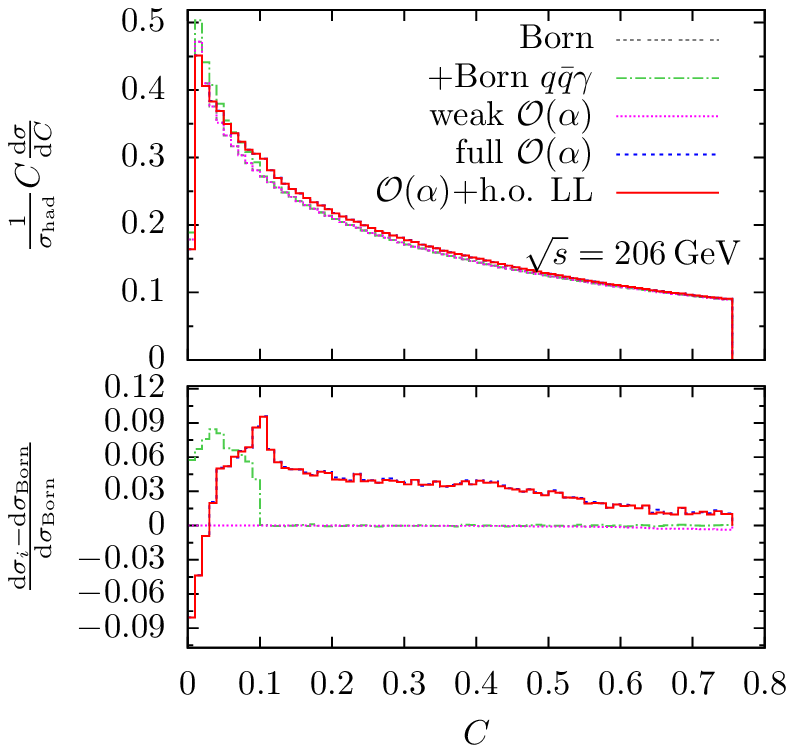,width=6.3cm}
\quad
\epsfig{file=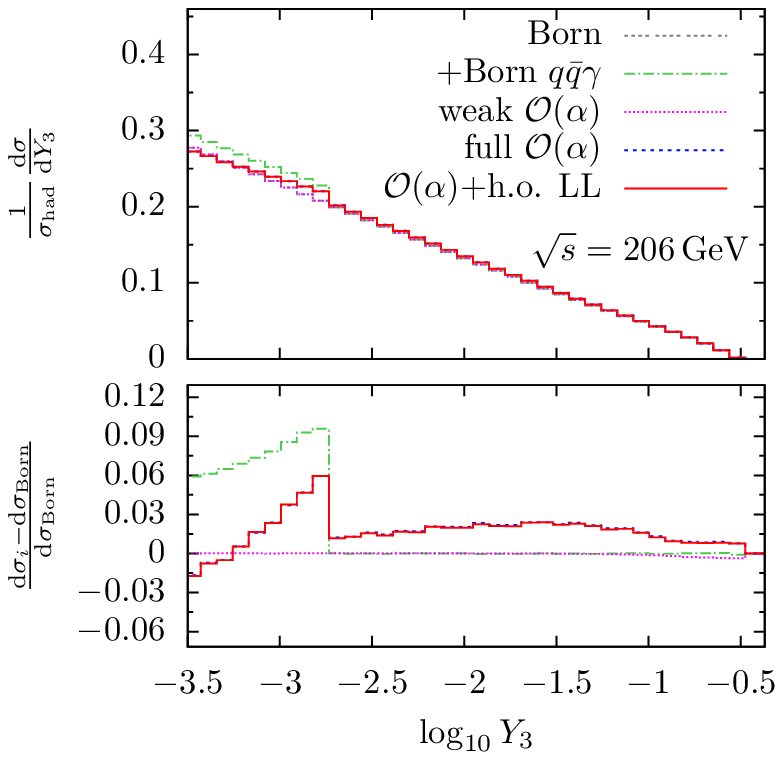,width=6.3cm}
\end{center}
\vspace*{-2em}
\caption{The event-shape distributions normalised 
to $\sigma_{\mathrm{had}}$ at $\sqrt{s}=206\GeV$.}
\label{fig:distri_206_1}
\end{figure}
\begin{figure}
\begin{center}
\epsfig{file=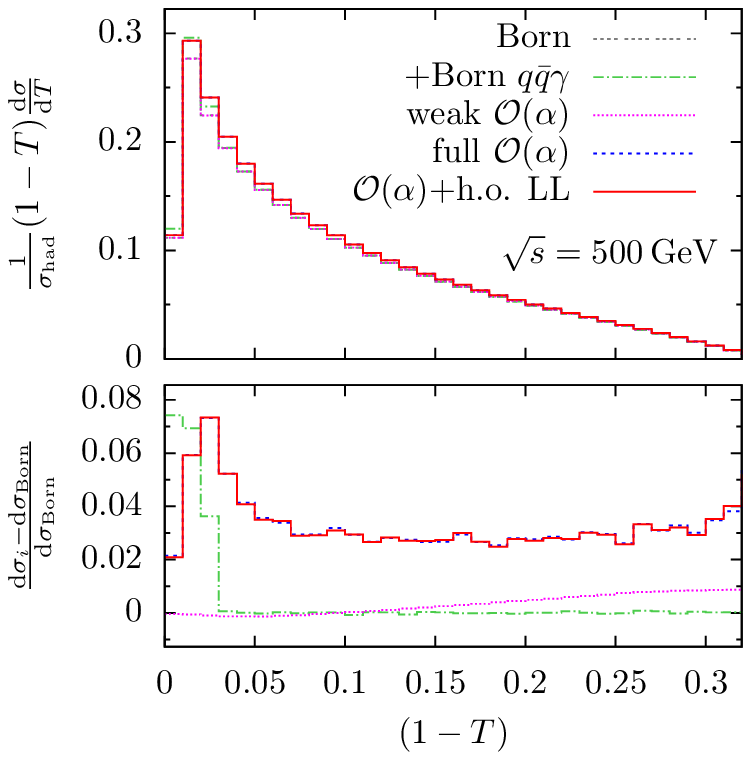,width=6.3cm}
\quad
\epsfig{file=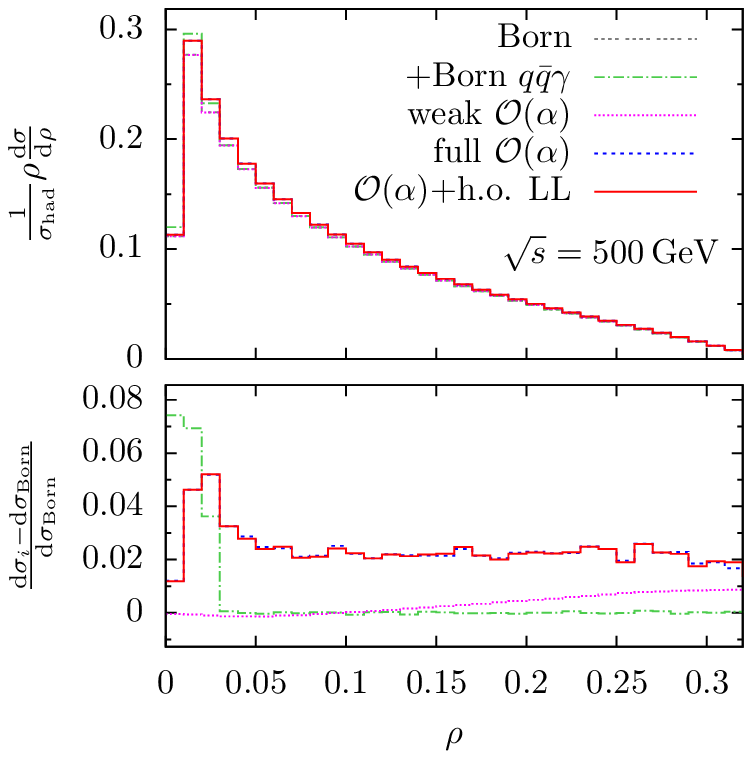,width=6.3cm}\\[2mm]
\epsfig{file=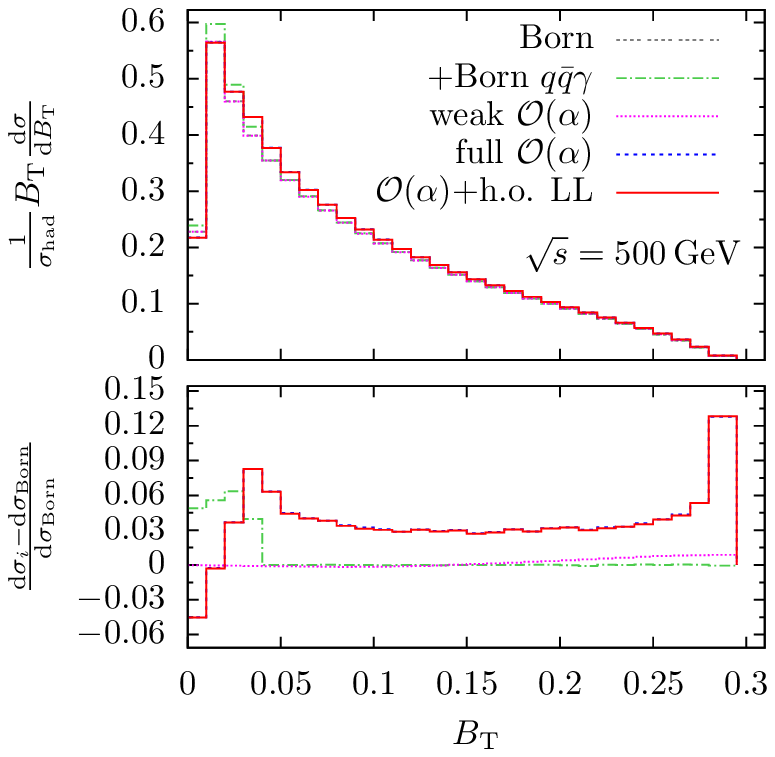,width=6.3cm}
\quad
\epsfig{file=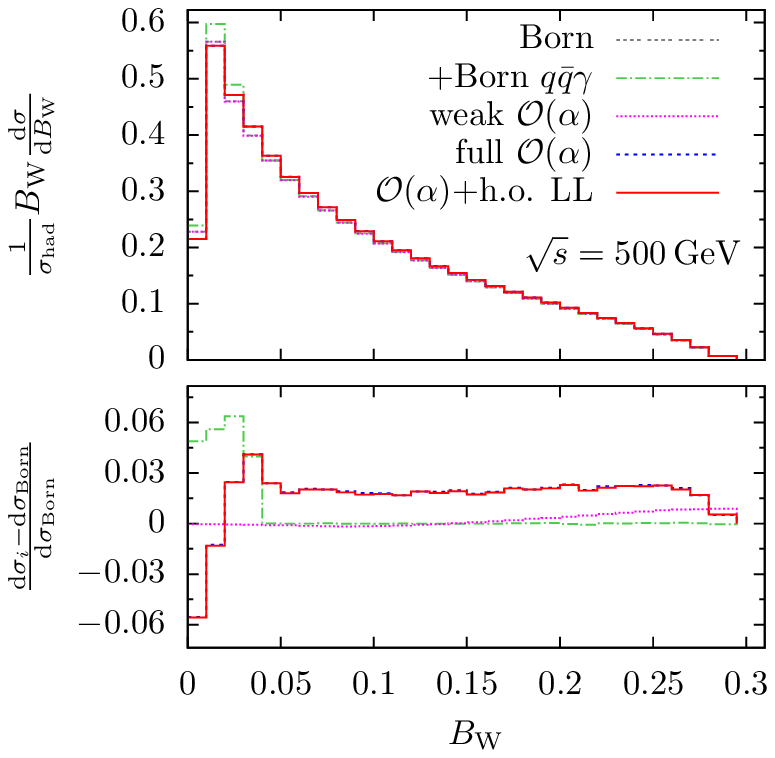,width=6.3cm}\\[2mm]
\epsfig{file=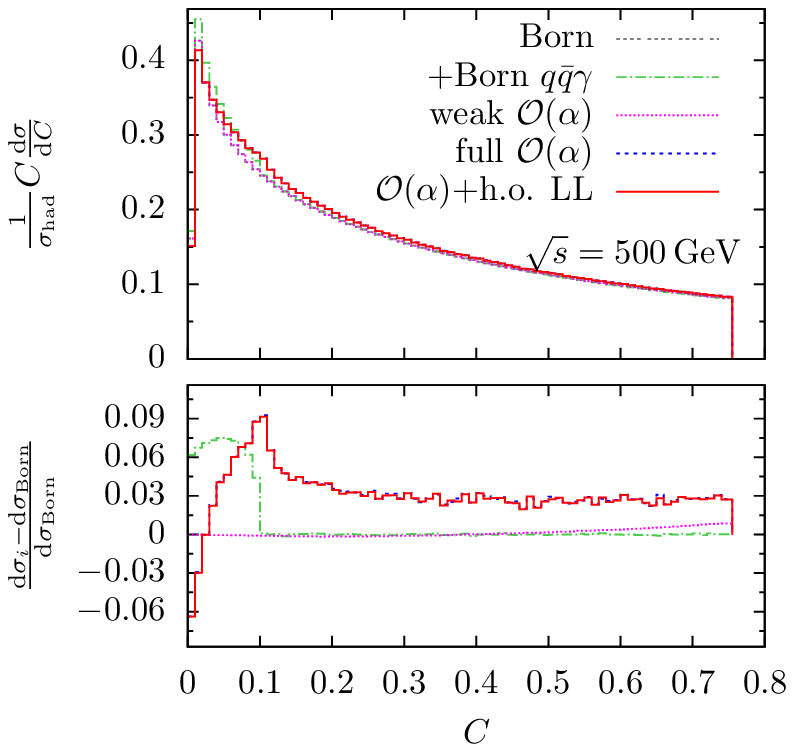,width=6.3cm}
\quad
\epsfig{file=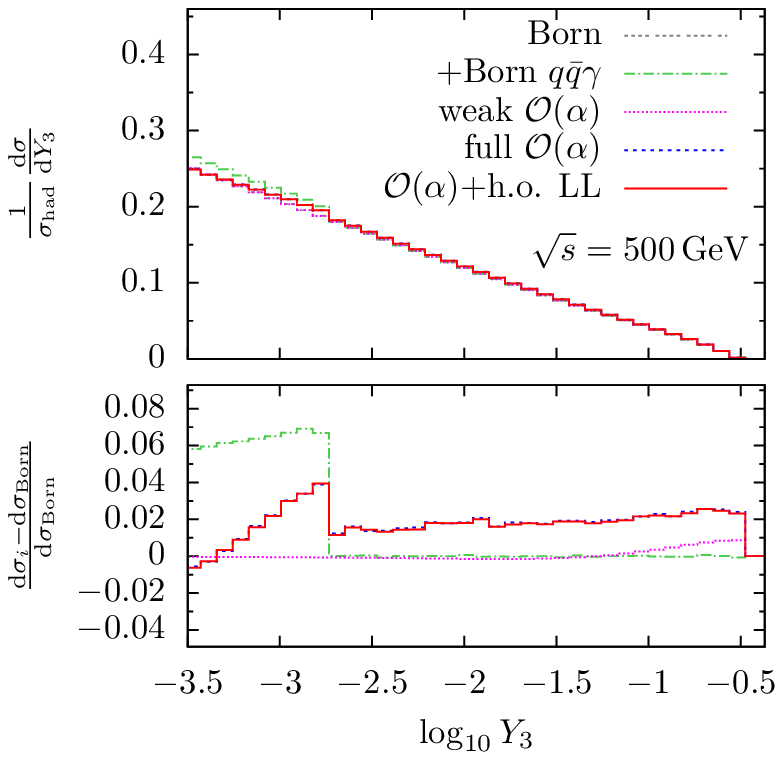,width=6.3cm}
\end{center}
\vspace*{-2em}
\caption{The event-shape distributions normalised 
to $\sigma_{\mathrm{had}}$ at $\sqrt{s}=500\GeV$.}
\label{fig:distri_500_1}
\end{figure}

In expanding the corrections according to \refeq{dsdyhad_EW}, and
retaining only terms up to LO in $\alphas$, we obtain the genuine
electroweak corrections to normalised event-shape distributions, which
we display at $\sqrt{s}=\MZ$ in \reffig{fig:distri_MZ_1}.  The Born
contribution is given by the $A$ term of \refeq{dsdyhad_EW}, while the
${\cal O}(\alpha)$ corrections are now given by $\delta_{\mathrm{EW}}$
of \refeq{deltaEW}. It can be seen very clearly that the large ISR
corrections cancel between the event-shape distributions and the
hadronic cross section when expanding the normalised distributions
properly, resulting in electroweak corrections of a few per cent.
Moreover, effects from ISR resummation are largely reduced as well,
and the difference between $\mathcal{O}{\lrb\alpha\rrb}$ and
$\mathcal{O}{\lrb\alpha\rrb}+{}$ h.o. LL is very small. 
The purely weak corrections are below the per-mille level.

For the thrust distribution, the full $\mathcal{O}{\lrb\alpha\rrb}$
corrections are almost constant around $0.5\%$ for $(1-T)>0.05$. The
coefficient $\delta_\gamma$ starts to emerge for $(1-T)=0.04$ and
contributes up to $2.6\%$. The full $\mathcal{O}(\alpha)$ corrections
peak for $(1-T)=0.02$ at $2.5\%$ and amount to $1.8\%$ for
$(1-T)=0.01$ in the first bin.  The full $\mathcal{O}(\alpha)$
corrections are flat and around $0.5\%$ for $\rho>0.05$,
$B_{\mathrm{W}}>0.05$, $C>0.1$, and $Y_3>0.002$.  For the
$B_{\mathrm{T}}$ distribution they are around $1\%$ and almost flat
for $B_{\mathrm{T}}>0.05$. The full $\mathcal{O}{\lrb\alpha\rrb}$
corrections reach a maximum between 1\% and 3\% typically for
small values of the event-shape variables and drop towards the first
bin down to between $1.8\%$ and  $-4.1\%$. Only for $B_{\mathrm{T}}$
we find also a maximum (of $4\%$) in the last bin. The LO
$q\bar{q}\gamma$ channel contributes only for small $y$ and can amount
up to 4\%.

In \reffig{fig:distri_172_1} we show our results for
$\sqrt{s}=172\GeV$.  The behaviour is similar as for $\sqrt{s}=\MZ$,
but in the middle of the distributions a peaking structure emerges. It
is located at $1-T,\rho,B_\mathrm{W}\approx 0.2$, $B_\mathrm{T}\approx
0.25$, $C\approx 0.65$, and $Y_3\approx 0.1$. 
We investigate this behaviour in detail in the next section.  In all
event-shape distributions the LO $q\bar q\gamma$ contribution ranges
between $3\%$ and $8\%$.  Outside the two-jet region and apart from
the domain of the peaking structure, the full
$\mathcal{O}{\lrb\alpha\rrb}$ corrections are flat and of the order of
$5\%$.
They peak near the onset of the $q\bar q\gamma$ final states between
$4\%$ and $10\%$, and drop in the first bin down to between $1.5\%$
and $-10\%$.

Results for $\sqrt{s}=206\GeV$, are displayed in
\reffig{fig:distri_206_1}.  In all event-shape distributions the LO
$q\bar q\gamma$ contribution ranges between $4\%$ and $9\%$. Outside
the two-jet region and outside the domain where the peaking structure
is located, the full $\mathcal{O}{\lrb\alpha\rrb}$ corrections are
flat between $0.1\%$ and $2\%$, they peak near the onset of the $q\bar
q\gamma$ final states between $5\%$ and $9\%$, and drop in the first
bin down to between $2\%$ and $-8\%$.  The peaking structure is now
situated at smaller values of $y$, it is less pronounced and,
especially for $B_\mathrm{W}$, $B_\mathrm{T}$, $C$, and $Y_3$, it
extends over a larger range of $y$.  Additionally, for large values of
$y$, the weak contribution slightly increases.

Finally, in \reffig{fig:distri_500_1} we show our results for
$\sqrt{s}=500\GeV$.  In the event-shape distributions the LO $q\bar
q\gamma$ contribution ranges between $2\%$ and $8\%$.  Outside the
two-jet region, the full $\mathcal{O}{\lrb\alpha\rrb}$ corrections are
flat between $2\%$ and $3\%$, they peak near the onset of the $q\bar
q\gamma$ final states between $2\%$ and $9\%$, and drop in the first
bin down to between $2\%$ and $-6\%$.  The weak corrections here
contribute up to $+1\%$ in all observables for large values of $y$.
The peaking structure as observed for $\sqrt{s}=172\GeV$ and
$\sqrt{s}=206\GeV$ has completely disappeared.

\begin{figure}[t]
\begin{center}
\epsfig{file=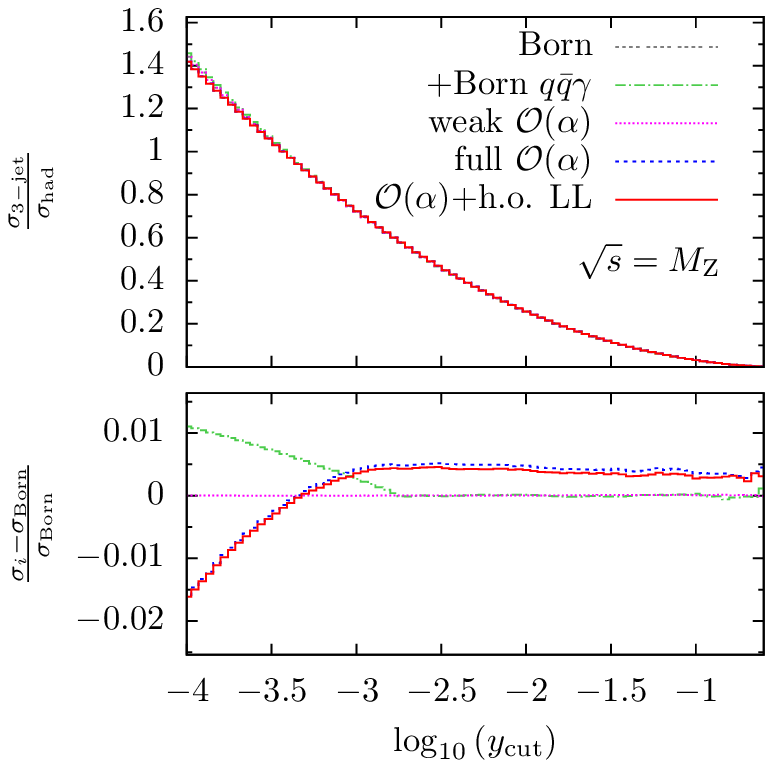,width=6.3cm}
\quad
\epsfig{file=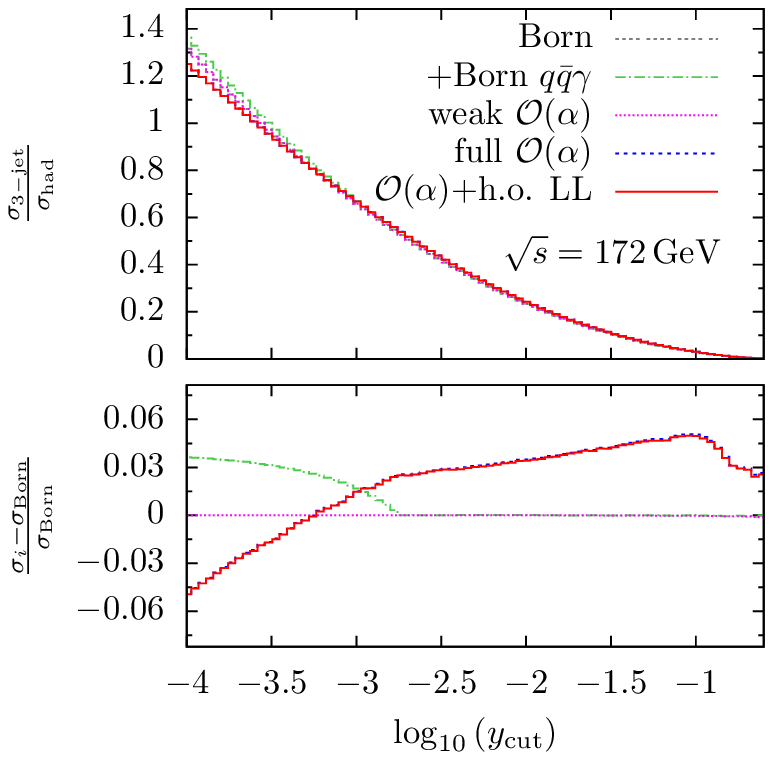,width=6.3cm}\\[2mm]
\epsfig{file=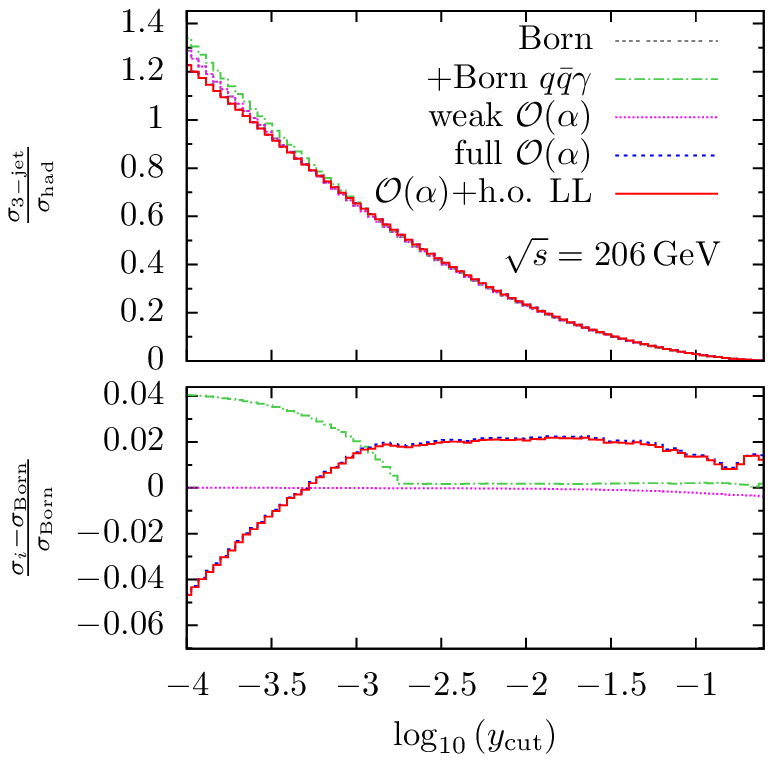,width=6.3cm}
\quad
\epsfig{file=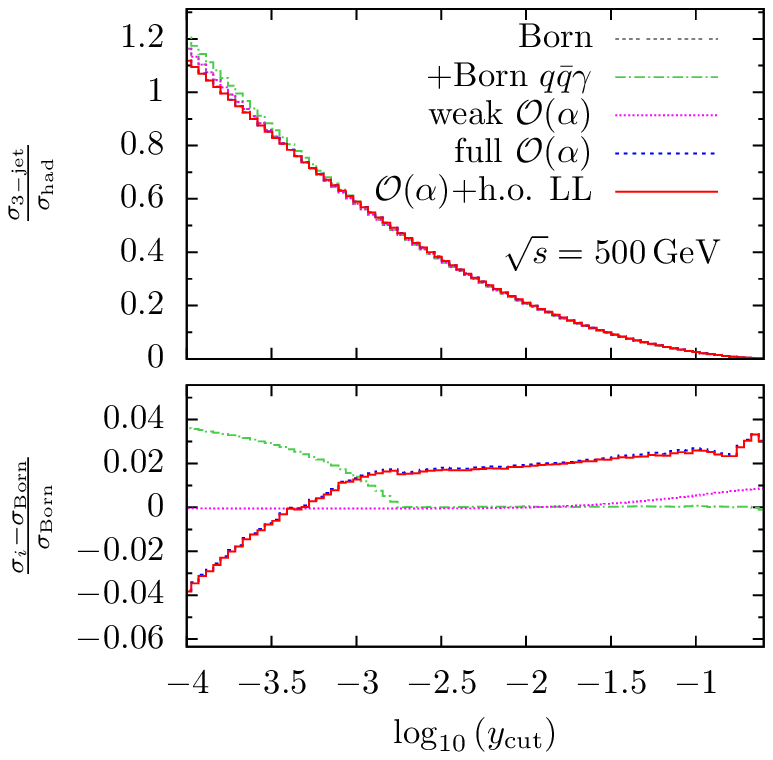,width=6.3cm}
\end{center}
\vspace*{-2em}
\caption{The three-jet rate normalised 
to $\sigma_{\mathrm{had}}$ at different CM energies.}
\label{fig:sig3}
\end{figure}
Figure~\ref{fig:sig3} displays the corrections to the three-jet rate
at various CM energies. As above, the corrections are
appropriately normalised to $\sigma_{\mathrm{had}}$.  At
$\sqrt{s}=M_\PZ$, the full $\mathcal{O}{\lrb\alpha\rrb}$ corrections
to the three-jet rate are about $0.5\%$ for $y_\mathrm{cut}\gtrsim
0.002$. Because we use $y_\mathrm{cut}=0.002$ in the event selection,
$q\bar q\gamma$ final states contribute only if
$y_\mathrm{cut}\lesssim 0.002$. The LO $q\bar q\gamma$ contribution
amounts to $1\%$. For $y_\mathrm{cut}< 0.002$, the full
$\mathcal{O}{\lrb\alpha\rrb}$ corrections become negative, reaching
$-1.5\%$ in the first bin. For $y_\mathrm{cut}\lesssim 0.0005$ the
three-jet rate becomes larger than $\sigma_{\mathrm{had}}$. This
behaviour indicates the breakdown of the perturbative expansion in
$\alphas$ due to large logarithmic corrections proportional to
$\log(y_\mathrm{cut})$ at all orders. Inclusion of higher-order QCD
corrections, which are large and negative
\cite{our3j,weinzierl3j}
in this region, yields a ratio of $\sigma_{\mbox{\scriptsize 3-jet}}$ to
$\sigma_{\mathrm{had}}$ less than unity for an extended range of
$\log(y_\mathrm{cut})$.  At the higher CM energies, the
corrections to the three-jet rate are larger than at $\sqrt{s}=M_\PZ$.
For $y_\mathrm{cut}\gtrsim 0.002$, they are almost constant and amount
to about 4\% at $172\GeV$, and about 2\% at $206\GeV$ and $500\GeV$.
In the region $y_\mathrm{cut}< 0.002$, we find a negative contribution
of up to $-5\%$ for very small values of $y_\mathrm{cut}$.

\subsection{Impact of the event-selection cuts on the event-shape distributions}
\label{se:results_paradep}
In the above results, we could clearly observe that the electroweak
corrections to event-shape distributions are not flat, but display
peak structures. These structures are most pronounced at $\sqrt{s} =
172\GeV$. They are discussed here for thrust as an example.  In
\reffig{fig:distri_172_1}, we see that the relative corrections show a
peaking structure for $1-T\approx 0.2$. To understand the origin of
these structures, we extensively studied how variations of the
event-selection cuts, especially the hard-photon cut, influence the
event-shape distributions.

We employ three different cuts in our calculation which
depend on four parameters:
\begin{itemize}
\item[1)] A cut on the production angle $\theta_i$ of all particles,
  such that only particles $i$ with
  $\cos\theta_i>\cos\theta_\mathrm{cut}$ are used in the
  reconstruction of the event-shape variables.  By default, we use the
  value $\cos\theta_\mathrm{cut}=0.965$.
\item[2)] A cut on the visible energy squared $s'$ of the final state,
  such that only events with $s'>s_{\mathrm{cut}}$ are accepted. By
  default, we use the value $s_{\mathrm{cut}}=0.81s$.
\item[3)] The maximum photon energy in a jet $z$, for which we require
  $z<z_{\mathrm{cut}}$, where the particles are clustered according to
  the Durham jet algorithm with parameter $y_{\mathrm{cut}}$.  By
  default, we use the values $z_\mathrm{cut}=0.9$ and
  $y_{\mathrm{cut}}=0.002$.
\end{itemize}

\begin{figure}[t]
\begin{center}
\epsfig{file=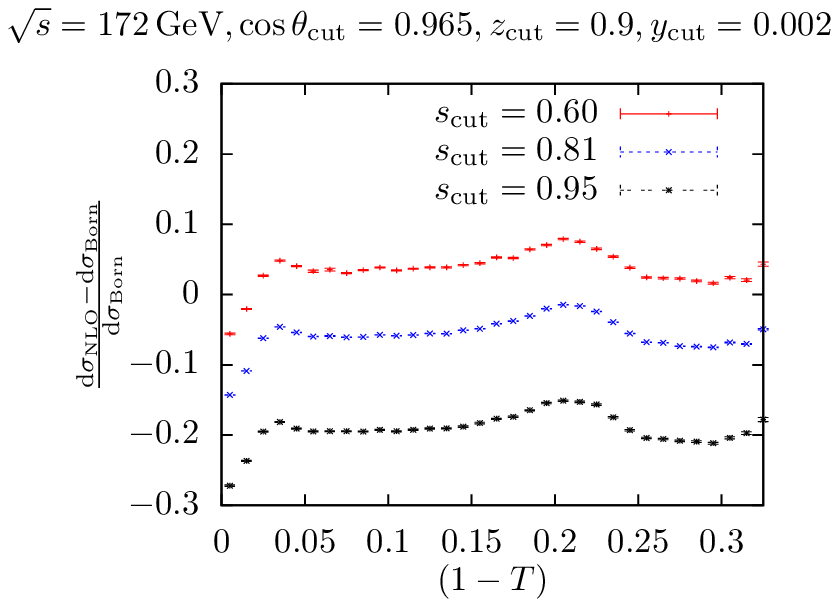,width=7.2cm}\quad
\epsfig{file=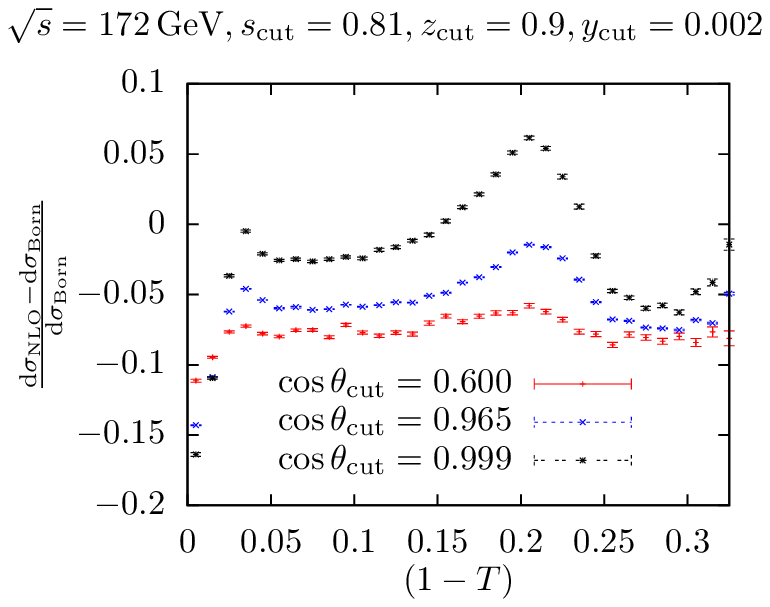,width=7.2cm}\\[2mm]
\epsfig{file=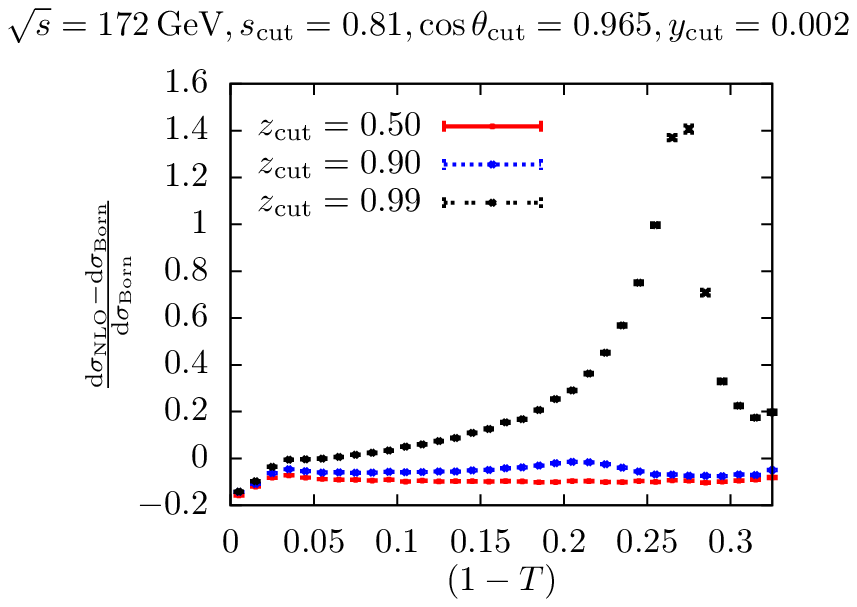,width=7.2cm}\quad
\epsfig{file=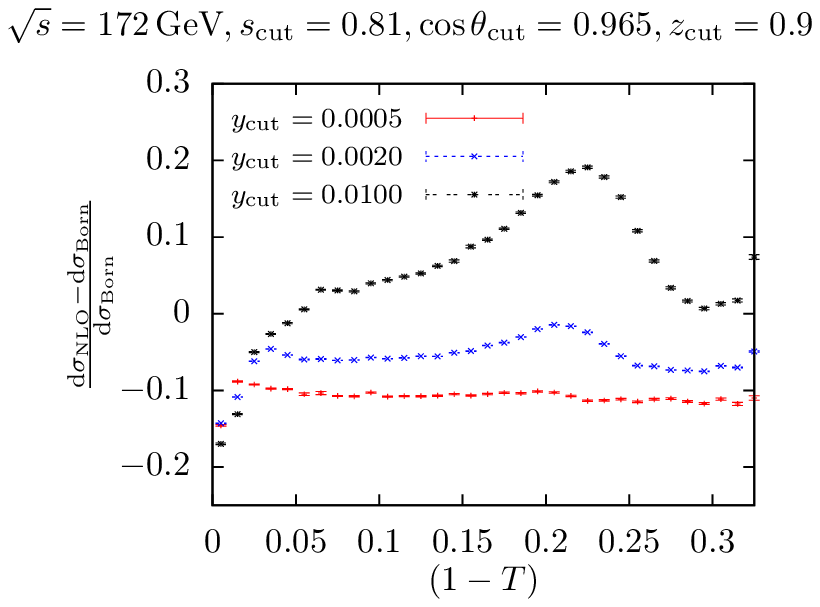,width=7.2cm}
\end{center}
\vspace*{-2em}
\caption{Dependence of the thrust distribution on different values of the phase-space cuts at $\sqrt{s}=172\GeV$.}
\label{fig:para_172}
\end{figure}
In \reffig{fig:para_172} we show the full
$\mathcal{O}{\lrb\alpha\rrb}$ corrections to the thrust distribution
normalised to the Born contribution for $\sqrt{s}=172\GeV$, where the
peak structures are most striking.  We plot the results for three
different values of a single cut parameter while we set the other
three cut parameters to their default value.  Going from left to right
and top to bottom, we vary $s_{\mathrm{cut}}$,
$\cos\theta_\mathrm{cut}$, $z_{\mathrm{cut}}$, and $y_{\mathrm{cut}}$.
By changing $s_{\mathrm{cut}}$ we observe a change in normalisation of
about $25\%$, while the shape of the distribution stays the same.
Varying $\cos\theta_\mathrm{cut}$ from larger to smaller values leads
to a more and more pronounced peak at $(1-T)\simeq0.2$. The
corrections grow below the peak but are only slightly changed above.
The different slopes for $1-T<0.03$ result from the changing
acceptance of ISR photons. Modifying $z_{\mathrm{cut}}$ has a dramatic
effect on the peak. For $z_{\mathrm{cut}}=0.99$ we find a very
pronounced resonance for $(1-T)\simeq0.28$. By reducing the cut, the
resonance gets strongly suppressed and moves to smaller values of
$(1-T)$.  By increasing $y_{\mathrm{cut}}$, we observe an enhancement
of the peak, as well as a slight shift towards larger values of $1-T$.

\begin{figure}[t]
\begin{center}
\epsfig{file=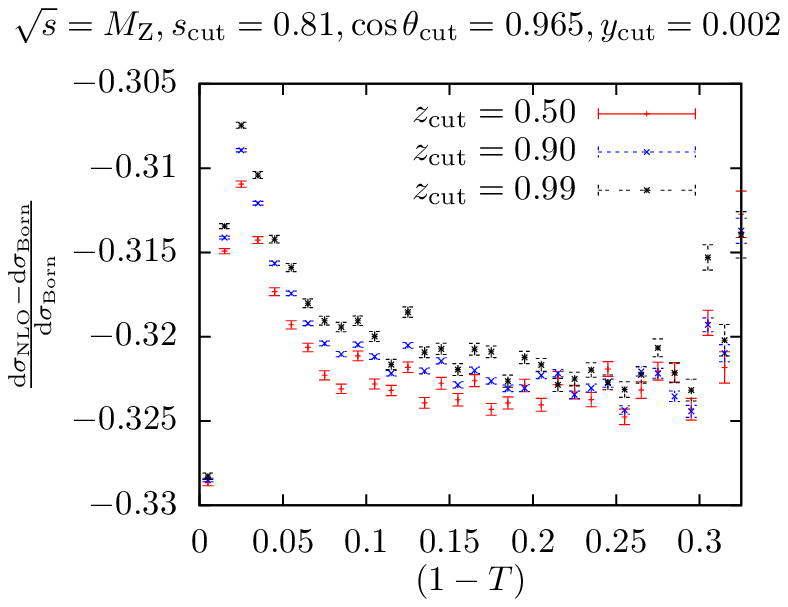,width=7.2cm}\quad
\epsfig{file=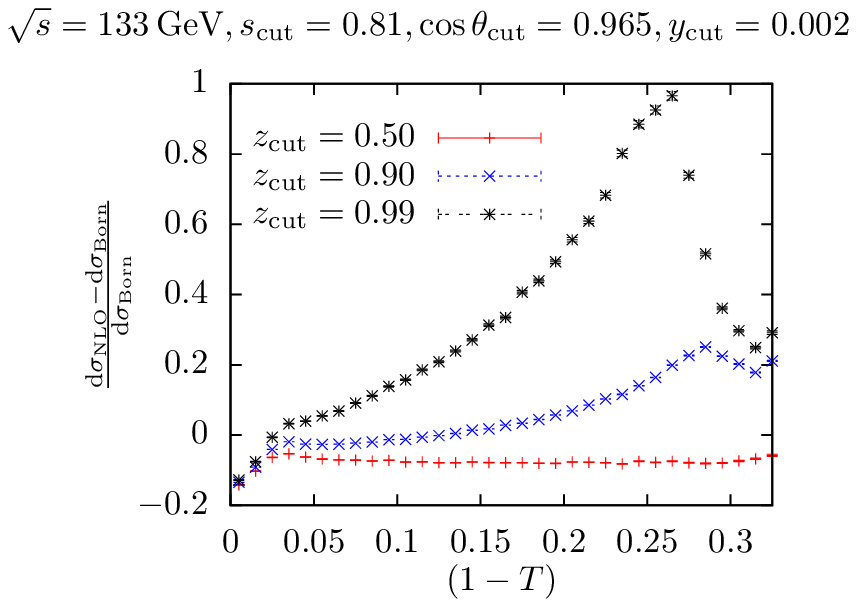,width=7.2cm}\\[2mm]
\epsfig{file=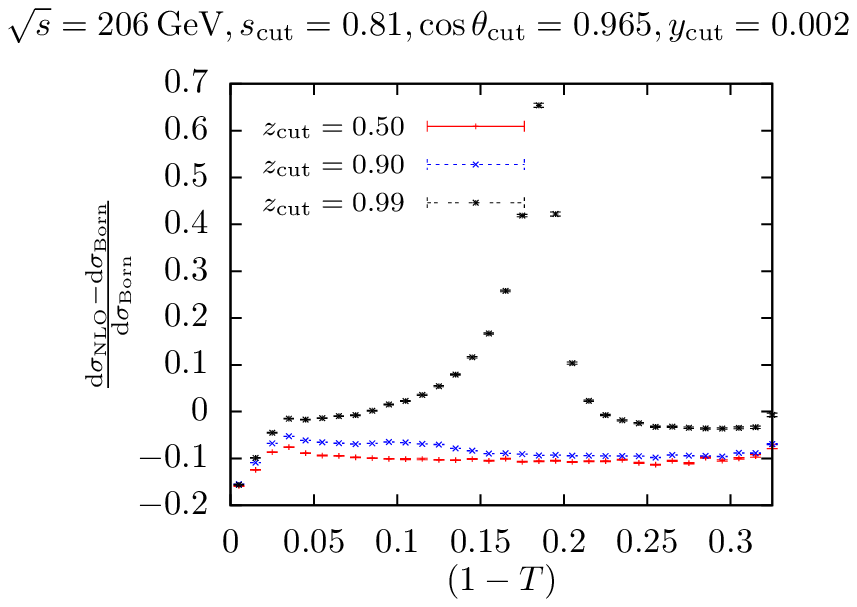,width=7.2cm}\quad
\epsfig{file=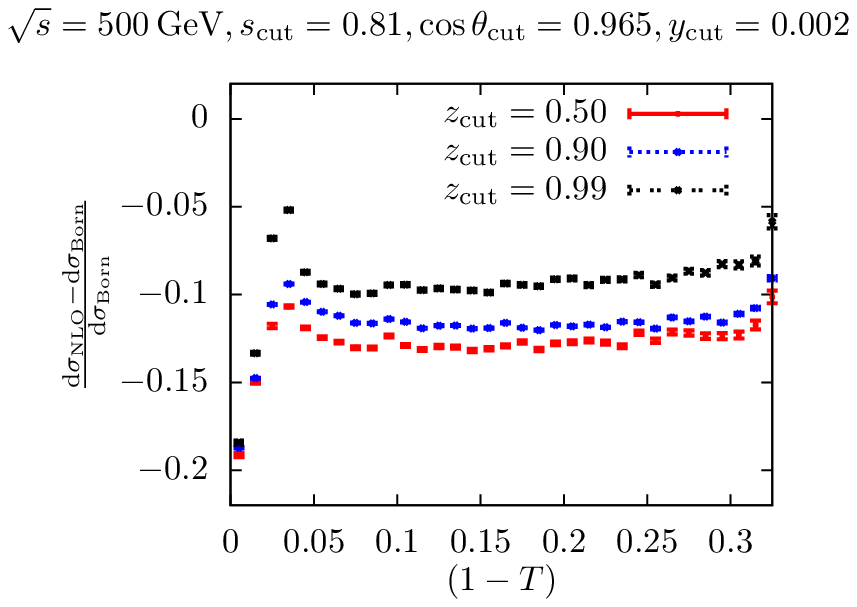,width=7.2cm}
\end{center}
\vspace*{-1em}
\caption{Dependence of the thrust distribution on different values of
  the cut $z_{\mathrm{cut}}$ at $\sqrt{s}=M_\PZ$, $133\GeV$,
  $206\GeV$, and $500\GeV$.}
\label{fig:para_z}
\end{figure}

In \reffig{fig:para_z} we study the change of the 
$\mathcal{O}{\lrb\alpha\rrb}$ corrections to the thrust distribution
normalised to the Born contribution with $z_{\mathrm{cut}}$
for $\sqrt{s}=\MZ$, $133\GeV$, $206\GeV$, and $500\GeV$.%
Varying $z_{\mathrm{cut}}$ for $\sqrt{s}=M_\PZ$ leaves the
distribution basically unchanged.  By increasing $z_{\mathrm{cut}}$
from 0.5 to 0.99 for $\sqrt{s}=133\GeV$ we find a growth of the peak
to almost $100\%$. We also observe that the peak moves from
$1-T\approx0.28$ for $z_{\mathrm{cut}}=0.9$ to $1-T\approx0.25$ for
$z_{\mathrm{cut}}=0.99$.  For $\sqrt{s}=206\GeV$ we see the same
features as for $\sqrt{s}=172\GeV$, but now with the peak at
$(1-T)\simeq0.19$ for large $z_{\mathrm{cut}}$. Finally, for
$\sqrt{s}=500\GeV$ a peaking structure seems to appear at
$(1-T)\simeq0.03$, while varying $z_{\mathrm{cut}}$
basically changes only the normalisation.

Through analysing events at the level of the Monte Carlo
generator, we find that the enhancement in the region of the peaking
structure always stems from $q\bar q\mathrm{g}\gamma$ final states,
where a soft gluon is clustered with a hard photon. Increasing
$\cos\theta_{\mathrm{cut}}$ leads to a logarithmic enhancement of
collinear ISR photons, increasing $z_{\mathrm{cut}}$ generally results
in a larger acceptance of photons inside jets, and increasing
$y_{\mathrm{cut}}$ causes more photons to be clustered together with
other partons, resulting in less events with isolated photons being
removed. We can therefore conclude that the peaking structure results
from the ISR photon contribution, where a soft gluon is clustered
together with the photon.

More precisely, the peaking structure can be explained by the
radiative-return phenomenon. Since we do not remove all energetic
photons, it is possible that a hard photon and a soft gluon are
clustered together, such that the energy fraction of the photon in the
jet does not exceed $z_\mathrm{cut}$ and the invariant mass of the
quark--antiquark--gluon system $s_{q{\bar q}g}$ is equal to the mass
of the $\PZ$ boson.  Such a configuration leads on the one hand to an
enhancement due to radiative return but also to a logarithmic
enhancement due to the soft gluon.

In order to analyse this effect further, we consider events where the
photon and the soft gluon are clustered together, such that we have a
three-particle final state that consists of a quark, an antiquark, and
a photonic jet. Assume that the quark, antiquark, and the photonic jet
have the three-momenta $\vec{p}_q,\vec{p}_{\bar q},\vec{p}_\gamma$ and
the energies $E_q,E_{\bar q},E_\gamma$, respectively. We use
energy-momentum conservation
\bea
\vec{p}_q+\vec{p}_{\bar q}+\vec{p}_\gamma&=&0,\nn\\
E_q+E_{{\bar q}}+E_\gamma&=&\sqrt{s}.
\label{res_mom_cons}
\eea
and demand that the invariant mass of the quark--antiquark pair is
equal to $\MZ$, such that
\be
s_{q{\bar q}}=\lrb E_q+E_{\bar q}\rrb^2-\lrb\vec{p}_q+\vec{p}_{\bar q}\rrb^2=
\lrb E_q+E_{\bar q}\rrb^2-\lrb\vec{p}_\gamma\rrb^2=\lrb E_q+E_{\bar q}\rrb^2-E_\gamma^2=\MZ^2.
\label{res_inv_mass_qq}
\ee
Energy conservation \refeq{res_mom_cons} and the mass-shell condition
\refeq{res_inv_mass_qq} imply
\be
E_q+E_{\bar q}=\frac{s+\MZ^2}{2\sqrt{s}},\qquad E_\gamma=\frac{s-\MZ^2}{2\sqrt{s}}.
\ee
It can be shown that in a three-jet configuration with massless
partons, thrust is always determined by the energy of the most
energetic particle $E_{\mathrm{max}}$
\cite{Dissertori:2003pj}, i.e.~
\be
T=\frac{2 E_{\mathrm{max}}}{\sqrt{s}}.
\ee
If we now assume that the quark and the antiquark carry the same
energy, we can calculate the energies of all three jets in the final
state at different energies:
\bea
E_q\lrb 133\GeV \rrb= E_{\bar{q}}\lrb 133\GeV \rrb\approx ~49\GeV,\quad E_\gamma\lrb 133\GeV \rrb&\!\approx\! &~35\GeV,\nn\\
E_q\lrb 172\GeV \rrb= E_{\bar{q}}\lrb 172\GeV \rrb\approx ~55\GeV,\quad E_\gamma\lrb 172\GeV \rrb&\!\approx\! &~62\GeV,\nn\\
E_q\lrb 206\GeV \rrb= E_{\bar{q}}\lrb 206\GeV \rrb\approx ~61\GeV,\quad E_\gamma\lrb 206\GeV \rrb&\!\approx\! &~84\GeV,\nn\\
E_q\lrb 500\GeV \rrb= E_{\bar{q}}\lrb 500\GeV \rrb\approx 129\GeV,\quad E_\gamma\lrb 500\GeV \rrb&\!\approx\! &242\GeV.
\eea
This leads to the following thrust values where the radiative-return
phenomena should appear:
\bea
(1-T_{\mathrm{RR}})(\sqrt{s}=133\GeV)\approx 0.27,\nn\\
(1-T_{\mathrm{RR}})(\sqrt{s}=172\GeV)\approx 0.28,\nn\\
(1-T_{\mathrm{RR}})(\sqrt{s}=206\GeV)\approx 0.19,\nn\\
(1-T_{\mathrm{RR}})(\sqrt{s}=500\GeV)\approx 0.03.
\eea
These values coincide perfectly with the peaks in
\reffigs{fig:para_172} and \ref{fig:para_z}.  Relaxing the assumption
that the quark and the antiquark carry the same energy only results in
a broadening of the peaking structure. Varying the value of
$z_{\mathrm{cut}}$ leads to different energies of the photonic jet and
therefore changes the allowed energies in the above analysis.  For
decreasing values of $z_{\mathrm{cut}}$, we therefore either only
observe the tail of the peak or cut it away completely which
effectively looks like a shift of the position or the disappearance of
the peak.  For $\sqrt{s}=133\GeV$ we observe the tail of the peak for
$E_q<E_{q,\mathrm{peak}}$ such that for decreasing $z_{\mathrm{cut}}$
the peak seems to move to larger values of $(1-T)$. For
$\sqrt{s}=172\GeV$ and $\sqrt{s}=206\GeV$ we observe the tail of the
peak for $E_\gamma>E_{\gamma,\mathrm{peak}}$ such that for decreasing
$z_{\mathrm{cut}}$ the peak seems to move to smaller values of
$(1-T)$.

The study in this section clearly illustrates the non-trivial effect
of realistic photon isolation criteria to the electroweak corrections
to jet observables.  The accidental clustering of a soft gluon with a
hard photon results in a photon jet with a photon energy fraction
below the rejection cut. In these events, the distribution of the
final-state jets, and their reconstructed pair invariant masses do no
longer reflect the underlying parton dynamics.  A similar
misreconstruction could also happen for electroweak corrections to
final states involving jets at hadron colliders, and clearly deserves
further study.

\begin{figure}[t]
\begin{center}
\epsfig{file=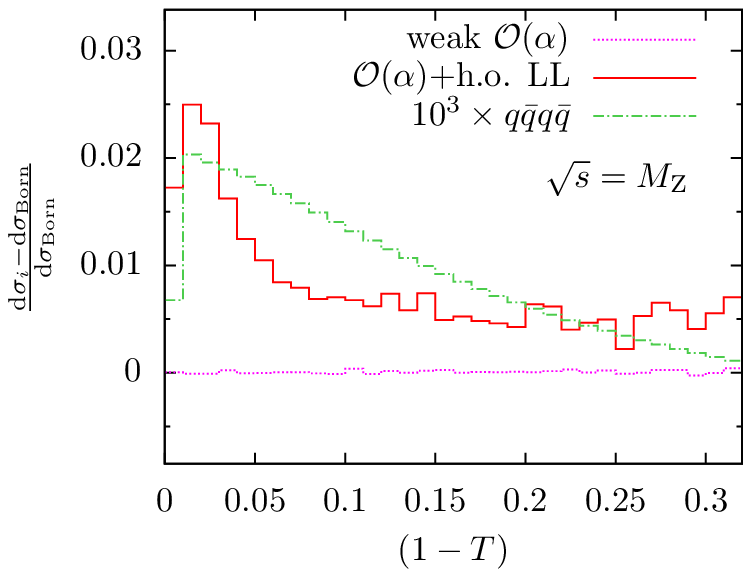,width=6.5cm}\hspace{1cm}
\epsfig{file=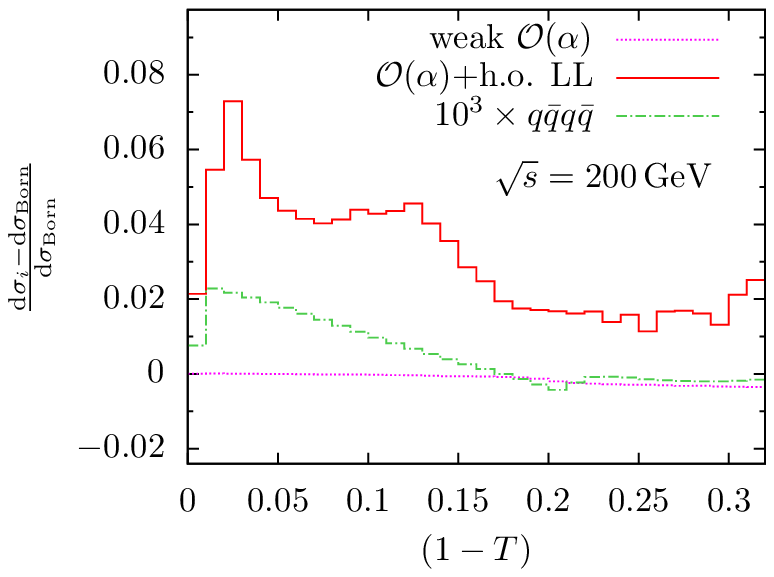,width=6.5cm}\\[2mm]
\epsfig{file=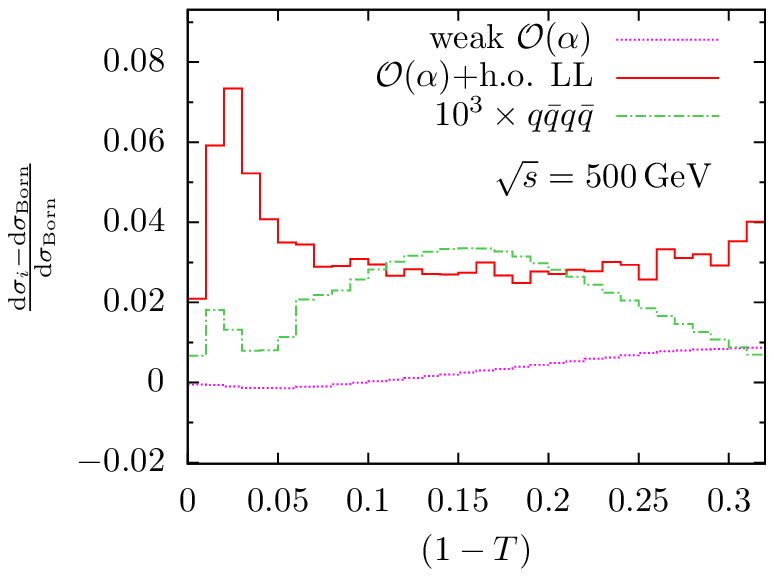,width=6.5cm}\hspace{1cm}
\epsfig{file=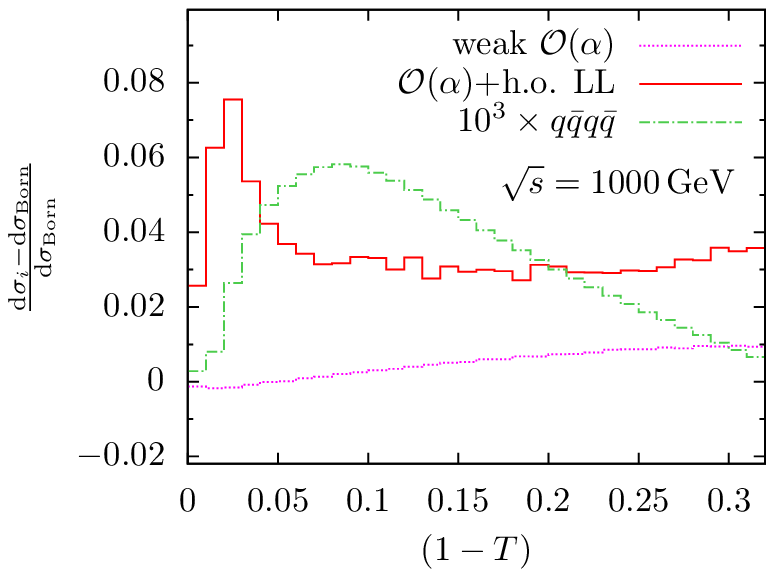,width=6.5cm}
\end{center}
\vspace*{-1em}
\caption{Electroweak corrections to the thrust distribution at
 different CM energies. The interference contribution between
 electroweak and QCD diagrams for the four-quark final state is
 scaled by a factor 1000.}
\label{fig:fourq}
\end{figure}
\subsection{Contribution from four-quark final states}
At ${\cal O}(\alpha^3\alphas)$, event-shape distributions and jet
cross sections receive a contribution from the process $\Pe^+\Pe^- \to
q\bar q q\bar q$ through the interference of QCD and electroweak
amplitudes [see (\ref{process-4q})]. Compared to other contributions at
this order, this four-quark interference contribution is very small.
Its typical magnitude amounts to about one per mille of the
electroweak correction, and is thus within the integration error of
the four-particle contribution. To illustrate the smallness of this
contribution, \reffig{fig:fourq} compares the four-quark
contribution (scaled by a factor 1000) to the total electroweak
correction to the normalised thrust distribution at different
CM energies. The relative magnitude of the four-quark
contribution is always at the per-mille level.

\section{Conclusions and outlook}
\label{sec:conc}
\setcounter{equation}{0}

In this paper, we have derived the NLO electroweak corrections to three-jet 
production and event-shape distributions in $\Pe^+\Pe^-$ annihilation. At this 
order, contributions arise from virtual corrections from weak gauge bosons
(which were evaluated in the complex-mass scheme to take proper account of 
the gauge-boson widths), from fermionic loops, from real and virtual photonic 
corrections and from interferences between electroweak and QCD diagrams for
four-quark final states.

Our calculation is one of the first to address electroweak corrections
to jet production observables. For this type of observables, one has
to take proper account of the experimental event-selection cuts,
which are aiming to reject events containing isolated photons. An
infrared-safe definition of isolated photons must permit some amount
of hadronic energy around the photon; for jet observables, this can be
realized by a democratic clustering procedure, used by the LEP
experiments.  In this approach, the photon is clustered like any other
hadron by the jet algorithm, resulting in a photon jet. If the photon
carries more than a predefined large fraction of the jet energy, it is
called isolated and the event is discarded.  In our calculation, we
have implemented this isolated-photon-veto procedure.  Since it
involves cuts on a specific, identified particle in the final state,
the resulting observable is no longer collinear-safe, and the
left-over collinear singularity is compensated by a further
contribution from the quark-to-photon fragmentation function.  We have
documented this part of the calculation in detail in different
schemes.

The NLO electroweak corrections to absolute cross sections and
event-shape distributions turn out to be numerically substantial: for
example for thrust at
$\sqrt{s}=\MZ$, they amount to a correction of $-32\%$, which is
largely dominated by initial-state radiation.  Beyond the NLO
electroweak corrections, we also included higher-order leading
logarithmic corrections, which are sizable. Their inclusion at
$\sqrt{s}=\MZ$ results in a total correction of $-28\%.$ Normalizing
these results to the total hadronic cross section (as is done in the
experimental measurement), corrected to the same order, only very
moderate corrections remain in the normalized jet cross sections and
event distributions, and practically no difference is observed between
the fixed-order NLO electroweak results and the results including
higher-order logarithmic corrections.

At LEP1, we find that NLO electroweak effects to event-shape
distributions amount to a correction of about one to two per cent. The
corrections are not uniform over the kinematical range, but tend to
increase towards the two-jet region, where the isolated-photon veto
becomes less efficient. The corrections to the three-jet rate are
below one per cent. Purely weak contributions form a gauge-invariant
subset of the electroweak corrections. At LEP1, these corrections are
below the per-mille level.

At LEP2 energies, the NLO electroweak corrections to event-shape
distributions and to the three-jet rate are typically at the level of two
to eight per cent. 
The largest contribution comes again from the photonic
corrections.  These are
very sensitive to the precise photon isolation cuts applied to select 
the events, and are not uniform over the range of event-shape variables.
The NLO electroweak event-shape distributions 
display peaks at LEP2 energies. These peaks are due to 
a remant of the radiative-return phenomenon, which is not fully suppressed 
by the photon isolation cuts. The position and energy dependence of these
peaks can be explained quantitatively. 

Event-shape and jet cross section data from LEP1 and LEP2 have been
corrected for photonic radiation effects using multi-purpose
leading-logarithmic event generator programs. To compare our results
with the experimental data would first require to undo these
corrections. A further complication in the comparison with data arises
from the fact that event-shape distributions at LEP2 were determined
in the $\Pe^+\Pe^-$ centre-of-mass frame for all events, including
initial-state-radiation events. Most of the event-shape variables are
not boost invariant, and should thus be reconstructed in the
centre-of-mass frame of the observed hadrons.  If reconstructed in the
$\Pe^+\Pe^-$ centre-of-mass frame, ideal two-jet events with
initial-state radiation will not be placed at the kinematical edge of
the distribution, thus violating infrared-safety criteria. Within a
perturbative calculation, it is not possible to apply the event
reconstruction in the $\Pe^+\Pe^-$ centre-of-mass frame. The purely
weak corrections at LEP1 and LEP2 were previously not accounted for in
the interpretation of event-shape and jet cross section data. Our
study shows that they are at the level of one per mille or below for
appropriately normalized distributions. At the current level of
experimental and theoretical precision, they are thus not yet relevant
to precision QCD studies of LEP data.

While the magnitude of electroweak corrections decreases towards
higher LEP2 energies, we observe them to increase again when going to
even higher energies, corresponding to a future linear collider. In
part, this increase comes from the fact that purely weak corrections
(which were negligible throughout the LEP2 energy range) become
sizable at high energies. At $\sqrt{s}=500$~GeV, NLO electroweak
corrections to event-shape distributions and jet cross sections amount
to two to four per cent, and have thus a potentially sizable impact on
precision QCD studies at a future linear collider. The purely weak
corrections reach up to one per cent at this energy. Most importantly,
our findings on the interplay of photon isolation and event-selection
cuts, and on the appropriate frame for the reconstruction of
initial-state radiation events will help to optimize precision QCD
studies at future high-energy $\Pe^+\Pe^-$ colliders from event shapes
and jet cross sections.

\section*{Acknowledgements}
This work was supported in part by the Swiss National Science
Foundation (SNF) under contracts 200020-116756, 200020-124773 and
200020-126691 and by the European Community's Marie-Curie Research
Training Network HEPTOOLS under contract MRTN-CT-2006-035505.


\begin{thebibliography}{100}

\bibitem{qcd}
D.J.\ Gross and F.\ Wilczek, Phys.\ Rev.\ D \textbf{8} (1973) 3633;\\
H.D.\ Politzer, Phys.\ Rept.\ \textbf{14} (1974) 129;\\
H.~Fritzsch, M.~Gell-Mann and H.~Leutwyler,
  Phys.\ Lett.\  B {\bf 47} (1973) 365.

\bibitem{tasso}
  R.~Brandelik {\it et al.}  [TASSO Collaboration],
  Phys.\ Lett.\  B {\bf 86} (1979) 243;\\
P.~S\"oding, B.~Wiik, G.~Wolf and S.L.~Wu, Talks given at Award Ceremony 
of the 1995 EPS High Energy and Particle Physics Prize, Proceedings of 
the {\it EPS High Energy Physics Conference}, Brussels, 1995, (World
Scientific), p.~3.

\bibitem{ellis}
J.~Ellis, M.K.~Gaillard and G.G.~Ross, Nucl.~Phys.~{\bf B111} (1976) 253;
{\bf B130} (1977) 516(E).


\bibitem{reviews}
O.~Biebel,
  Phys.\ Rept.\  {\bf 340} (2001) 165;\\
S.\  Kluth,
  {Rept.\ Prog.\ Phys.\ }  {\bf 69} (2006) 1771;\\
S.~Bethke,
  Prog.\ Part.\ Nucl.\ Phys.\  {\bf 58} (2007) 351.

\bibitem{alephqcd}
  D.~Buskulic {\it et al.}  [ALEPH Collaboration],
  Z.\ Phys.\  C {\bf 73} (1997) 409;\\
  A.~Heister {\it et al.}  [ALEPH Collaboration],
  Eur.\ Phys.\ J.\  C {\bf 35} (2004) 457.

\bibitem{opal}
  P.~D.~Acton {\it et al.}  [OPAL Collaboration],
  Z.\ Phys.\  C {\bf 59} (1993) 1;\\
  G.~Alexander {\it et al.}  [OPAL Collaboration],
  Z.\ Phys.\  C {\bf 72} (1996) 191;\\
  K.~Ackerstaff {\it et al.}  [OPAL Collaboration],
  Z.\ Phys.\  C {\bf 75} (1997) 193;\\
  G.~Abbiendi {\it et al.}  [OPAL Collaboration],
  Eur.\ Phys.\ J.\  C {\bf 16} (2000) 185
  [hep-ex/0002012];\\
  G.~Abbiendi {\it et al.}  [OPAL Collaboration],
  Eur.\ Phys.\ J.\  C {\bf 40} (2005) 287
  [hep-ex/0503051].

\bibitem{l3}
  M.~Acciarri {\it et al.}  [L3 Collaboration],
  Phys.\ Lett.\  B {\bf 371} (1996) 137;\\
  M.~Acciarri {\it et al.}  [L3 Collaboration],
  Phys.\ Lett.\  B {\bf 404} (1997) 390;\\
  M.~Acciarri {\it et al.}  [L3 Collaboration],
  Phys.\ Lett.\  B {\bf 444} (1998) 569;\\
  P.~Achard {\it et al.}  [L3 Collaboration],
  Phys.\ Lett.\  B {\bf 536} (2002) 217
  [hep-ex/0206052];\\
  P.~Achard {\it et al.}  [L3 Collaboration],
  Phys.\ Rept.\  {\bf 399} (2004) 71
  [hep-ex/0406049].


\bibitem{delphi}
  P.~Abreu {\it et al.}  [DELPHI Collaboration],
  Phys.\ Lett.\  B {\bf 456} (1999) 322;\\
  J.~Abdallah {\it et al.}  [DELPHI Collaboration],
  Eur.\ Phys.\ J.\  C {\bf 29} (2003) 285
  [hep-ex/0307048];\\
  J.~Abdallah {\it et al.}  [DELPHI Collaboration],
  Eur.\ Phys.\ J.\  C {\bf 37} (2004) 1
  [hep-ex/0406011].

\bibitem{sld}
  K.~Abe {\it et al.}  [SLD Collaboration],
  Phys.\ Rev.\  D {\bf 51} (1995) 962
  [hep-ex/9501003].

\bibitem{jade}
P.~A.~Movilla Fernandez, O.~Biebel, S.~Bethke, S.~Kluth and P.~Pfeifenschneider
                  [JADE Collaboration],
  Eur.\ Phys.\ J.\  C {\bf 1} (1998) 461
  [hep-ex/9708034];\\
 P.~Pfeifenschneider {\it et al.}  [JADE collaboration and OPAL
                  Collaboration],
  Eur.\ Phys.\ J.\  C {\bf 17} (2000) 19
  [hep-ex/0001055].



\bibitem{ERT}
R.K.~Ellis, D.A.~Ross and A.E.~Terrano,
Nucl.\ Phys.\ B {\bf 178} (1981) 421.

\bibitem{kunszt}
 Z.~Kunszt,
  Phys.\ Lett.\  B {\bf 99} (1981) 429;
J.A.M.~Vermaseren, K.J.F.~Gaemers and S.J.~Oldham,
  Nucl.\ Phys.\  B {\bf 187} (1981) 301;\\
K.~Fabricius, I.~Schmitt, G.~Kramer and G.~Schierholz, Z.~Phys.~C {\bf
  11} (1981) 315.


\bibitem{event}
Z.\ Kunszt and P.\ Nason, in {\it Z Physics at LEP 1}, CERN Yellow Report
89-08, Vol.~1, p.~373;\\
  W.~T.~Giele and E.W.N.~Glover,
  Phys.\ Rev.\  D {\bf 46} (1992) 1980;\\
 S.~Catani and M.~H.~Seymour,
  Phys.\ Lett.\  B {\bf 378} (1996) 287.

\bibitem{nlla}
  S.~Catani, L.~Trentadue, G.~Turnock and B.R.~Webber,
  Nucl.\ Phys.\  B {\bf 407} (1993) 3;\\
  A.~Banfi, G.P.~Salam and G.~Zanderighi,
  JHEP {\bf 0201} (2002) 018
  [hep-ph/0112156];\\
R.W.L.~Jones, M.~Ford, G.P.~Salam, H.~Stenzel and D.~Wicke,
  JHEP {\bf 0312} (2003) 007 [hep-ph/0312016].


\bibitem{ourevent}
  A.~Gehrmann-De Ridder, T.~Gehrmann, E.W.N.~Glover and G.~Heinrich,
  Phys.\ Rev.\ Lett.\  {\bf 99} (2007) 132002
  [arXiv:0707.1285 [hep-ph]];
  JHEP {\bf 0712} (2007) 094
  [arXiv:0711.4711 [hep-ph]];
 JHEP {\bf 0905} (2009) 106  [arXiv:0903.4658 [hep-ph]].

\bibitem{weinzierlevent}
 S.~Weinzierl,
  JHEP {\bf 0906} (2009) 041
  [arXiv:0904.1077 [hep-ph]];
  Phys.\ Rev.\  D {\bf 80} (2009)  094018 
  [arXiv:0909.5056 [hep-ph]].




\bibitem{our3j}
  A.~Gehrmann-De Ridder, T.~Gehrmann, E.W.N.~Glover and G.~Heinrich,
  JHEP {\bf 0711} (2007) 058
  [arXiv:0710.0346];
  Phys.\ Rev.\ Lett.\  {\bf 100} (2008) 172001
  [arXiv:0802.0813 [hep-ph]].

\bibitem{weinzierl3j}
S.~Weinzierl,
 Phys.\ Rev.\ Lett.\  {\bf 101} (2008) 162001
  [arXiv:0807.3241 [hep-ph]];
  JHEP {\bf 0907} (2009) 009
  [arXiv:0904.1145 [hep-ph]].




\bibitem{gionata}
 T.~Gehrmann, G.~Luisoni and H.~Stenzel,
  Phys.\ Lett.\  B {\bf 664} (2008) 265 [arXiv:0803.0695 [hep-ph]].


\bibitem{asevent}
G.~Dissertori, {\it et al.},
  JHEP {\bf 0802} (2008) 040
  [arXiv:0712.0327 [hep-ph]];
  JHEP {\bf 0908} (2009) 036
  [arXiv:0906.3436 [hep-ph]];\\
R.A.~Davison and B.R.~Webber,
  Eur.\ Phys.\ J.\  C {\bf 59} (2009) 13
  [arXiv:0809.3326 [hep-ph]];\\
S.~Bethke, S.~Kluth, C.~Pahl and J.~Schieck  [JADE Collaboration],
  Eur.\ Phys.\ J.\  C {\bf 64} (2009) 351
  [arXiv:0810.1389 [hep-ph]];\\
 T.~Gehrmann, M.~Jaquier and G.~Luisoni,
  arXiv:0911.2422 [hep-ph].


\bibitem{asjets}
G.~Dissertori, {\it et al.},
 Phys.\ Rev.\ Lett.\  {\bf 104} (2010) 072002
[arXiv:0910.4283 [hep-ph]].

\bibitem{becherschwartz}
T.~Becher and M.~D. Schwartz,
\newblock JHEP {\bf 07} (2008)  034 [arXiv:0803.0342 [hep-ph]].


\bibitem{CarloniCalame:2008qn}
  C.~M.~Carloni-Calame, S.~Moretti, F.~Piccinini and D.~A.~Ross,
  JHEP {\bf 0903} (2009) 047
  [arXiv:0804.3771 [hep-ph]];
  Eur.\ Phys.\ J.\  C {\bf 62} (2009) 355
  [Erratum-ibid.\  C {\bf 62} (2009) 453]
  [arXiv:0811.4758 [hep-ph]];
  arXiv:0903.0490 [hep-ph].

\bibitem{Denner:2009gx}
  A.~Denner, S.~Dittmaier, T.~Gehrmann and C.~Kurz,
  Phys.\ Lett.\  B {\bf 679} (2009) 219
  [arXiv:0906.0372 [hep-ph]].

\bibitem{Koller:1978kq}
K.~Koller, T.~F. Walsh, and P.~M. Zerwas,
\newblock Zeit. Phys. {\bf C2} (1979) 197.


\bibitem{Bethke:1988zc}
JADE, S.~Bethke {\em et~al.},
\newblock Phys. Lett. {\bf B213} (1988) 235.


\bibitem{Dissertori:2003pj}
G.~Dissertori, I.~G. Knowles, and M.~Schmelling,
\newblock Oxford, UK: Clarendon (2003) 538 p.



\bibitem{Brown:1990nm}
N.~Brown and W.~J. Stirling,
\newblock Phys. Lett. {\bf B252} (1990) 657.

\bibitem{Catani:1991hj}
S.~Catani, Y.~L. Dokshitzer, M.~Olsson, G.~Turnock, and B.~R. Webber,
\newblock Phys. Lett. {\bf B269} (1991) 432.

\bibitem{Brandt:1964sa}
S.~Brandt, C.~Peyrou, R.~Sosnowski, and A.~Wroblewski,
\newblock Phys. Lett. {\bf 12} (1964) 57 .

\bibitem{Farhi:1977sg}
E.~Farhi,
\newblock Phys. Rev. Lett. {\bf 39} (1977) 1587.

\bibitem{Clavelli:1981yh}
L.~Clavelli and D.~Wyler,
\newblock Phys. Lett. {\bf B103} (1981) 383.

\bibitem{Rakow:1981qn}
P.~E.~L. Rakow and B.~R. Webber,
\newblock Nucl. Phys. {\bf B191} (1981) 63.

\bibitem{Catani:1992jc}
S.~Catani, G.~Turnock, and B.~R. Webber,
\newblock Phys. Lett. {\bf B295} (1992) 269.

\bibitem{Parisi:1978eg}
G.~Parisi,
\newblock Phys. Lett. {\bf B74} (1978) 65.

\bibitem{Donoghue:1979vi}
J.~F. Donoghue, F.~E. Low, and S.-Y. Pi,
\newblock Phys. Rev. {\bf D20} (1979) 2759.

\bibitem{Brown:1991hx}
N.~Brown and W.~J. Stirling,
\newblock Z. Phys. {\bf C53} (1992) 629.


\bibitem{Bethke:1991wk}
S.~Bethke, Z.~Kunszt, D.~E. Soper, and W.~J. Stirling,
\newblock Nucl. Phys. {\bf B370} (1992) 310.

\bibitem{Catani:1991kz}
S.~Catani, G.~Turnock, B.~R. Webber, and L.~Trentadue,
\newblock Phys. Lett. {\bf B263} (1991) 491.

\bibitem{Korchemsky:1994is}
G.~P. Korchemsky and G.~Sterman,
\newblock Nucl. Phys. {\bf B437}5 (1995) 415  [hep-ph/9411211].

\bibitem{Dokshitzer:1995zt}
Y.~L. Dokshitzer and B.~R. Webber,
\newblock Phys. Lett. {\bf B352} (1995) 451 [hep-ph/9504219].

\bibitem{Dokshitzer:1997ew}
Y.~L. Dokshitzer and B.~R. Webber,
\newblock Phys. Lett. {\bf B404} (1997) 321 [hep-ph/9704298].

\bibitem{Dokshitzer:1998pt}
Y.~L. Dokshitzer, A.~Lucenti, G.~Marchesini, and G.~P. Salam,
\newblock JHEP {\bf 05} (1998) 003 [hep-ph/9802381].

\bibitem{Anastasiou:2004qd}
C.~Anastasiou, K.~Melnikov, and F.~Petriello,
\newblock Phys. Rev. Lett. {\bf 93} (2004) 032002 [hep-ph/0402280].

\bibitem{GehrmannDeRidder:2004tv}
A.~Gehrmann-De~Ridder, T.~Gehrmann, and E.~W.~N. Glover,
\newblock Nucl. Phys. {\bf B691} (2004) 195 [hep-ph/0403057].

\bibitem{Weinzierl:2006ij}
S.~Weinzierl,
\newblock Phys. Rev. {\bf D74} (2006) 014020 [hep-ph/0606008].

\bibitem{Signer:1996bf}
A.~Signer and L.~J. Dixon,
\newblock Phys. Rev. Lett. {\bf 78} (1997) 811 [hep-ph/9609460].

\bibitem{Dixon:1997th}
L.~J. Dixon and A.~Signer,
\newblock Phys. Rev. {\bf D56} (1997) 4031 [hep-ph/9706285].

\bibitem{Nagy:1997yn}
Z.~Nagy and Z.~Trocsanyi,
\newblock Phys. Rev. Lett. {\bf 79} (1997) 3604 [hep-ph/9707309].

\bibitem{Campbell:1998nn}
J.~M. Campbell, M.~A. Cullen, and E.~W.~N. Glover,
\newblock Eur. Phys. J. {\bf C9} (1999) 245 [hep-ph/9809429].

\bibitem{Weinzierl:1999yf}
S.~Weinzierl and D.~A. Kosower,
\newblock Phys. Rev. {\bf D60} (1999) 054028 [hep-ph/9901277].

\bibitem{Maina:2002wz}
E.~Maina, S.~Moretti, and D.~A. Ross,
\newblock JHEP {\bf 04} (2003) 056 [hep-ph/0210015].



\bibitem{Barate:1996fi}
ALEPH collaboration, R.~Barate {\em et~al.},
\newblock Phys. Rept. {\bf 294} (1998) 1.

\bibitem{Bohm:1986rj}
M.~B\"ohm, H.~Spiesberger, and W.~Hollik,
\newblock Fortsch. Phys. {\bf 34} (1986) 687.


\bibitem{Denner:1999gp}
A.~Denner, S.~Dittmaier, M.~Roth, and D.~Wackeroth,
\newblock Nucl. Phys. {\bf B560} (1999) 33 [hep-ph/9904472].


\bibitem{Denner:2005fg}
A.~Denner, S.~Dittmaier, M.~Roth, and L.~H. Wieders,
\newblock Nucl. Phys. {\bf B724} (2005) 247 [hep-ph/0505042].


\bibitem{Hahn:2000kx}
T.~Hahn,
\newblock Comput. Phys. Commun. {\bf 140} (2001) 418 [hep-ph/0012260].

\bibitem{Hahn:1998yk}
T.~Hahn and M.~Perez-Victoria,
\newblock Comput. Phys. Commun. {\bf 118} (1999) 153 [hep-ph/9807565].

\bibitem{Denner:1991kt}
A.~Denner,
\newblock Fortschr. Phys. {\bf 41} (1993) 307.

\bibitem{Bredenstein:2008zb}
  A.~Bredenstein, A.~Denner, S.~Dittmaier and S.~Pozzorini,
  JHEP {\bf 0808} (2008) 108
  [arXiv:0807.1248 [hep-ph]].



\bibitem{Denner:2003iy}
A.~Denner, S.~Dittmaier, M.~Roth, and M.~M. Weber,
\newblock Nucl. Phys. {\bf B660} (2003) 289 [hep-ph/0302198].

\bibitem{Denner:2003zp}
A.~Denner, S.~Dittmaier, M.~Roth, and M.~M. Weber,
\newblock Nucl. Phys. {\bf B680} (2004) 85 [hep-ph/0309274].

\bibitem{Kublbeck:1990xc}
  J.~K\"ublbeck, M.~B\"ohm and A.~Denner,
  Comput.\ Phys.\ Commun.\  {\bf 60} (1990) 165;
  H.~Eck and J.~K\"ublbeck, {\it Guide to FeynArts 1.0\/},
  University of W\"urzburg, 1992.


\bibitem{Dittmaier:1998nn}
S.~Dittmaier,
\newblock Phys. Rev. {\bf D59} (1999) 016007 [hep-ph/9805445].



\bibitem{Denner:2005es}
A.~Denner, S.~Dittmaier, M.~Roth, and L.~H. Wieders,
\newblock Phys. Lett. {\bf B612} (2005) 223 [hep-ph/0502063].



\bibitem{'tHooft:1978xw}
G.~'t~Hooft and M.~J.~G. Veltman,
\newblock Nucl. Phys. {\bf B153} (1979) 365.

\bibitem{Beenakker:1988jr}
W.~Beenakker and A.~Denner,
\newblock Nucl. Phys. {\bf B338} (1990) 349.

\bibitem{Denner:1991qq}
A.~Denner, U.~Nierste, and R.~Scharf,
\newblock Nucl. Phys. {\bf B367} (1991) 637.

\bibitem{Dittmaier:2003bc}
  S.~Dittmaier,
  Nucl.\ Phys.\  B {\bf 675} (2003) 447
  [hep-ph/0308246].

\bibitem{Denner:2002ii}
A.~Denner and S.~Dittmaier,
\newblock Nucl. Phys. {\bf B658} (2003) 175 [hep-ph/0212259].

\bibitem{Denner:2005nn}
A.~Denner and S.~Dittmaier,
\newblock Nucl. Phys. {\bf B734} (2006) 62 [hep-ph/0509141].

\bibitem{Passarino:1978jh}
G.~Passarino and M.~J.~G. Veltman,
\newblock Nucl. Phys. {\bf B160} (1979) 151.


\bibitem{Baer:1988ux}
H.~Baer, J.~Ohnemus, and J.~F. Owens,
\newblock Z. Phys. {\bf C42} (1989) 657.

\bibitem{Bohm:1993qx}
M.~B\"ohm and S.~Dittmaier,
\newblock Nucl. Phys. {\bf B409} (1993) 3.

\bibitem{Dittmaier:1993da}
S.~Dittmaier and M.~B\"ohm,
\newblock Nucl. Phys. {\bf B412} (1994) 39.

\bibitem{Giele:1991vf}
W.~T. Giele and E.~W.~N. Glover,
\newblock Phys. Rev. {\bf D46} (1992) 1980.

\bibitem{Giele:1993dj}
W.~T.~Giele, E.~W.~N.~Glover and D.~A.~Kosower,
\newblock Nucl. Phys.  {\bf B403} (1993) 633.




\bibitem{Wackeroth:1996hz}
D.~Wackeroth and W.~Hollik,
\newblock Phys. Rev. {\bf D55} (1997) 6788 [hep-ph/9606398].

\bibitem{Baur:1998kt}
U.~Baur, S.~Keller, and D.~Wackeroth,
\newblock Phys. Rev. {\bf D59} (1999) 013002 [hep-ph/9807417].

\bibitem{Catani:1996vz}
S.~Catani and M.~H. Seymour,
\newblock Nucl. Phys. {\bf B485} (1997) 291 [hep-ph/9605323].


\bibitem{Denner:2000bj}
A.~Denner, S.~Dittmaier, M.~Roth, and D.~Wackeroth,
\newblock Nucl. Phys. {\bf B587} (2000) 67 [hep-ph/0006307].




\bibitem{Yennie:1961ad}
D.~R. Yennie, S.~C. Frautschi, and H.~Suura,
\newblock Ann. Phys. {\bf 13} (1961) 379.


\bibitem{Glover:1993xc}
E.~W.~N. Glover and A.~G. Morgan,
\newblock Z. Phys. {\bf C62} (1994) 311.

\bibitem{Poulsen:2006}
A.~Gehrmann-De Ridder, T.~Gehrmann and E.~Poulsen,
  Phys.\ Rev.\ Lett.\  {\bf 96} (2006) 132002
  [hep-ph/0601073];
  Eur.\ Phys.\ J.\  C {\bf 47} (2006) 395
  [hep-ph/0604030];\\
E.~Poulsen,
\newblock Diploma Thesis, University of Zurich  (2006).


\bibitem{Dittmaier:1999mb}
S.~Dittmaier,
\newblock Nucl. Phys. {\bf B565} (2000) 69 [hep-ph/9904440].

\bibitem{Dittmaier:2008md}
S.~Dittmaier, A.~Kabelschacht, and T.~Kasprzik,
\newblock Nucl. Phys. {\bf B800} (2008) 146, [arXiv:0802.1405 [hep-ph]].


\bibitem{GehrmannDeRidder:1997wx}
  A.~Gehrmann-De Ridder, T.~Gehrmann and E.~W.~N.~Glover,
  Phys.\ Lett.\  B {\bf 414} (1997) 354
  [hep-ph/9705305];\\
A.~Gehrmann-De Ridder and E.~W.~N.~Glover,
  Nucl.\ Phys.\  B {\bf 517} (1998) 269
  [hep-ph/9707224].


\bibitem{aleph}
  D.~Buskulic {\it et al.}  [ALEPH Collaboration],
  Z.\ Phys.\  C {\bf 69} (1996) 365.

\bibitem{Altarelli:1996gh}
{W.~Beenakker et al., in }{Altarelli, G.}, T.~Sj\"ostrand, and F.~Zwirner,
\newblock p.~79,
\newblock REPORT-NUM--CERN-96-01.


\bibitem{Amsler:2008zzb}
Particle Data Group, C.~Amsler {\em et~al.},
\newblock Phys. Lett. {\bf B667} (2008) 1.


\bibitem{Bardin:1988xt}
D.~Y. Bardin, A.~Leike, T.~Riemann, and M.~Sachwitz,
\newblock Phys. Lett. {\bf B206} (1988) 539.


\bibitem{Chetyrkin:2000yt}
  K.G.~Chetyrkin, J.H.~K\"uhn and M.~Steinhauser,
  Comput.\ Phys.\ Commun.\  {\bf 133} (2000) 43
  [hep-ph/0004189].



\end{thebibliography}
\end{document}